\PassOptionsToPackage{pdfpagelabels=false}{hyperref}
\PassOptionsToPackage{nonamebreak}{natbib}   
\documentclass[fleqn,usenatbib]{mnras}
\usepackage{newtxtext,newtxmath}
\usepackage[T1]{fontenc}
\usepackage{ae,aecompl}

\usepackage{graphicx}	
\usepackage{amsmath}	
\usepackage{amssymb}	
\usepackage{subfigure}

\newcommand\altaffilmark[1]{$^{#1}$}
\newcommand\altaffiltext[1]{$^{#1}$}

\defcitealias{ElBadry_2017}{E18}

\title[APOGEE MS Binaries]{Discovery and Characterization of 3000+ Main-Sequence Binaries from APOGEE Spectra}

\author[El-Badry et al.]{
\parbox[t]{\textwidth}{ 
Kareem El-Badry\thanks{E-mail: kelbadry@berkeley.edu}\altaffilmark{1,2},
Yuan-Sen Ting\altaffilmark{3,4,5,6}, 
Hans-Walter Rix\altaffilmark{2},
Eliot Quataert\altaffilmark{1},
Daniel R. Weisz\altaffilmark{1}, 
Phillip Cargile\altaffilmark{7},
Charlie Conroy\altaffilmark{7},
David W. Hogg\altaffilmark{2,8,9,10},
Maria Bergemann\altaffilmark{2},
and Chao Liu\altaffilmark{9}
} 
\vspace*{6pt} \\
\altaffiltext{1}{Department of Astronomy and Theoretical Astrophysics Center, University of California Berkeley, Berkeley, CA 94720, USA} \\
\altaffiltext{2}{Max Planck Institute for Astronomy, D-69117 Heidelberg, Germany} \\
\altaffiltext{3}{Research School of Astronomy \& Astrophysics, The Australian National University, Canberra ACT 0200, Australia} \\
\altaffiltext{4}{Institute for Advanced Study, Princeton, NJ 08540, USA} \\
\altaffiltext{5}{Observatories of the Carnegie Institution of Washington, Pasadena, CA 91101, USA} \\
\altaffiltext{6}{Department of Astrophysical Sciences, Princeton University, Princeton,
NJ 08544, USA} \\
\altaffiltext{7}{Department of Astronomy, Harvard University, Cambridge, MA, 02138, USA} \\
\altaffiltext{8}{Center for Computational Astrophysics, Flatiron Institute, New York, NY 10010, USA} \\
\altaffiltext{9}{Center for Cosmology and Particle Physics, Department of Physics, New York University, New York, NY 10003, USA} \\
\altaffiltext{10}{Center for Data Science, New York University, New York, NY 10011, USA} \\
\altaffiltext{11}{Kay Lab of Optical Astronomy, National Astronomical Observatories, Chinese Academy of Sciences, Beijing 100012, China} \\
}

\date{Accepted to MNRAS, January 26, 2018}
\pubyear{2018}

\begin{document}
\label{firstpage}
\pagerange{\pageref{firstpage}--\pageref{lastpage}}
\maketitle

\begin{abstract}
We develop a data-driven spectral model for identifying and characterizing spatially unresolved multiple-star systems and apply it to APOGEE DR13 spectra of main-sequence stars. Binaries and triples are identified as targets whose spectra can be significantly better fit by a superposition of two or three model spectra, drawn from the same isochrone, than any single-star model. From an initial sample of $\sim$20,000 main-sequence targets, we identify $\sim$2,500 binaries in which both the primary and secondary star contribute detectably to the spectrum, simultaneously fitting for the velocities and stellar parameters of both components. We additionally identify and fit $\sim$200 triple systems, as well as $\sim$700 velocity-variable systems in which the secondary does not contribute detectably to the spectrum. Our model simplifies the process of simultaneously fitting single- or multi-epoch spectra with composite models and does not depend on a velocity offset between the two components of a binary, making it sensitive to traditionally undetectable systems with periods of hundreds or thousands of years. In agreement with conventional expectations, almost all the spectrally-identified binaries with measured parallaxes fall above the main sequence in the color-magnitude diagram. We find excellent agreement between spectrally and dynamically inferred mass ratios for the $\sim$600 binaries in which a dynamical mass ratio can be measured from multi-epoch radial velocities. We obtain full orbital solutions for 64 systems, including 14 close binaries within hierarchical triples. We make available catalogs of stellar parameters, abundances, mass ratios, and orbital parameters.
\end{abstract}



\section{Introduction}
About half of solar-type stars are in binary or higher order multiple-star systems \citep{Raghavan_2010, Moe_2017}. Beyond the Solar neighborhood, most binaries are too close on the sky to be spatially resolved; they appear as single photometric point sources, and both components of binary systems contribute to the spectra observed by spectroscopic surveys.

Spectroscopically identifying such unresolved binaries is straightforward only if the period is relatively short ($P\lesssim 5$ years). In this case, spectra exhibit split or ``double'' lines if the two components have comparable luminosities (so-called ``SB2'' systems), and two peaks can be identified in the cross-correlation function \citep{Pourbaix_2004, Fernandez_2017, Merle_2017}. Even if the secondary is faint and does not contribute significantly to the spectrum, short-period binaries can be identified from radial velocity variability when multi-epoch spectra are available \citep[``SB1'' systems;][]{Minor_2013, Troup_2016, PriceWhelan_2017, Badenes_2017}.

However, about half of solar-type binaries have periods exceeding 200 years \citep{Duquennoy_1991, Duchene_2013}. The typical line-of-sight velocity separation between the two stars in such systems is of order $1\,\rm km\,s^{-1}$, while the typical change in the stars' individual velocities over a one year baseline is of order $0.01\,\rm km\,s^{-1}$. Such systems will be missed by binary-detection methods based on the Doppler shift. 

Unresolved binarity in main-sequence stars presents both a nuisance and an opportunity for spectroscopic surveys of the Milky Way. Because spectral morphology is a strong function of effective temperature, contamination from a cooler secondary star\footnote{We adopt the convention that the secondary is the less-massive of the two stars \citep[e.g.][]{Duchene_2013}. For the equal-age, equal-composition main-sequence binaries that we model, the secondary is always cooler and less luminous.} makes the observable spectrum of an unresolved binary different from that of the primary, and in many cases, different from that of any single star. This means that, if binarity is ignored and all spectra are simply fit with single-star models, biases can be introduced in the stellar parameters and abundances inferred for unrecognized binaries \citep{ElBadry_2017}.

On the other hand, binarity-induced features in stellar spectra can be exploited to detect binaries that could not be detected based on velocity shifts alone: binaries can be identified as systems whose spectrum can be significantly better fit by a binary spectral model (i.e., a sum of two single-star models) than any single-star model. This approach, if it can successfully be applied to large spectroscopic surveys, will make possible systematic study of the Galactic binary population on an unprecedented scale. 

\citet[][hereafter E18]{ElBadry_2017} recently demonstrated that fitting a binary model to \textit{synthetic} APOGEE-like spectra makes it possible to spectroscopically identify many binaries and to simultaneously recover the atmospheric parameters and abundances of both component stars. In this paper, we apply the method described in \citetalias{ElBadry_2017} to real spectra from DR13 of the APOGEE survey \citep{Majewski_2017}. We focus on main-sequence stars, for which the effects of unresolved binarity on the spectrum are typically larger than in giants. We demonstrate that, although the spectral signatures of binarity are strongest in close systems with a large velocity offset between the two stars, binaries with mass ratios $0.4 \leq q \leq 0.8$ can be detected with high fidelity even in the absence of any detectable velocity offset (where $q=m_2/m_1$).    

This paper is organized as follows. We describe our spectral model for single and binary stars in Section~\ref{sec:methods} and its application to the combined APOGEE spectra in Section~\ref{sec:combined}. In Section~\ref{sec:visit}, we extend the model to fit multi-epoch spectra of close binaries with detectable velocity changes between visits, calculating dynamical mass ratios from the relative velocities of the two components. We identify and derive parameters for close binaries, triples, and systems with unusual velocity shifts in Section~\ref{sec:visit_results} and derive orbital solutions for the subset of binaries with sufficient visits and phase coverage in Section~\ref{sec:orbits}. We discuss our results and conclude in Section~\ref{sec:discussion}. 

We provide many of the underlying model details in the Appendices.  Specifically, in Appendix~\ref{sec:neural_network}, we describe the spectral model; model selection and tests with semi-empirical synthetic binary spectra are described in Appendix~\ref{sec:model_selection}; shortcomings of the model and false positives are discussed in Appendix~\ref{sec:false_positive}, and diagnostics of orbit-fitting convergence are presented in Appendix~\ref{sec:orbit_convergence}. Available catalogs are described in Appendix~\ref{sec:data}.

\section{Methods}
\label{sec:methods}
Our binary spectral model depends on two steps: (a) creating a data-driven generative model for single-star spectra (Section~\ref{sec:single_star}), and (b) combining the spectra of two single star models, with a suitable velocity offset (Section~\ref{sec:binary_spectra}). To find candidate binaries, we fit spectra with both single-star and binary models (Section~\ref{sec:fitting}) and identify systems that can be significantly better fit by a binary model (Appendix~\ref{sec:model_selection}).

In this work, we only attempt to fit main sequence stars; i.e., targets with $\log g \geq 4$. We do not attempt to identify binaries in which one star is a giant because in most giant-dwarf binaries, the dwarf secondary will contribute a negligible fraction of the total light, while in giant-giant binaries, two components with the same age will necessarily have similar masses, and thus, quite similar spectra. We note that short-period binaries containing giants can be straightforwardly detected from radial velocity variability \citep{Troup_2016, Badenes_2017}, and some giant-subgiant binaries can likely be detected spectroscopically (Section~\ref{sec:prev_work}). 

\subsection{Single-star spectral model}
\label{sec:single_star}
We model APOGEE spectra of single stars using a data-driven\footnote{We also experimented with using synthetic, ab-initio spectral models, but we found them ill-suited for identifying binaries because systematic shortcomings in synthetic models cause almost \textit{all} spectra to be significantly better fit by a sum of two models than a single model.} generative model to predict the rest-frame normalized flux density at a given wavelength as a function of a set of ``labels,'' $\vec{\ell}$, which determine the spectrum. Our approach is very similar to that employed by \textit{The Cannon} \citep{Ness_2015}: the spectral model is a fitting function that maps labels to normalized spectra, and the free parameters of this fitting function are determined by optimization on a training set, whose labels are obtained separately or known a priori, e.g., from ab-initio fitting.

The primary difference between our method and existing implementations of \textit{The Cannon} is that, as in \citet{Ting_2017b} and \citetalias{ElBadry_2017}, we model the normalized flux density at a particular wavelength pixel using an artificial neural network rather than a polynomial function. We find a neural network model to be more flexible than a polynomial and to typically produce smaller errors in model spectra during cross-validation; this formalism, which we refer to as \textit{The Payne}, is described further in Appendix~\ref{sec:neural_network} and will be explained in detail in Ting et al. (in prep). The full spectral model then consists of all the individual neural networks for all wavelength pixels stitched together.

We predict rest-frame spectra with a single-star model that depends on five labels,
\begin{equation}
\label{eqn:single}
\vec{\ell}=\left(T_{{\rm eff}},\log g,\left[{\rm Fe/H}\right],\,{\rm \left[Mg/Fe\right]},v_{{\rm macro}}\right).
\end{equation}
We use [Mg/Fe] as a proxy for all ``$\alpha-$elements.'' We experimented with including more elemental abundances as labels, including C, N, O, and Si. We found that this did not substantially change our identification of likely binary targets or their inferred mass ratios, so we opted to use a relatively simple model in the interests of reduced complexity. $v_{\rm macro}$ primarily accounts for the effects of stellar rotation, and is small ($< 10\,\rm km\,s^{-1}$) for most stars with $T_{\rm eff} \lesssim 6000\, \rm K$. In practice, spectra are not observed in the rest frame, so an additional label $v_{\rm Helio}$ also determines the model spectrum and must be included in fitting. However, our neural network model always predicts spectra in the rest frame; Doppler shifts are applied subsequently.

An ideal training set would contain only stars known to be single a priori. Unfortunately, it is nearly impossible to conclusively rule out the possibility that an unresolved system is a binary.\footnote{The only exception is in the immediate Solar neighborhood ($d\lesssim 8\,\rm pc$), where a combination of direct imaging and speckle interferometry can resolve nearly all systems where a velocity offset is not detectable \citep{Simons_1996, Reid_1997}. However, there are only $\sim 66$ stars in the Solar neighborhood for which binarity can be ruled out with high confidence; of these, only the Sun and Arcturus have been observed by APOGEE.} We therefore construct a training set by beginning with a random sample of main sequence APOGEE stars and then iteratively removing stars whose spectra can be significantly better fit by the binary model described in Section~\ref{sec:binary_spectra}. The ASPCAP pipeline does not derive reliably calibrated abundances for dwarfs. ``Ground truth'' labels for stars in the training set were derived from ab-initio fitting with single-star models, following a procedure similar to that used by \citet{Ting_2017b}; see Ting et al. (in prep) for details. For the initial training set, we randomly selected 2000 targets distributed throughout the region of label space within which a spectral model was desired, namely $4200\,{\rm K} < T_{\rm eff} < 7000\,{\rm K}$, $4.0 < \log g < 5.0$, $-1 < \rm [Fe/H] < 0.5$, $-0.4 < \rm [Mg/Fe] < 0.6$, and $0\, {\rm km\,s^{-1}} < v_{\rm macro} < 45\,{\rm km\,s^{-1}}$. We only attempt to fit targets for which the labels determined from ab-initio fitting lie within this region of parameter space, as (a) we are only interested in main-sequence stars, and (b) the labels determined from ab-initio fitting are less reliable outside this range (Ting et al., in prep). 

There is of course no guarantee that the targets in our initial training set are actually single stars. After training the initial model, we therefore fit all spectra in the training set both with the initial single-star model and a binary model (as described in Section~\ref{sec:binary_spectra}) based on this single-star model. We then removed from the training set the $\sim$300 targets that could be significantly better fit by a binary-model than a single-star model\footnote{Here, we quantified ``significantly better fit'' as having $\chi^2_{\rm single\,star} - \chi^2_{\rm binary}>1000$. We develop a more detailed threshold for model selection in Appendix~\ref{sec:model_selection}.} and re-trained the single-star model on the resulting ``cleaned'' training set. We repeated this cleaning and re-training procedure until none of the targets in the training set could be significantly better fit by a binary model. This approach converges quickly: after the second iteration, fewer than 10 targets in the cleaned training set could be significantly better fit by a binary model; after the third iteration, no additional targets in the training set could be significantly better fit by a binary model.

This iterative cleaning procedure likely does not remove all unresolved binaries from the training set: only binaries whose combined spectrum is significantly different from any single-star star spectrum can be identified. For APOGEE-like spectra of solar-type stars with negligible velocity offsets, the range of mass ratios over which binarity is detectable is $0.4 \lesssim q \lesssim 0.85$ \citepalias{ElBadry_2017}. Binaries in the training set with mass ratios outside this range will not contaminate the spectral model, since their spectra are not significantly different from the spectrum of a single star with the labels of the primary. 

Our approach would likely not work if binaries dominated the training set, or if the functional form of the spectral model were sufficiently complex to incorporate spectral features due to binarity in the single-star model. Because binaries with spectra that are significantly better fit by a binary model constitute only $\sim$15\% of the initial training set and the spectral model is not very complex (we use a small neural network with only 1 hidden layer of 5 neurons), detectable binary spectra are essentially treated as outliers and removed during iterative cleaning, preventing the model from overfitting the signature of unresolved binarity into the single-star model. 

\subsection{Binary spectral model}
\label{sec:binary_spectra}
We assume that both components of a binary system have the same age and composition. Fitting a binary model thus adds three free parameters compared to the single-star model: the mass ratio, $q=m_2/m_1$, which determines $T_{\rm eff}$ and $\log g$ of the secondary, and $v_{\rm macro}$ and $v_{\rm Helio}$ of the secondary. To model the normalized spectrum of a binary with a particular mass ratio, we estimate $T_{\rm eff}$ and $\log g$ of the secondary using MIST isochrones \citep{Choi_2016},\footnote{In practice, we predict $T_{\rm eff}$ and $\log g$ of the secondary from $T_{\rm eff}$ and $\log g$ of the primary, $\rm [Fe/H]$, and $q$ using a neural network trained on a large grid of binary isochrones with $0.01\,\leq{\rm (age/Gyr)}\leq13.5$ and $-1\leq\left[{\rm Fe/H}\right]\leq0.5$. We have verified through cross validation that typical errors in the thus-estimated parameters of the secondary are small ($\sim$20 K in $T_{\rm eff}$ and $\sim$0.01 dex in $\log g$).} predict the single-star spectra of the primary and secondary in unnormalized space, apply a Doppler shift, add the two spectra, and finally pseudo-continuum normalize the total spectrum; see \citetalias{ElBadry_2017} for details. 

Since the data-driven model for single stars operates on normalized spectra, predicting unnormalized spectra for the primary and secondary requires a model for the pseudo-continuum by which the normalized spectra can be multiplied. We obtain the pseudo-continuum for a single star at a particular point in label space by applying our pseudo-continuum fitting procedure (see Section~\ref{sec:fitting}) to a spectrum produced by a synthetic spectral model trained on Kurucz spectra \citep{Kurucz_1970, Kurucz_1979, Kurucz_1993}. Synthetic spectra are first produced with units of surface flux density and are then multiplied by the surface area of the star in question, using radii estimated from \texttt{MIST} isochrones. The unnormalized flux density of an unresolved binary system viewed from a distance $D$ is then given by 
\begin{align}
f_{\lambda,{\rm binary}}=\frac{1}{D^{2}}\left(R_{1}^{2}f_{\lambda,1}+R_{2}^{2}f_{\lambda,2}\right),
\end{align}
where $R_1$ and $R_2$ represent the radii of the primary and secondary star, and $f_{\lambda,1}$ and $f_{\lambda,2}$ represent their individual flux densities. Because we subsequently normalize $f_{\lambda,\,{\rm binary}}$ prior to fitting, the distance $D$ is an arbitrary scaling factor and does not enter our analysis. In practice, $R_1$ and $R_2$ are estimated from $T_{\rm eff}$, $\log g$, and $\rm [Fe/H]$ using a neural network trained on a large grid of \texttt{MIST} isochrones.

Our results are not sensitive to the choice of synthetic model spectra, which sets only the relative flux contribution of the primary and the secondary, because the total binary spectrum is again normalized prior to fitting. We have verified that we obtain similar results by simply defining a continuum for each star as a blackbody with appropriate $T_{\rm eff}$ scaled by the surface area of the star. 

For long-period systems with negligible velocity shifts, our model cannot detect binaries with mass ratios $q \lesssim 0.4$, because the secondary contributes a negligible fraction of the total light, or $q \gtrsim 0.85$, because the spectra of the primary and secondary are too similar. In practice, another, often more stringent limit on the lowest detectable mass ratio is set by our spectral model's minimum $T_{\rm eff}$ of $4200\,\rm K$. For systems with a hot primary star ($T_{\rm eff} \gtrsim 6500\,\rm K$), this limit is not important, since a secondary with $T_{\rm eff} < 4200\rm\,K$ would be too faint to contribute significantly to the spectrum anyway. However, the model's minimum $T_{\rm eff}$ reduces the range of detectable mass ratios for systems with cooler primaries: for a primary with $T_{\rm eff} = 5800\,\rm K$, the effective minimum $q$ that can be modeled is $q_{\rm min} \approx 0.62$, while for a primary with $T_{\rm eff} = 5000\,\rm K$, $q_{\rm min} \approx 0.75.$ We discuss this further in Appendix~\ref{sec:semi_empirical}.

\subsection{Model fitting}
\label{sec:fitting}
Best-fit labels for binary and single-star models are determined through full-spectrum fitting of normalized spectra in vacuum wavelengths. Pseudo-continuum normalization is carried out using the \textit{Cannon}-type normalization routine from the \texttt{APOGEE} package \citep{Bovy_2016b}, which fits a 4th order Chebyshev polynomial to pixels in which the gradient of the data-driven spectral model with respect to the labels is small. Bad pixels and pixels with poor sky subtraction, as flagged in the bitmasks produced by the APRED pipeline \citep{Nidever_2015}, are masked during normalization and fitting. 

Fitting is carried out using the Scipy \texttt{curve\_fit} routine, which implements the ``trust region reflective'' algorithm \citep{Branch_1999} for $\chi^2$ minimization. When fitting a single spectrum with a single-star model,  we find that the optimization essentially always converges on the true global minimum, irrespective of the location in label space where it is initialized. However, for the binary model there is an obvious degeneracy: the normalized spectrum of a $q=1$ binary model is identical to that of a $q=0$ model in the limit of no velocity offset. Hence, the posterior for a binary model is often bimodal in $q$, and minimization can sometimes converge on a false local minimum. We therefore initialize $\sim$10 separate optimizers with different initial values of $q$ when fitting a binary model. If these do not all converge to the same model, we take as the best model the one that reaches the lowest global $\chi^2$. We have verified by fitting semi-empirical synthetic binary spectra that this approach converges on the true global minimum in $\sim$99\% of all cases (see Appendix~\ref{sec:semi_empirical}).

Most APOGEE targets are observed more than once, with time baselines between individual visits ranging from $\sim$1 hour to $\sim$1200 days.\footnote{The APOGEE observing strategy aims to observe most targets 3 times, over a minimum time baseline of 1 month. Some targets, primarily faint stars, are visited more often to accumulate S/N; some targets in unfavorable locations, such as the Galactic Bulge, are visited only once \citep{Zasowski_2013}. Most targets with baselines longer than 1 year, as well as those with multiple visits within 1 night, are targets which were observed initially during the survey commissioning period and again during the main survey.} Spectra from individual visits are shifted to rest frame and coadded to produce a single combined spectrum with higher S/N than the individual visit spectra by the APSTAR pipeline \citep{Nidever_2015}. It is these combined spectra that are fit by the ASPCAP pipeline to derive the stellar parameters and abundances published for the main survey \citep{Holtzman_2015, GarciaPerez_2016}, but the reduced spectra from individual visits are also made publicly available. 

Combined spectra are easier to work with than individual visit spectra both because they have higher S/N and because stars are often observed with a different fiber and with a different barycentric velocity at each visit, so that the combined spectrum is less affected by bad pixels, poor sky subtraction, and telluric absorption than the individual visit spectra. We therefore fit the combined spectra rather than spectra from individual visits when possible. However, if a system is an unresolved close binary, the orbital configuration and relative radial velocities of the primary and secondary will change between visits, so that the morphology of the total binary spectrum is different in each visit. In such cases, the combined spectrum does not represent any real physical system, and fitting it can yield biased labels.

For this reason, we attempt to fit all targets that may be close binaries using the individual visit spectra rather than the combined spectrum. We identify potential close binaries as targets for which (a) the best-fit model to the combined spectrum is a binary model in which the line-of-sight velocity separation of the two components, $\Delta v_{\rm los},$ is greater than $10\,\rm km\,s^{-1}$, or (b) the $V_{\rm scatter}$ term calculated from the radial velocities determined by the APSTAR pipeline \citep{Nidever_2015} is greater than $1\, \rm km\,s^{-1}$, indicating potential radial velocity variability. Some of these targets, particularly stars with high $T_{\rm eff}$ or low S/N, are single stars with poorly constrained radial velocities, but many are close binary systems. Fitting individual visit spectra for targets with $V_{\rm scatter} > 1\,\rm km\,s^{-1}$ also protects against the possibility of a single star erroneously appearing to be a binary if the radial velocities are calculated incorrectly while creating the combined spectrum; otherwise, coadding two visit spectra with different Doppler shifts could produce a combined spectrum bearing erroneous signatures binarity with $q=1$. 

The number of free parameters to be optimized increases substantially when we fit spectra from many visits simultaneously, since the radial velocities at each visit are all free parameters. This can make the fit more susceptible to convergence on an erroneous local minimum in $\chi^2$; we discuss the measures taken to ensure global convergence in this case in Section~\ref{sec:visit}. 

For both single-visit and combined spectra, we inflate the uncertainties of pixels with $\rm S/N > 200$ to 0.5\% (i.e., S/N of 200) during fitting because empirical S/N diagnostics based on the variation in a given pixel across visits show that the noise model underestimates uncertainties for bright stars and is likely limited by systematics at this level \citep{Nidever_2015}. We also find that our fitting approach often performs poorly at low S/N, primarily due to poor continuum normalization. We therefore do not attempt to fit any visit spectra with median S/N $< 30\,\rm pixel^{-1}$. Since most APOGEE targets are bright, this restriction excludes less than 20\% of the targets in our sample; for these targets, we report labels obtained by fitting the combined spectrum, which has higher S/N, but we caution that results for targets with large $V_{\rm scatter}$ and low S/N are likely less reliable. 

We do not report uncertainties on labels for individual targets. Formal fitting uncertainties based on the concavity of the likelihood function in the vicinity of the global maximum can be computed with \texttt{curve\_fit} \citep[e.g.][]{Ness_2015, Ho_2017}, and comparable uncertainties can be obtained by MCMC sampling. However, the thus-obtained uncertainties are typically unrealistically small for high- S/N spectra (e.g., $\sigma(T_{\rm eff}) < 10\,\rm K$ for typical APOGEE spectra) because they do not properly account for systematic errors in the spectral model. Systematic errors cam arise if (a) the spectral model is not sufficiently complex to account for all the variance in the dataset, (b) there are unaccounted-for errors in the labels assigned to the training set, or (c) the adopted set of labels does not fully characterize all the variance in the dataset. We investigate the typical precision of our best-fit labels in Appendix~\ref{sec:cross_validation}.

\subsection{Fitting Multi-Epoch Spectra}
\label{sec:visit}

We attempt to fit the individual visit spectra rather than the combined spectra of all stars that were visited more than once and are flagged as potential close binaries. In order to fully exploit the  information contained in the spectra, we fit all single-visit spectra for each system \textit{simultaneously}, requiring the physical parameters of the component stars to be the same at all epochs. Because we fit all visit spectra with the same spectral model, we implicitly treat the instrumental line spread function as constant across all fibers and visits. For the single-star model, we also require the line-of-sight velocity to be the same at each epoch; in this case, the model is no more complex than when fitting a single combined spectrum.

For an isolated binary system, the line-of-sight velocities of the two components are not independent: in the center-of-mass frame, conservation of linear momentum requires that the radial velocity of the primary along any line of sight, $v_{1}$, and that of the secondary, $v_{2}$, are related by $v_{2} = -v_{1}/q_{\rm dyn}$, where $q_{\rm dyn}$ is the dynamical mass ratio of the system. If the center-of-mass heliocentric velocity of the binary is $\gamma$, then 
\begin{align}
\label{eqn:vr1_vr2}
v_{{\rm Helio,}2}=\gamma+\left(\gamma-v_{{\rm Helio,1}}\right)/q_{{\rm dyn}}.
\end{align}
Here $v_{\rm Helio}$ denotes a velocity at a single epoch, measured in the frame of the center of mass of the Solar system.

For true, isolated binary systems containing two main sequence stars, $q_{\rm dyn}$ should be equal to the \textit{spectral} mass ratio $q$, which determines the contribution of the secondary star to the binary spectrum. We will use $q$ and $q_{\rm spec}$ interchangeably in the rest of this paper. However, we fit $q_{\rm dyn}$ and $q_{\rm spec}$ separately to allow for the possibility of companions whose contribution to the spectrum is different from what is predicted by the dynamical mass ratio. This could occur, for example, if there are biases in the isochrones used in the spectral model, if the secondary falls near the edge of the APOGEE fiber and only a fraction of its flux contributes to the spectrum, or if a third object is present in the system. Comparing the best-fit $q_{\rm dyn}$ and $q_{\rm spec}$ provides a useful diagnostic of the accuracy of our spectral model.  

Our basic ``SB2'' binary model does not allow the velocities of both stars to vary freely, but instead enforces the restriction that the velocities at all epochs follow Equation~\ref{eqn:vr1_vr2} when two or more visit spectra are fit simultaneously. In most cases, this leads to best-fit velocities that are similar (within $\sim 200\,\rm m\,s^{-1}$ on average, and nearly always within a few $\rm km\,s^{-1}$) to those obtained when Equation~\ref{eqn:vr1_vr2} is not enforced. However, there are some targets for which the best-fit velocities are very different -- and produce a much better fit -- when Equation~\ref{eqn:vr1_vr2} is not enforced than when it is. Such systems have velocities inconsistent with being a simple two-body system and likely contain a third component. To avoid mischaracterizing these systems, we also fit all targets with a binary model in which the velocities of both components are allowed to vary freely; systems that are significantly better-fit by this model are classified as SB2s with an unseen third component (see Section~\ref{sec:summary_labels} for details).

We also find systems in which there is a clear radial velocity shift in the spectrum between different visits but no individual visit spectrum is better-fit by a binary model; i.e., the existence of a companion can be inferred from its gravitational effects on the primary, but the companion does not significantly contribute to the observed spectrum. Most of these single-line binary (``SB1'') systems are probably ordinary main-sequence binaries with low mass ratios and relatively short periods; some are likely binaries in which the companion is a stellar remnant. To distinguish between SB1s and SB2s, we fit all potential close binary systems with an SB1 model, which is identical to the single-star model, except that the radial velocity is allowed to vary between visits. We designate systems as SB1s if the SB1 and SB2 models converge on essentially the same fit; i.e., if there is no detectable contribution to the spectrum from the secondary. 

Finally, we find some systems whose visit spectra cannot be well-fit by any single star or binary model: the binary model provides a better fit than the single-star model, but many lines are poorly fit or are missing entirely from the best-fit binary model. We find that many of these systems can be much better fit by a \textit{triple} model: i.e. three stars with independent velocities and masses, restricted to lie on the same isochrone. 

\begin{figure*}
    \includegraphics[width=\textwidth]{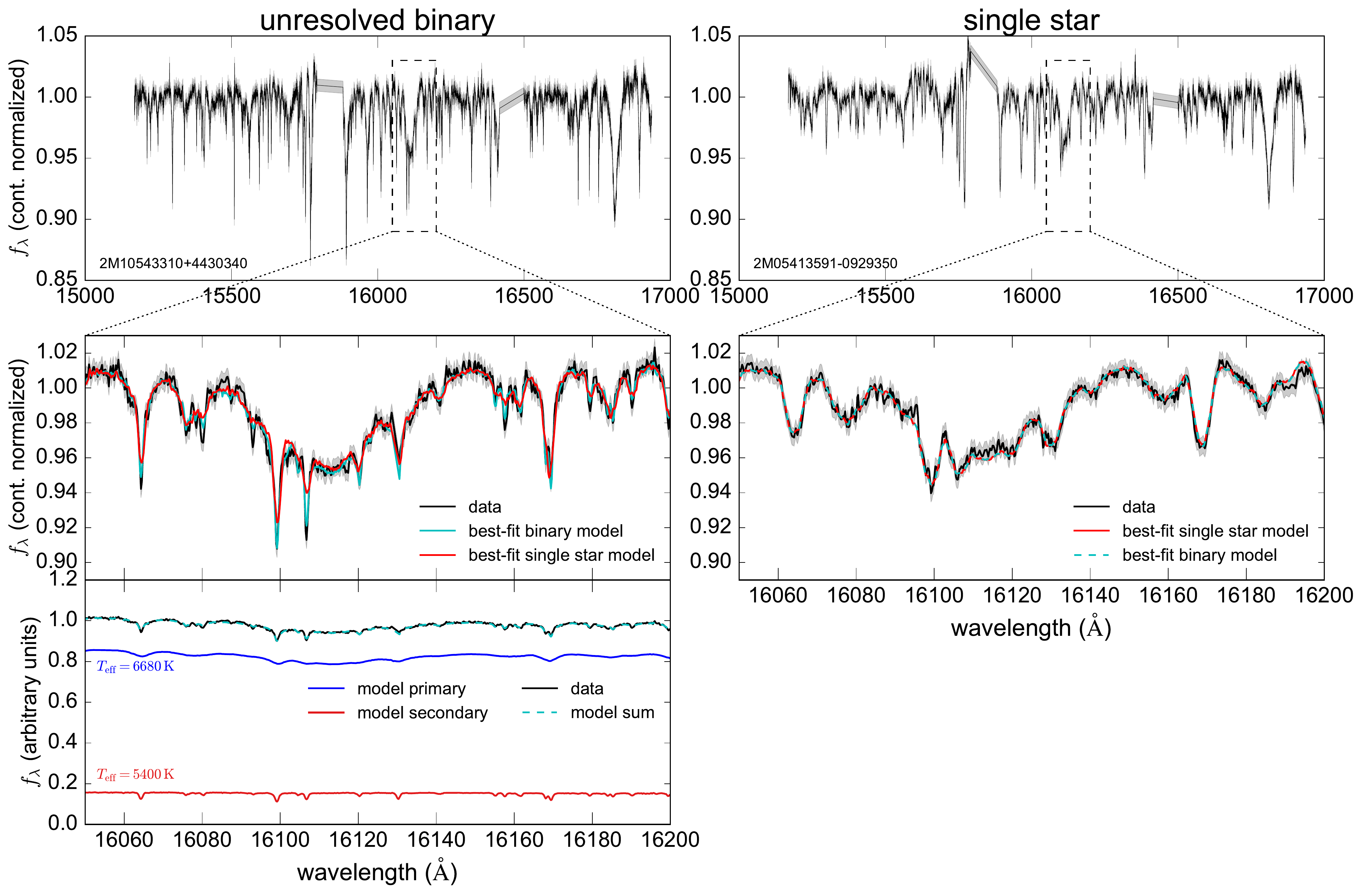}
    \caption{\textbf{Left}: Spectrum of an unresolved main-sequence binary with $q=m_2/m_1 \approx 0.7$ as observed by APOGEE. Top panel shows the full normalized spectrum. Middle panel shows the spectrum and best-fit binary and single-star models, zoomed-in on a narrow wavelength range enclosing a hydrogen Brackett line. The binary model fits the data significantly better than the single-star model. Bottom panel shows the two components of the best-fit binary model. The spectrum's broad features are due primarily to the hotter star, which contributes $>80\%$ of the total light but has no strong narrow lines; the shape of the sharp line profiles is primarily due to the cooler star. Our method makes it possible to identify many long-period binaries like this one, in which the velocity offset between the two stars is negligible.
    \textbf{Right}: Spectrum of a presumed single star with similar parameters to the primary in the system shown in the left panels. In this case, the best-fit binary and single-star models are identical.}
    \label{fig:hotstar}
\end{figure*}

\begin{figure*}
    \includegraphics[width=\textwidth]{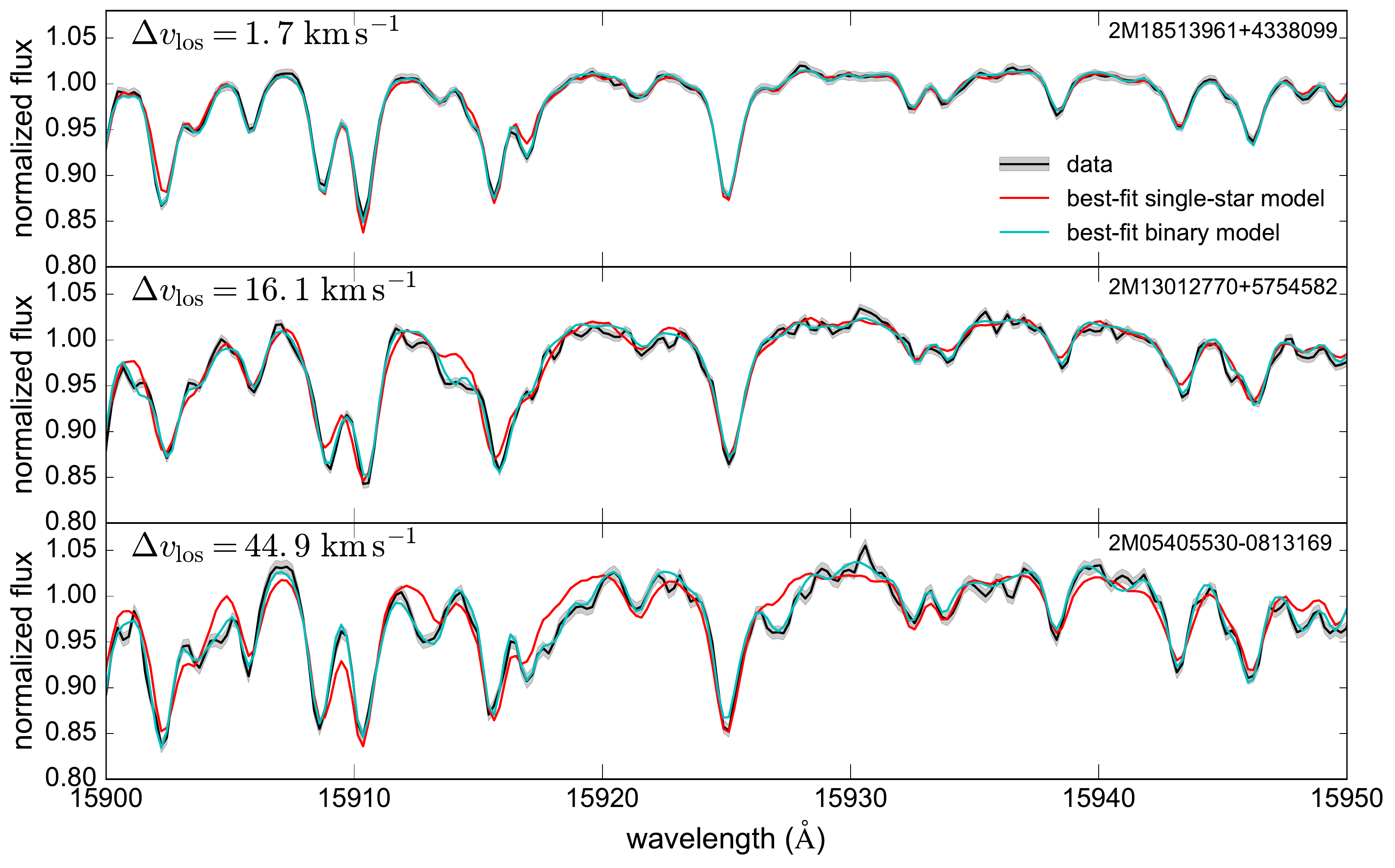}
    \caption{Examples of single-star and binary model fits to binary systems with a negligible (top), intermediate (middle), and large (bottom) line-of-sight velocity offset between the two stars. All three systems have a mass ratio $q=m_2/m_1\sim 0.7$, a primary star with $T_{\rm eff}\sim 5400\,\rm K$ and $\log g \sim 4.5$, and $\rm [Fe/H]\sim 0$. Detecting binarity in systems with a large velocity offset ($\Delta v_{\rm los} \gtrsim 15\,\rm km\,s^{-1}$) is straightforward, because the two stars' lines become separated in velocity space. However, binarity can also be detected in many systems where the line-of-sight velocity offset is negligible, as in the top panel, because the two component stars have different temperatures and ionization states, so their combined spectrum cannot be well-fit by any single-star model.}
    \label{fig:three_dv_specs}
\end{figure*}

\subsubsection{Summary of models fit to visit spectra}
\label{sec:summary_labels}
We simultaneously fit the $N$ visit spectra for each object in the ``potential close binary'' subsample with a total of five different models, which we summarize here. We classify systems based on the total $\chi^2$ of each model, preferring the least complex model when different models have similar $\chi^2$. 

\begin{enumerate}
\item \textit{Single-star}: The single-star model has 6 free parameters, regardless of the number of visit spectra: 
\begin{equation}
\label{eqn:singe_visit_labels}
\vec{\ell}_{{\rm single\,{\rm star}}}=\left(T_{{\rm eff}},\log g,\left[{\rm Fe/H}\right],{\rm \left[Mg/Fe\right],}v_{{\rm macro}}, v_{{\rm Helio}}\right).
\end{equation}
In particular, this model forces the heliocentric velocity of the star to be the same in all visits. 
\item \textit{SB1}: The SB1 model is identical the single-star model, except that the heliocentric velocity is allowed to vary between the $N$ visits. The $5+N$ free parameters are:
\begin{equation}
\label{eqn:SB1_labels}
\vec{\ell}_{{\rm SB1}}=\left(T_{{\rm eff}},\log g,\left[{\rm Fe/H}\right],{\rm \left[Mg/Fe\right],}v_{{\rm macro,}}v_{{\rm Helio},i}\right),
\end{equation}
where $i$ enumerates the visits. 
\item \textit{SB2}: The SB2 model fits two stars, with different velocities at each visit, but with the restriction that the velocity satisfy Equation~\ref{eqn:vr1_vr2}. The $9+N$ free parameters are 
\begin{equation}
\label{eqn:SB2_labels}
\begin{aligned}
\vec{\ell}_{{\rm SB2}}={} & \bigl (T_{{\rm eff}},\log g,\left[{\rm Fe/H}\right],{\rm \left[Mg/Fe\right],}q, \bigr. \\
& \bigl. v_{{\rm macro1}},v_{{\rm macro2}}, q_{{\rm dyn}},\gamma,v_{{\rm Helio1},i} \bigr).
\end{aligned}
\end{equation}

\item \textit{SB2 with unseen 3rd object}: This model fits two stars but allows their velocities to vary freely, without enforcing Equation~\ref{eqn:vr1_vr2}. If it provides a significantly better fit than the SB2 model, the relative radial velocity shifts are inconsistent with being a simple Keplerian two-body system. The $7 + 2N$ free parameters are:
\begin{equation}
\label{eqn:SB2_dv_labels}
\begin{aligned}
\vec{\ell}_{{\rm SB2,\,unseen\,3rd\,object}}={} & \bigl (T_{{\rm eff}},\log g,\left[{\rm Fe/H}\right],{\rm \left[Mg/Fe\right],}q, \bigr. \\
& \bigl. v_{{\rm macro1}},v_{{\rm macro2}}, v_{{\rm Helio1},i},  v_{{\rm Helio2},i} \bigr).
\end{aligned}
\end{equation}

\item \textit{SB3}: The SB3 model fits three stars and imposes no restrictions on their velocities. The $9 + 3N$ free parameters are:
\begin{equation}
\label{eqn:SB3}
\begin{aligned}
\vec{\ell}_{{\rm SB3}}={} & \bigl (T_{{\rm eff}},\log g,\left[{\rm Fe/H}\right],{\rm \left[Mg/Fe\right],}q_2, q_3, v_{{\rm macro1}}, \bigr. \\
& \bigl. v_{{\rm macro2}}, v_{{\rm macro3}}, v_{{\rm Helio1},i}, v_{{\rm Helio2},i}, v_{{\rm Helio3},i} \bigr),
\end{aligned}
\end{equation}
where $q_2 = m_2/m_1$ and $q_3 = m_3/m_1$. 
\end{enumerate}

We note that the SB2 models are in principle identical to the SB1 model (and the SB3 model to the SB2 models) in the limit where $q=0$. We keep these models separate in practice because our model does not transition smoothly from the minimum possible $q$ that can be modeled (corresponding to $T_{\rm eff}=4200\,\rm K$) to $q=0$.

Fitting many visits simultaneously increases the number of labels to be fit, increasing the risk of the optimizer's convergence on a local minimum. For example, for a target with 10 visits, fitting the SB2 (SB3) model requires optimization of the likelihood in 19 (39) dimensions, which is computationally demanding. In tests with synthetic binary spectra, we find that convergence on the global best-fit is nearly always achieved \textit{as long as the optimizer is initialized reasonably close to the global minimum}; i.e., with all velocities within $\pm \sim 20 \,\rm km\,s^{-1}$ of their true values at all epochs. We therefore first fit individual visit spectra one at a time to estimate the velocity of each component at each epoch, and then use the resulting best-fit labels to initialize the global optimizer during simultaneous fitting. Because the velocity offsets at each epoch are nearly uncorrelated with those in other epochs -- i.e., changing $v_{\rm Helio,1,i}$ only shifts the spectrum predicted for the $i$th visit -- the optimization remains tractable in many dimensions.

\section{Results}
\label{sec:combined}
We fit the spectra of 20,142 targets from APOGEE DR13 that ab-initio fitting with single-star models (Ting et al., in prep) found to (a) lie on the main sequence ($\log g > 4$), (b) fall within the region of label space where the synthetic spectral model is reliable ($4200\,{\rm K}\leq T_{\rm eff} \leq 7000\,{\rm K}$ and $-1 \leq \rm [Fe/H] \leq 0.5$), and (c) be acceptably fit, in a $\chi^2$ sense, by synthetic spectral models. From this initial sample, we identify 2645 targets in which more than one star contributes significantly to the spectrum and an additional 663 targets with time-variable radial velocities but no detectable spectral contribution from the secondary. Catalogs of targets classified as single stars, binaries, and triples are presented in Appendix~\ref{sec:data}.

Figure~\ref{fig:hotstar} illustrates how our model identifies systems that are likely binaries but show no significant radial velocity 
variability or split lines due to a velocity offset between the two components. Panels on the left show the spectrum of a target that can be significantly better fit by a binary model than a single-star model; those on the right show one that cannot. 

We fit the full spectrum simultaneously, but we zoom-in on a small region to show the qualitative signatures of binarity. The spectrum in the left-hand panels contains features of both hot and cool stars: wide hydrogen lines and rotationally-broadened line profiles at the wing of all lines, and deeper, narrow line cores that do not show rotational broadening. No single-star model can achieve a good fit: the absorption lines in the best-fit single star model are too shallow, and some lines in the data spectrum are blended in the best-fit single star model or are missing altogether. On the other hand, the binary model can provide a good fit and reproduces the line profiles of the observed spectrum. The decomposition of the binary model spectrum in the bottom panel shows that the broad features are all due to the hot primary star, while the sharper features originate in the spectrum of the cooler secondary. 

In the right panels of Figure~\ref{fig:hotstar}, we show the spectrum of a typical single star with stellar parameters and abundances similar to the primary in the left panels; as expected, it is similar to the spectrum in the left panels with the sharper, narrow lines removed. In this case, the binary and single-star models converge on what is essentially the same spectrum, so there is no reason to prefer the binary model. 

The binary spectrum in the left panels of Figure~\ref{fig:hotstar} illustrates why it is often possible to spectrally identify binaries even when one star is much brighter than the other: although the secondary star in the binary system contributes less than 20\% of the total \textit{light}, it contributes a large fraction of the total \textit{absorption} because lines in hotter stars are often intrinsically weaker than those in cool stars. For many binaries containing a hot primary and cool secondary, the spectrum and binary model exhibit lines that are completely absent from the spectrum of the primary because the relevant species are ionized at its higher $T_{\rm eff}$. 

\subsection{Effect of a velocity offset}
\label{sec:vel_offset}

Although a line-of-sight velocity difference between the primary and secondary star is in many cases not required to identify binaries with our model, a velocity offset makes the signatures of unresolved binarity more obvious and extends the range of detectable mass ratios. This is illustrated in Figure~\ref{fig:three_dv_specs}, which compares the spectra of three binary systems with similar stellar parameters, abundances, and mass ratios, but a range of velocity offsets between the primary and secondary component. The system shown in the top panel has a small line-of-sight velocity offset, similar to the system in the left panels of Figure~\ref{fig:hotstar}. In this case, the effects of binarity are quite subtle, and binarity can likely only be detected with detailed spectral modeling. As the velocity offset increases (middle and bottom panels), binarity-induced changes to the spectrum become more obvious. In all three panels of Figure~\ref{fig:three_dv_specs}, we plot the APSTAR combined spectrum, not the spectra from individual visits. However, the target in the bottom panel, which is the only target of the three for which we might expect a large velocity change between visits, was only visited once.

For APOGEE spectra with $R = 22,500$, one resolution element corresponds to a radial velocity difference of $\delta v\sim c/R \sim 13.5\,\rm km\,s^{-1}$. The traditional method of identifying binaries as systems in which the cross-correlation function of an observed spectrum with a synthetic template exhibits two peaks can only reliably detect binaries in which the line-of-sight velocity offset is of order 1-3 resolution elements; such systems are usually referred to as ``SB2'' systems. For example, \citet{Fernandez_2017} found that binaries could only be reliably detected in APOGEE spectra when the maximum line-of-sight velocity separation exceeded $\Delta v_{\rm los} = 30\,\rm km\,s^{-1}$.\footnote{Other surveys find similar sensitivity to spectroscopic binaries; e.g., \citet{Merle_2017} found that binaries could be detected down to $\Delta v_{\rm los} = 15\,\rm km\,s^{-1}$ in UVES spectra ($R=47000$) from the Gaia-ESO survey, and \citet{Matijevic_2010} found a minimum $\Delta v_{\rm los}$ for reliable detection of $50\,\rm km\,s^{-1}$ in the RAVE survey ($R=7500$).} Figure~\ref{fig:three_dv_specs} shows that even a small velocity offset can substantially strengthen the signatures of binarity. How much a velocity offset improves detectability for our method depends on the stellar parameters and abundances of the primary, because it is easier to detect velocity offsets in stars with many deep, narrow lines. For most stars with $T_{\rm eff}\lesssim 6500\,\rm K$, a velocity offset of $\Delta v_{\rm los}\gtrsim (5-10)\,\rm km\,s^{-1}$ makes it possible to identify binaries from single-epoch spectra even when the mass ratio is close to $q=1$; such systems are not otherwise detectable with our method (see Appendix~\ref{sec:semi_empirical}).

\subsection{Results for multi-epoch spectra with velocity variability}
\label{sec:visit_results}
\begin{figure}
    \includegraphics[width=\columnwidth]{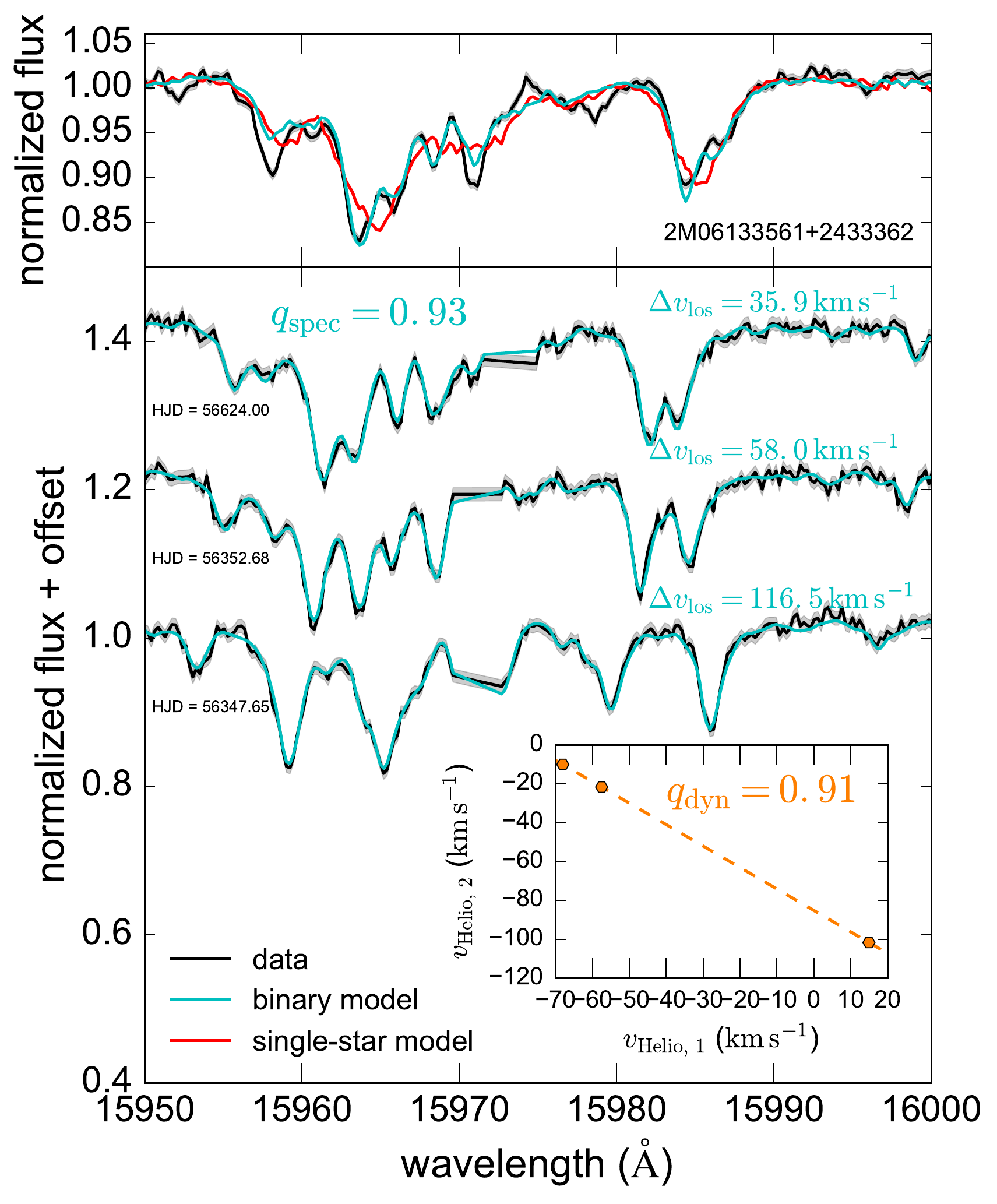}
    \caption{A binary system in which the stars' velocities change between visits. Top panel shows a small portion of the \textit{combined} spectrum (black), which is produced by coadding spectra from different visits, best-fit single-star model (red), and best-fit binary model (cyan). The binary model provides a better fit than the single-star model, but it cannot fully reproduce the combined spectrum. Bottom panel shows the individual visit and the best-fit SB2 model, which produces an excellent match to all the individual visit spectra. Inset shows heliocentric velocities of the primary and secondary star at each epoch; momentum conservation requires that these lie on a line with slope $-1/q_{\rm dyn}$, where $q_{\rm dyn}$ is the dynamical mass ratio. The spectrally-inferred mass ratio, $q_{\rm spec} = 0.93$, is in good agreement with the dynamical mass ratio, $q_{\rm dyn}=0.91$.}
    \label{fig:single_visit}
\end{figure}

\begin{figure*}
\includegraphics[width=\textwidth]{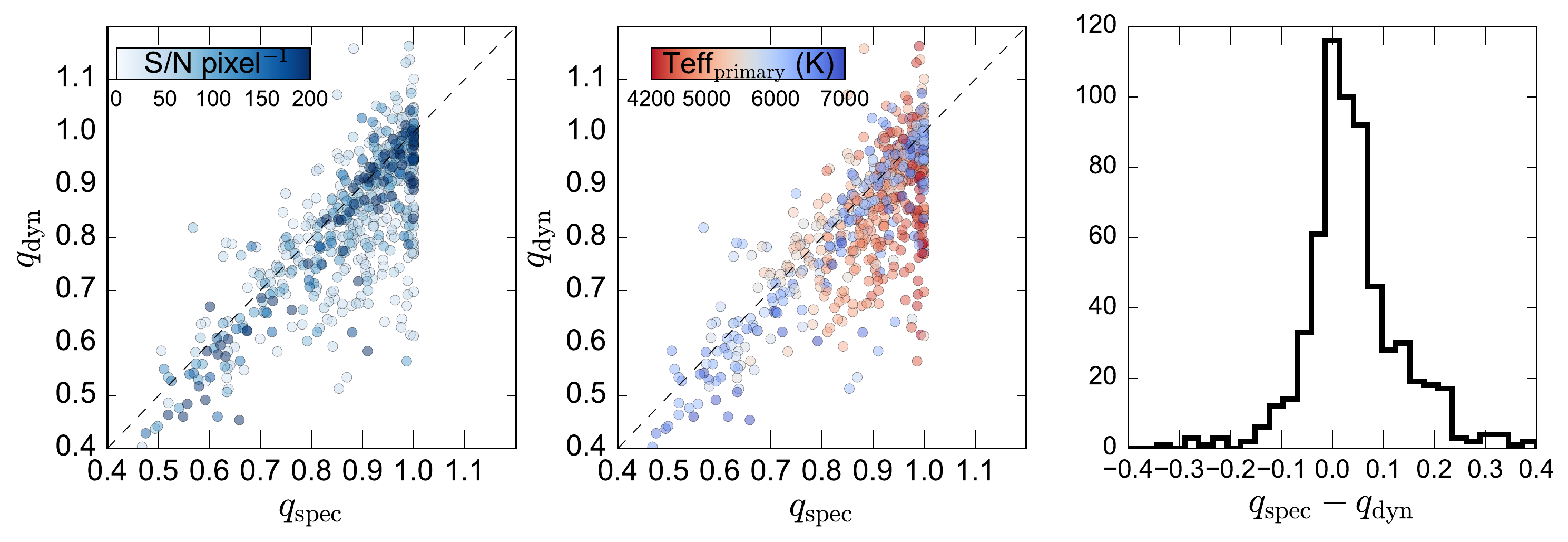}
\caption{Comparison of spectroscopically- and dynamically-inferred mass ratios for ``SB2'' binary systems in which a dynamical mass ratio can be measured. $q_{\rm spec}$ is measured from the relative contribution of each star to the spectrum, and $q_{\rm dyn}$, from the relative changes of the radial velocities of the primary and secondary across multiple epochs (see Figure~\ref{fig:single_visit}). The designation of primary and secondary components is based on their relative contribution to the spectrum: $q_{\rm spec}$ is bounded by 1, but $q_{\rm dyn}$ is not. 623 systems have sufficiently short periods to allow measurement of $q_{\rm dyn}$. Most systems for which the disagreement between $q_{\rm spec}$ and $q_{\rm dyn}$ is large have low S/N (left) and cool primaries (middle). Due to the spectral model's minimum $T_{\rm eff}$ of 4200 K, low mass ratio systems can only be detected if the primary is hot, and mass ratios are less accurate for cooler systems. The median absolute difference between $q_{\rm dyn}$ and $q_{\rm spec}$ is 0.048.}
\label{fig:q_spec_q_dyn}
\end{figure*}

\begin{figure}
  \includegraphics[width=\columnwidth]{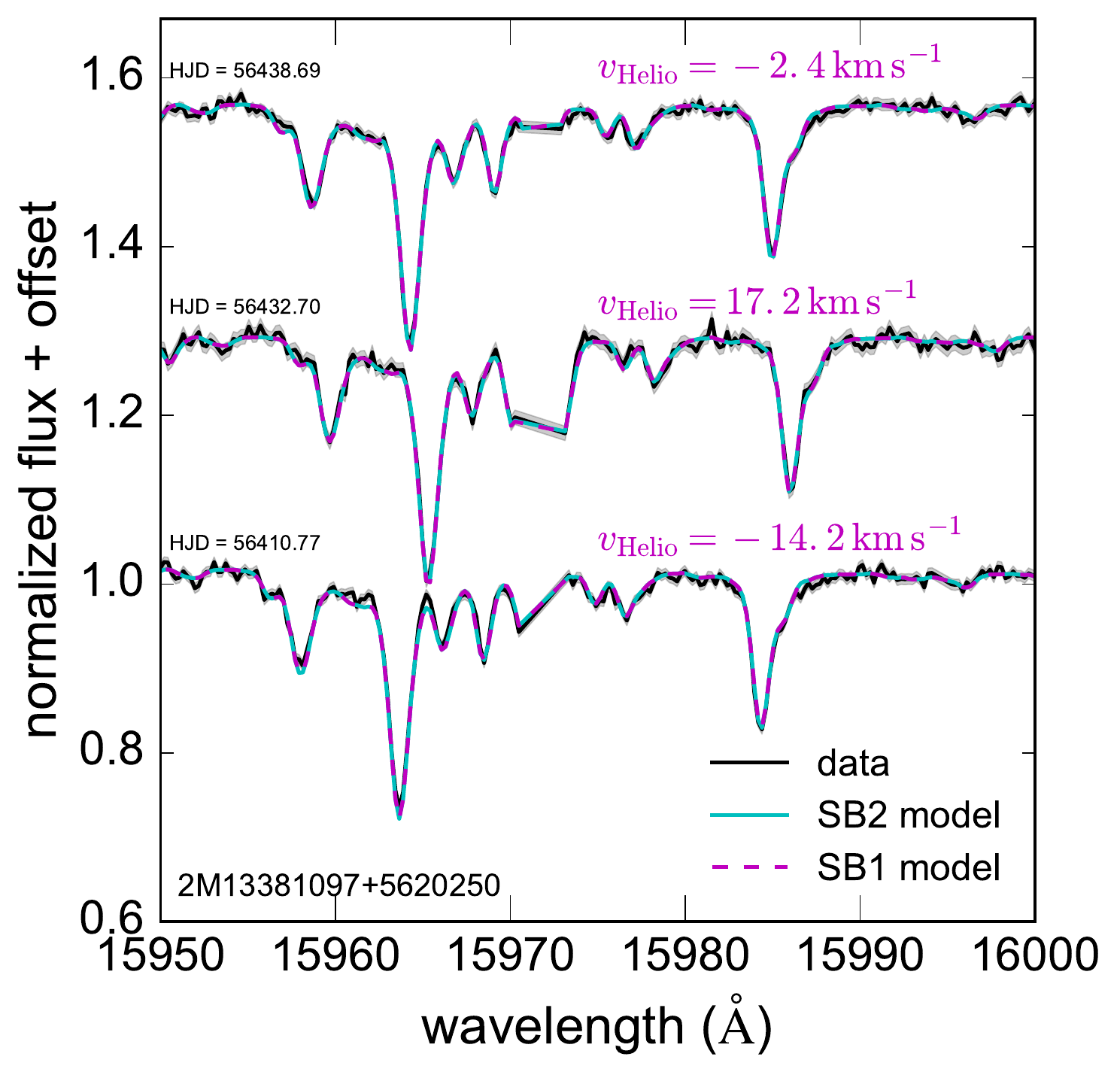}
    \caption{Visit spectra and best-fit models for an SB1 system. The SB1 model contains only a single star contributing to the spectrum, but its radial velocity can vary across visits. The SB2 model includes the possibility of a second star contributing to the spectrum. In this case, the best-fit SB1 and SB2 models are identical, indicating that there is no detectable contribution to the spectrum from the secondary. However, radial velocity variability of the primary clearly indicates that a companion is present. Assuming that the companion is a main-sequence star, an upper-limit of $q \lesssim 0.45$ can derived; if $q$ were larger, the secondary would contribute detectably to the spectrum, and the SB2 model would provide a better fit.}
    \label{fig:sb1}
\end{figure}

\begin{figure}
\includegraphics[width=\columnwidth]{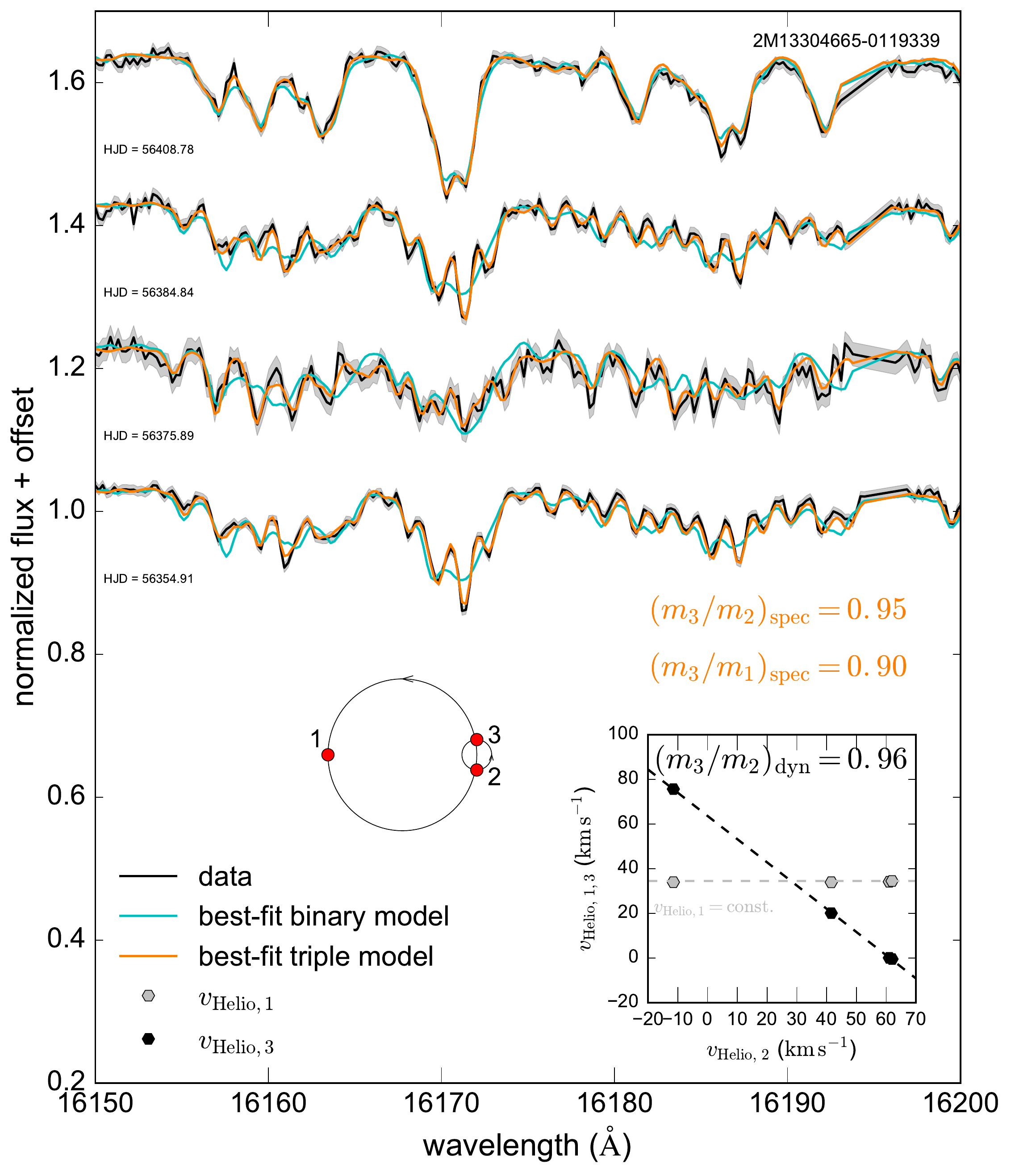}
\caption{Visit spectra of a target identified as a triple (SB3) system. The three components have different line-of-sight velocities, so many lines can be seen in triple, and an SB2 model cannot provide a good fit. Inset shows the line-of-sight velocities of each component at each epoch. The heliocentric velocity of the primary is consistent with being constant at $v_{\rm Helio,1} \approx 34.5 \rm\,km\,s^{-1}$, so no dynamical mass ratio can be estimated for $m_2/m_1$ or $m_3/m_1$. However, $v_{\rm Helio,2}$ and $v_{\rm Helio,3}$ fall on a line with an implied mass ratio $m_3/m_2$ consistent with the spectrally inferred one. This implies that the system is a hierarchical triple, as is illustrated in the orbit schematic.}
\label{fig:triple}
\end{figure}

Examples of targets whose spectra are best-fit by SB2, SB1, SB2 with an unseen 3rd object, and SB3 models are shown in Figures~\ref{fig:single_visit},~\ref{fig:sb1},~\ref{fig:bin_dv}, and~\ref{fig:triple}.

Figure~\ref{fig:single_visit} shows a system that is best-fit by the SB2 model (i.e., case (iii) from Section~\ref{sec:summary_labels}) and exhibits spectra that change substantially from one epoch to the next. In the upper panel, we plot the combined spectrum and the best-fit binary and single-star models obtained by fitting it. Although the binary model is a better fit (and our initial fit to the combined spectrum did flag the system as a likely binary), the fit is not very good: some features in the combined spectrum cannot be accommodated by either the single-star or the binary model. In the lower panel, we show the spectra obtained in the three individual visits, which are coadded to produce the combined spectrum, and the binary model obtained by simultaneously fitting them. The fit to the individual visit spectra is good. The poor fit to the combined spectrum is a consequence of the fact that the components' velocities change between visits, meaning that the combined spectrum is an unphysical superposition of different spectra. 

The inset in Figure~\ref{fig:single_visit} shows the heliocentric velocities of the primary and secondary star at each visit for the best-fit SB2 model. The slope and intercept of the line on which these velocities fall can be used to calculate the dynamical mass ratio, $q_{\rm dyn}$, and the center-of-mass velocity, $\gamma$. For binary systems in which the velocities of the two stars change significantly between visits, it is therefore possible to obtain a constraint on the mass ratio that is independent of the spectral label $q$. Such constraints will of course not be reliable if the orbital configuration does not change significantly between visits: in this case, all measurements of $v_{\rm Helio,1}$ and $v_{\rm Helio,2}$ will be clustered around one point, and the slope of the line is ill-constrained. We also emphasize that linear momentum conservation requires that the slope of the line on which $v_{\rm Helio,2}$ and $v_{\rm Helio,1}$ fall must be \textit{negative} for a true binary system. \citet{Fernandez_2017} attempted to infer $q_{\rm dyn}$ also from systems in which the slope of this line is positive or zero (e.g. their Figure 6), but mass ratios inferred in this way have no physical interpretation and indicate either inaccurate radial velocity measurements or the presence of a third, unseen component.

In Figure~\ref{fig:q_spec_q_dyn}, we compare the best-fit values of $q_{\rm spec}$ and $q_{\rm dyn}$ obtained for SB2 systems in which $q_{\rm dyn}$ can be reliably measured; we identify such systems as those in which the range of $v_{\rm Helio}$ spanned across visits is at least $10\,\rm km\,s^{-1}$ for both stars, corresponding to a velocity shift of slightly less than 1 resolution element. We color points by the median of S/N per pixel as reported in the allVisit catalog, where the median is over all visit spectra used in the fit. 

The agreement between $q_{\rm spec}$ and $q_{\rm dyn}$ is in general quite good, with a median absolute difference between $q_{\rm spec}$ and $q_{\rm dyn}$ of ${\rm med}(|q_{{\rm spec}}-q_{{\rm dyn}}|)=0.048$ and a corresponding middle 68\% range of ($0.012-0.14$). The agreement is on average better for targets whose spectra have higher S/N; most systems with significantly different $q_{\rm spec}$ and $q_{\rm dyn}$ have $\rm S/N \lesssim 50$. Particularly at lower mass ratios, $q_{\rm dyn}$ is on average slightly lower than $q_{\rm spec}$; i.e., assuming $q_{\rm dyn}$ is usually more accurate than $q_{\rm spec}$, the latter is biased to slightly higher $q$. This can be understood as a consequence of the minimum $T_{\rm eff}$ of our spectral model, which sets an effective minimum $q_{\rm spec}$. If a cool primary has a companion with $T_{\rm eff}$ cooler than 4200 K that cannot be fully accommodated by the spectral model, a better fit can often still be achieved by a binary model with $T_{\rm eff} = 4200\,\rm K$ and a too-high $q_{\rm spec}$ than a single star model which ignores the secondary entirely. This in part explains the substantial number of cool systems with $q_{\rm spec}$ near 1 and lower $q_{\rm dyn}$, though we note that most cool systems also have lower S/N. 

If the secondary is very faint compared to the primary, its contribution to the spectrum may be completely undetectable, in which case binary and single-star models will converge to the same model spectrum as long as the velocities are allowed to vary between visits. Such systems can be distinguished from isolated single stars by the fact that the ``SB1'' model provides a better fit than the single-star model, which requires a target's velocity to be the same at all visits. Figure~\ref{fig:sb1} shows an example of such a system. Our model makes it possible to set an upper limit on the mass ratio, under the assumption that the companion is a main sequence star: in this case, the SB2 model would provide a better fit than the SB1 model if the secondary had $T_{\rm eff} \gtrsim 4200\,\rm K$. This limit is likely conservative in practice for main-sequence secondaries, as discussed above. However, it will not apply for binaries in which the companion is a stellar remnant. 

We note that most SB1s and some close SB2s can be qualitatively identified as unlikely to be single based on the scatter across visits in the radial velocities measured by the APSTAR pipeline \citep[e.g.][]{Badenes_2017}. However, we find a nontrivial number of SB2 systems ($\sim$100 systems out of the $\sim$20000 targets studied in this work) that show clearly time-variable spectra, with changes in the velocities of both components larger than $30\,\rm km\,s^{-1}$, for which the APSTAR-derived $v_{\rm Helio}$ measurements change at the $< 1\,\rm km\,s^{-1}$ level. This indicates that APOGEE radial velocity measurements are likely problematic for these systems, and studies that flag short-period binaries based on velocity variability will miss some SB2 systems.  

Figure~\ref{fig:triple} shows an example of a spectrum classified as a triple. The SB2 model (cyan) clearly cannot provide a good fit to the observed spectra, which simply have too many lines; on the other hand, the triple model is a good fit to all visits. The inset shows the velocities of the three components at each epoch; note that these are all allowed to vary freely and are not restricted to follow any equivalent of Equation~\ref{eqn:vr1_vr2}. One component, the spectral primary, has effectively constant velocity (within $\pm 0.5\,\rm km\,s^{-1}$) across all visits. On the other hand, the velocities of the secondary and tertiary components vary a great deal between visits and fall on a line with negative slope, just as in the case of close binaries (Figure~\ref{fig:single_visit}). The most straightforward explanation for these kinematics is that the system is a hierarchical triple \citep[e.g.][]{Ford_2000, Toonen_2016}
consisting of a close binary orbiting a third system with a period much longer than that of the close binary, so that the velocity of the primary and the center-of-mass velocity of the close binary do not change significantly over the temporal baseline between visits (which is $\sim 54$ days for this target). This type of hierarchical orbital is illustrated schematically in Figure~\ref{fig:triple}. Consistent with this interpretation, the spectrally inferred mass ratio between the two components of the close binary is similar to the dynamical mass ratio inferred from the slope of the line on which their velocities fall. 

We find 114 triple systems, most of which have the same qualitative velocity configuration as the system in Figure~\ref{fig:triple}: they contain one component with effectively constant velocity over all visits and two components with variable velocities that fall on a line as expected by a close binary. This is not surprising, as hierarchical configurations are the natural stable end state of the dynamical evolution of (otherwise chaotic) triple systems \citep{Naoz_2013}. We also find systems in which the velocity of the third (long-period) component is not constant but changes approximately linearly with time; this is expected if the system's outer period  is long compared to the observation baseline but not so long that no change can be observed. In such cases, the heliocentric velocities of the other two components do not fall on a straight line but exhibit some intrinsic scatter; this scatter can be reduced if a constant multiple of the linear trend of the lone star is subtracted from the velocities of the other two stars. 

Such systems are almost certainly gravitationally bound triples, since the velocities of all three components are correlated. However, for triples in which the velocity of one component is consistent with being constant over the time baseline spanned by observations, there is no guarantee that the three stars are actually gravitationally bound: the observed velocities could also be explained by a chance alignment between a close binary system and a background or foreground star. Whether such chance alignments constitute a substantial fraction of the targets we identify as hierarchical binaries can be diagnosed by comparing the center-of-mass velocity of the close binaries to the velocity of the third component. For gravitationally bound triples, these should be reasonably similar, with offsets of order the orbital velocity of the long-period component. The typical offsets should be larger (at minimum, of order $30\,\rm km\,s^{-1}$, the velocity dispersion of the Milky Way's stellar disk) for chance alignments. 

\begin{figure}
\includegraphics[width=\columnwidth]{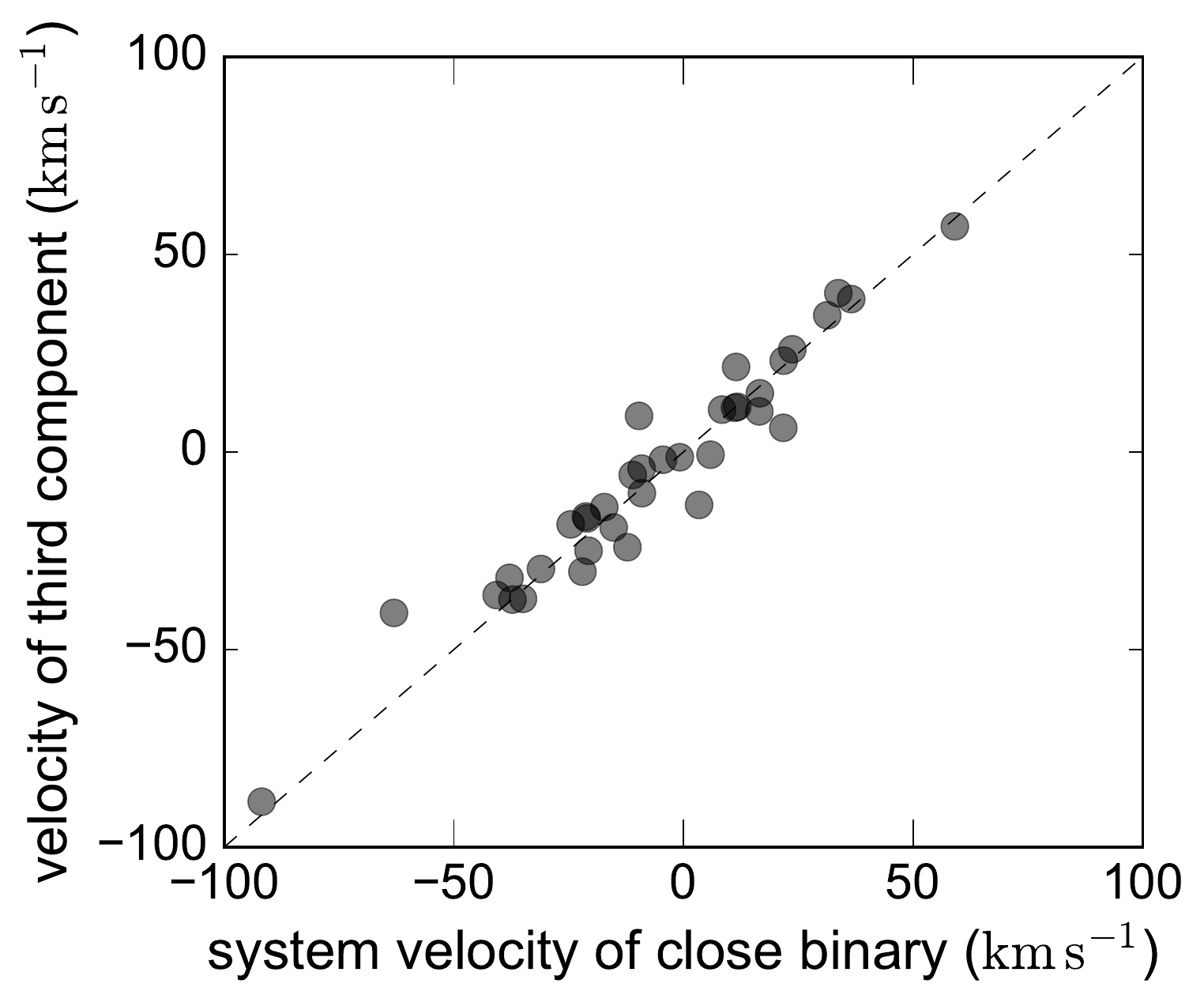}
\caption{Line-of-sight velocities for hierarchical triples containing a close binary and third component with a much larger separation (see schematic in Figure~\ref{fig:triple}). The tight correlation between the center-of-mass velocity of the close sub-binary and the velocity of the long-period component indicates that most systems are bona-fide gravitationally bound triples, not chance alignments between a close binary and a background or foreground star.}
\label{fig:triple_vsys}
\end{figure}

We investigate this explicitly in Figure~\ref{fig:triple_vsys}. Here we only plot systems that are consistent with the velocity of the of the long-period component being fixed over all epochs; we identify such cases as systems in which the change in the velocity of the long-period component across epochs is less than $2\,\rm km\,s^{-1}$ when all velocities are allowed to vary freely. Consistent with the expectation for bound triples, the system velocity of the close binary is in most cases within $10\,\rm km\,s^{-1}$ of that of the third component. There are $5$ systems in which the offset is larger, but due to the relatively short observational baselines, we find that none of these velocity offsets are large enough to rule out the possibility that all three stars are gravitationally bound. We discuss the possibility of contamination due to chance-alignments of stars further in Section~\ref{sec:chance_align}.

\begin{figure}
\includegraphics[width=\columnwidth]{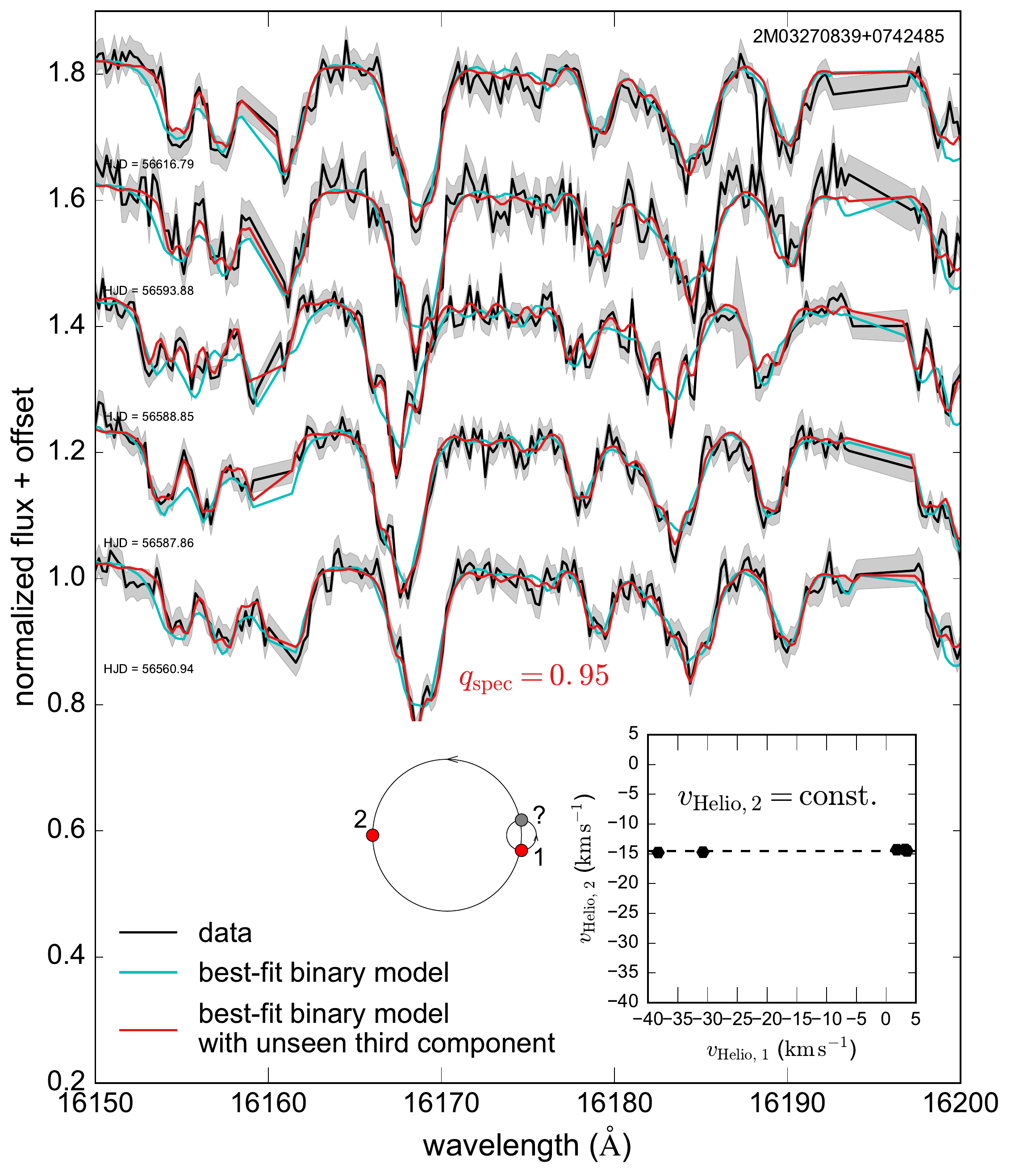}
\caption{Visit spectra of a triple system in which the third component does not contribute significantly to the spectrum but can be detected gravitationally. Cyan line shows best-fit SB2 model with the restriction that $v_{\rm Helio,1}$ and $v_{\rm Helio,2}$ fall on a line with negative slope (Equation~\ref{eqn:vr1_vr2}). Red line shows the best-fit binary model in which the velocities of the primary and secondary are allowed to vary freely. Inset shows the line-of-sight velocities corresponding to the red model. The velocity of the secondary is consistent with being constant at $v_{\rm Helio,2} \approx -14.5 \rm\,km\,s^{-1}$, while that of the primary varies substantially. This implies that the system is a hierarchical triple in which one component of the close binary does not contribute to the spectrum (i.e., it is a stellar remnant or faint M-dwarf); this is shown schematically in the orbital diagram.}
\label{fig:bin_dv}
\end{figure}

Along with SB1s, SB2s, and SB3s, we also identify a class of systems in which the presence of a third component can be deduced from radial velocity measurements, but only two star contribute significantly to the spectrum. Figure~\ref{fig:bin_dv} shows an example of such a target. The standard SB2 model, which enforces Equation~\ref{eqn:vr1_vr2} with $q_{\rm dyn} \geq 0.2$, cannot satisfactorily fit the spectrum. However, the ``SB2 with unseen third component'' model, which allows the velocities of both components to vary freely, provides a good fit, converging on a solution in which the velocity of one component is consistent with being fixed across epochs while that of the other varies. 

As illustrated in the orbital schematic in Figure~\ref{fig:bin_dv}, such a radial velocity pattern can be explained straightforwardly if the system is a hierarchical triple in which the close binary is an SB1; i.e., one component of the close binary does not contribute to the spectrum, either because its mass is low or because it is a compact remnant. No dynamical mass ratio can be inferred for these systems, because the acceleration of the variable velocity component is due primarily to the unseen component. We identify 108 SB2 systems in which the presence of a third component can be inferred dynamically; the majority of these systems have velocity configurations similar to Figure~\ref{fig:bin_dv}, with one component's velocity essentially constant. 

\subsection{Color-magnitude diagram}
\begin{figure*}
    \includegraphics[width=\textwidth]{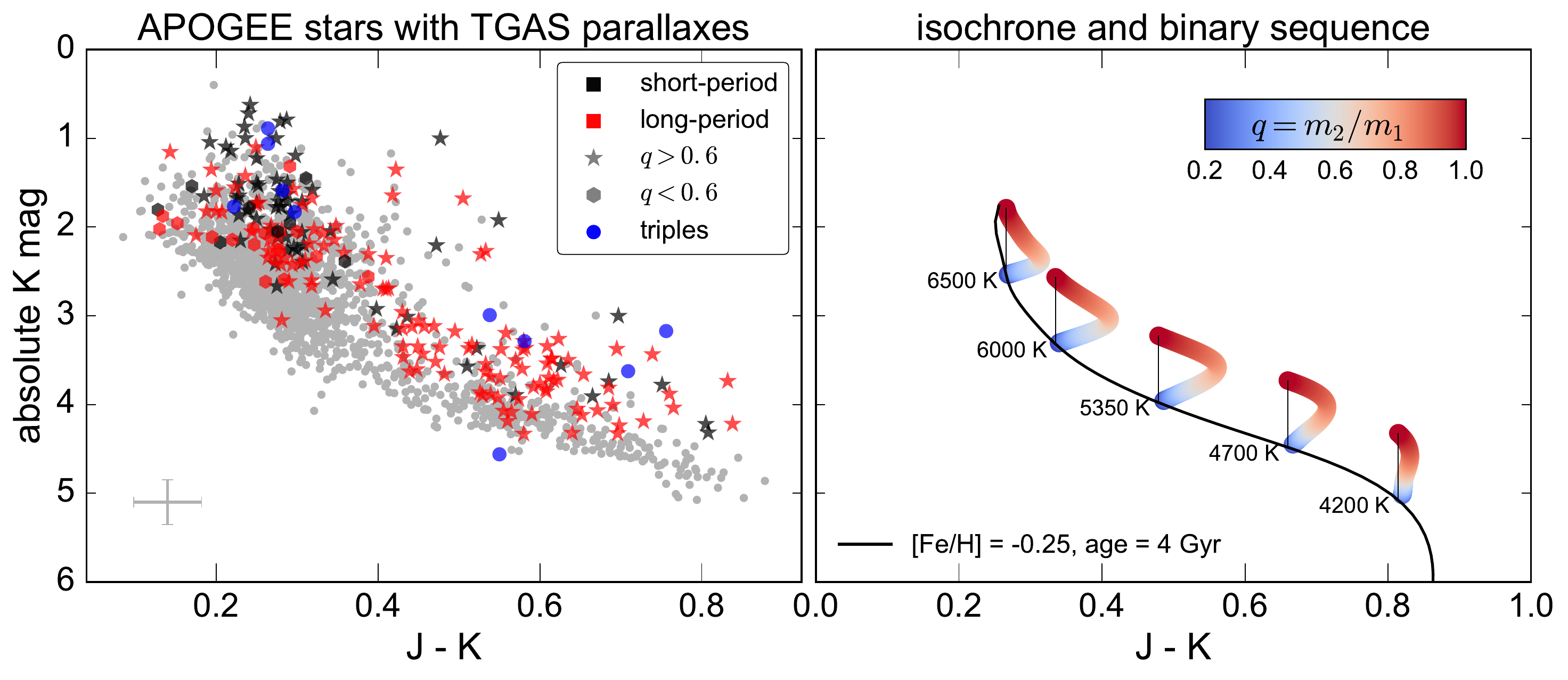}
    \caption{\textbf{Left}: Color-magnitude diagram of all stars in our sample with parallax errors of 10\% or less (gray circles). Black and red symbols represent spectroscopically identified binaries, with maximum line-of-sight velocity offsets between the two components greater (short-period) and less than (long-period) $30\,\rm km\,s^{-1}$. Binaries with mass ratios $q<0.6$ are marked with hexagons; those with $q>0.6$, with stars. Triples are marked with blue points. Most spectroscopically-identified binaries lie above the main sequence. Traditional SB2 identification methods  can only identify close binaries with large velocity offsets (black symbols); our method can identify many more long-period binaries with negligible velocity offsets (red symbols). \textbf{Right}: Schematic effect of unresolved binarity on the CMD. Black line shows a \texttt{MIST} isochrone for single stars with a single age and metallicity; colored loci show where binaries with different mass ratios and primary $T_{\rm eff}$ fall when they are spatially unresolved.}
    \label{fig:cmd}
\end{figure*}
A straightforward diagnostic to verify that the targets we spectroscopically identify as binaries are primarily true binaries, as opposed to single stars whose spectra contain unusual features that are not well accounted for in the spectral model, is to examine their distribution in a color-magnitude diagram (CMD). True binaries are expected to lie above the single-star main sequence of a CMD \citep[][]{Hurley_1998, Li_2013}: binaries with $q \sim 1$ will have the same color as would either single star but will be twice as luminous, while binaries with $0.5 \lesssim q \lesssim 0.9$ will be both brighter and redder than a single star with the parameters of the primary. 

Accurate measurements of absolute magnitude (and hence distance) are required to construct the CMD. To identify stars in our sample with accurate distance measurements, we cross-matched it with the Tycho-Gaia astrometric solution (TGAS) catalog \citep{Michalik_2015} using the \texttt{gaia\_tools.xmatch} routine written by Jo Bovy. This revealed 1925 stars in our sample with parallax errors of 10\% or better,\footnote{This corresponds to a magnitude error of $\pm 0.22$ mag, plus typical 2MASS photometric errors of $\pm 0.03$ mag. We do not attempt to correct for extinction or redding, which is expected to be modest in the near-infrared at the distances of the stars with accurate parallaxes (which have a median distance of 200 pc).} 217 of which were spectroscopically identified as multiple-star systems in which at least two components contribute detectably to the spectrum. 

We plot the CMD for these objects, based on 2MASS photometry, in the left panel of Figure~\ref{fig:cmd}. As expected, the majority of spectroscopically-identified binaries are scattered above the main sequence. We stress that our model for identifying binaries operates exclusively on normalized spectra and does not rely whatsoever on photometry; the fact that nearly all of the spectroscopically identified binaries populate the expected locus of the CMD above the main sequence is therefore a robust confirmation that our model is finding real binaries.

In the right panel of Figure~\ref{fig:cmd}, we show schematically how the presence of an unresolved companion is expected to change a star's position on the CMD. For a single stellar population, binaries with mass ratios $0.6 \lesssim q \lesssim 1$ form a coherent second sequence $\sim 0.75$ magnitudes above the main sequence; i.e., it is \textit{not} the case over this range of mass ratios that binaries with higher $q$ fall farther above the main sequence. This occurs because as $q$ is increased from 0.6 to 1 and $T_{\rm eff}$ of the secondary increases, unresolved binaries move blueward parallel to the main sequence. On the other hand, in binaries with $q\lesssim 0.4$, the secondary contribute so little light that the change in the unresolved system's location on the CMD is negligible. 

The lowest mass-ratio to which we are sensitive is $q\sim 0.4$, so the majority the binaries we identify should scatter above the main sequence for their respective isochrone. In the left panel of Figure~\ref{fig:cmd}, we plot separately binaries with $q \leq 0.6$ (which are only detectable around primaries with $T_{\rm eff} \gtrsim 5800\,\rm K$; see Appendix~\ref{sec:semi_empirical}) and those with higher mass ratios. As expected, the lower mass ratio systems are on average below the higher mass ratio systems on the CMD. With one exception,\footnote{We have investigated the spectra of this target (\texttt{2M07212735+2342096}) in detail and find it to be an unambiguous triple, with clear changes in spectral morphology between visits. Why it falls below the main sequence is not clear; one possibility is that marginally resolved multiplicity led to an overestimate of its parallax.} systems identified as triples (SB3) fall above the binary main sequence. We do not mark SB1s in Figure~\ref{fig:cmd}; their distribution on the CMD is similar to that of single stars, likely because most have low mass ratios. 

The sample of stars for which accurate parallaxes are available spans a wide range of metallicities and ages, so significant intrinsic scatter is expected in the distribution of both single stars and binaries on the CMD. We note that there are some stars that our model does not identify as binaries but which still scatter well above the main sequence. We suspect that most of these systems are binaries with $q\sim 1$ and small velocity offsets; these are not detectable in our current framework. 

We also divide suspected binaries into subsamples with large and small velocity offsets, corresponding approximately to systems which could and could not be detected with traditional CCF-based binary detection methods \citep{Fernandez_2017}. Only $\sim$30\% of systems have velocity offsets that are large enough to be detected with traditional methods. This highlights one of the primary advantages of the method introduced in this work: it is sensitive to a substantially larger fraction of the binary population than methods based on radial velocity separation or variability alone. 

\subsection{Deriving Orbital Parameters}
\label{sec:orbits}

\begin{figure*}
\centering 
\subfigure[radial velocity curve]{\label{fig:rv_curve}\includegraphics[width=\columnwidth]{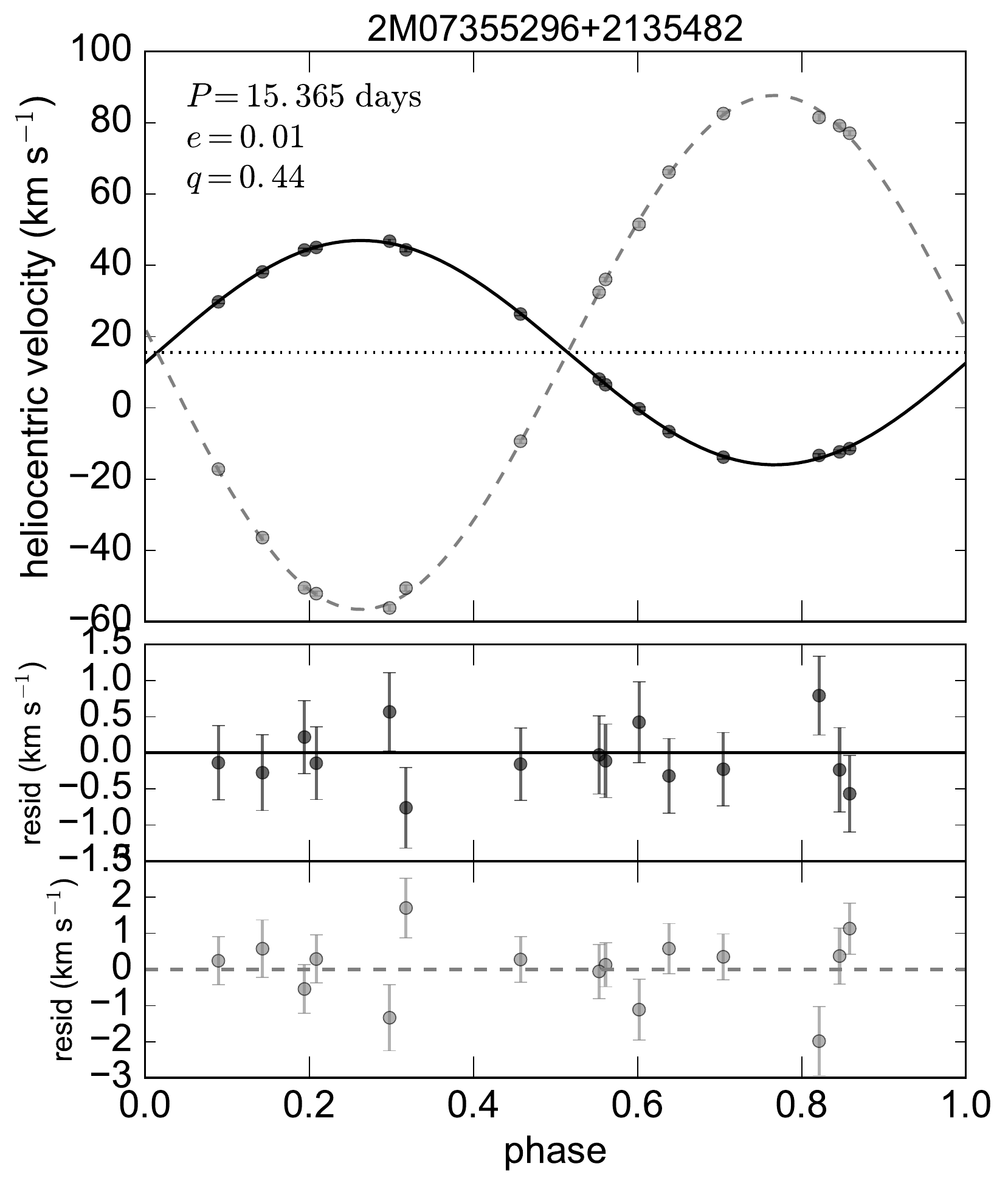}}
\subfigure[orbital parameter constraints]{\label{fig:mcmc_orbit}\includegraphics[width=\columnwidth]{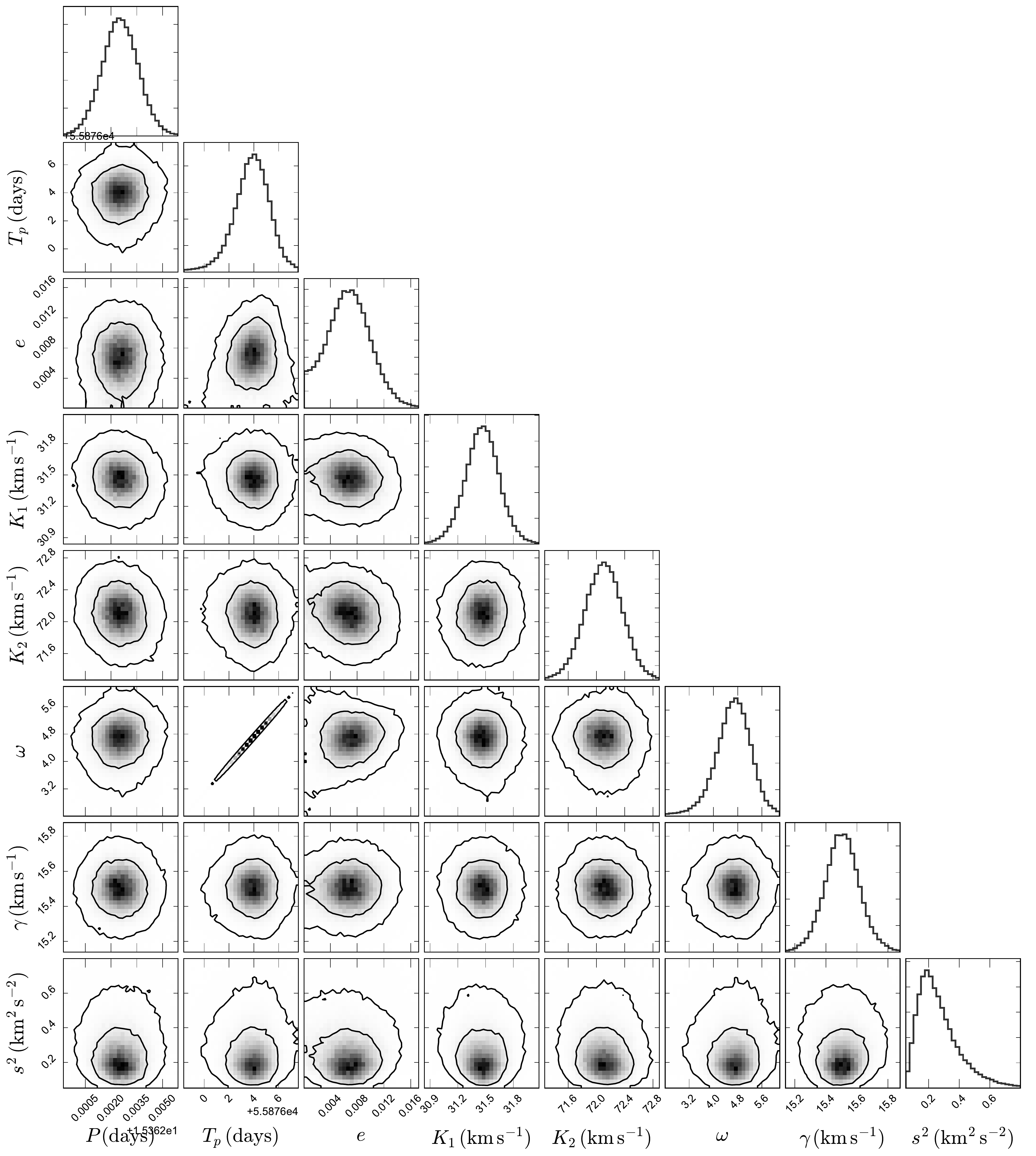}}
\caption{\textbf{Left}: Orbit fit to an SB2 systems with 15 epochs. In this relatively low-mass ratio system, the secondary contributes only $\sim$3\% of the light, but we can still determine its velocity to $\pm \sim 1\,{\rm km\,s^{-1}}$. Black and gray lines show the heliocentric velocity of the primary and secondary star. \textbf{Right}: 68 and 95\% marginalized probability regions for the system's orbital parameters. Because the eccentricity is very nearly 0, $T_{p}$ and $\omega$ are highly degenerate and are individually not well constrained, reflecting the fact that these quantities are meaningless for a circular orbit. We derive similar orbital solutions, shown in Table~\ref{tab:orbits}, for 64 systems.}
\end{figure*}

We derive full orbital solutions for 64 binary systems which have sufficient visits and phase coverage to constrain the orbit. Our criteria for determining whether the available velocity data are sufficient to constrain a system's orbit are discussed in Appendix~\ref{sec:orbit_convergence}. We only attempt to derive orbital solutions for systems in which at least two stars contribute to the spectrum; orbital solutions for SB1 systems in APOGEE can be found in \citet{Troup_2016}. 

Velocities for both components are returned as labels for the best-fit spectral model. An initial estimate of the radial velocity uncertainty at each visit is obtained through bootstrapping: Gaussian noise proportional to the best-fit model residual at each pixel is added to the data spectrum and the fit is repeated; the uncertainty on each radial velocity is taken to be the standard deviation of the best-fit velocity at each epoch when this procedure is repeated many times.

Fitting a Keplerian orbit amounts to simultaneously maximizing the likelihood of the radial velocity curves of the primary and secondary, where the model radial velocity for a given set of orbital parameters is obtained by solving the two-body problem \citep{Murray_2010}. We use a custom Python implementation of the adaptive simulated annealing algorithm \citep{IglesiasMarzoa_2015} to obtain an initial maximum-likelihood estimate of the best-fit orbital parameters and then sample the parameter space in the vicinity of the maximum likelihood with \texttt{emcee} \citep{ForemanMackey_2013} to estimate parameter uncertainties. We use non-informative, flat priors throughout. In addition to the 7 standard Keplerian orbital parameters,\footnote{These include the period, $P$, periastron time, $T_p$, eccentricity, $e$, argument of periastron, $\omega$, center-of-mass velocity, $\gamma$, and the velocity semi-amplitudes, $K_{1}$ and $K_{2}$.} we fit a ``jitter'' term, $s^2$, to allow for the possibility of intrinsic scatter in the radial velocities due to e.g. stellar pulsation or underestimated radial velocity uncertainties \citep[see e.g.][]{Baluev_2009, PriceWhelan_2017}. The effective total uncertainties in the radial velocities used in fitting are then $\sigma_{{\rm tot,}i}^{2}=\sigma_{i}^{2}+s^{2}$, where $\sigma_i$ are the radial velocity uncertainties at each epoch found from bootstrapping. Explicitly, the log-likelihood function is 
\begin{align}
\label{eqn:likelihood}
\begin{aligned}
\ln\mathcal{L}={} &-\frac{1}{2}\sum_{i}^{N}\biggl \{ \frac{[v_{r}(t_{i},\vec{\theta}_{1})-v_{{\rm Helio1,}i}]^{2}}{\sigma_{1,i}^{2}+s^{2}}+\ln[2\pi(\sigma_{1,i}^{2}+s^{2})] \biggr. \\ & \biggl. {}+ \frac{[v_{r}(t_{i},\vec{\theta}_{2})-v_{{\rm Helio2,}i}]^{2}}{\sigma_{2,i}^{2}+s^{2}}+\ln[2\pi(\sigma_{2,i}^{2}+s^{2})]\biggr\}.
\end{aligned}
\end{align}
Here the sum is over $N$ epochs at times $t_i$, $v_{r}(t,\vec{\theta})$ represents the predicted radial velocity at time $t$ for a system with orbital parameters $\vec{\theta}$ \citep{Murray_2010}, and $\vec{\theta}_{1}=\left(P,T_{p},e,\omega,K_{1},\gamma\right)$ and $\vec{\theta}_{2}=\left(P,T_{p},e,\omega+\pi,K_{2},\gamma\right)$ are the orbital parameters for each component. For most systems, the best-fit jitter is small ($s^2 \sim 0.1\,\rm km^2\,s^{-2}$), indicating that our estimates of $\sigma_i$ are reasonably accurate. However, for stars with large $v_{\rm macro}$, which indicates significant rotation, jitter is sometimes of order $1\,\rm km\,s^{-1}$. This suggests that our velocity measurements are less accurate for rapidly rotating stars, which is also supported by our experiments with semi-empirical binary spectra (Appendix~\ref{sec:semi_empirical}). 

For some systems with mass ratios near 1, we found that it was initially impossible to obtain a good fit to the measured radial velocities because the velocity assignments of the two stars were switched in the fits for one or more visits. We attempted to fit these systems by allowing the fitting algorithm to switch the assigned velocities for individual visits if doing so would improve the fit. In most cases, this solved the problem. A few systems ($\sim$5\% of those with sufficient coverage) remained with radial velocity curves that could not be well-fit by a Keplerian orbit, even with the possibility of switching the assigned velocities; these systems may have poorly measured radial velocities or contain an unseen component.

Figure~\ref{fig:rv_curve} shows an example orbital solution for a system with typical phase coverage, radial velocity errors, and number of epochs. This system has the lowest mass ratio, $q=0.44$, of the systems for which we derive orbital solutions. Because of the system's low mass ratio, the secondary contributes only a small fraction ($\sim $3\%) of the total light in the spectrum; it is not obvious from visual inspection that more than one star contributes to the spectrum. However, the secondary is detected unambiguously by our model, and the fact that the primary and secondary velocities all fall on a Keplerian orbit confirms the validity of the detection. 

Marginalized probabilities for the orbital parameters of this system are shown in Figure~\ref{fig:mcmc_orbit}. Most orbital parameters are well-constrained for this system, without strong parameter covariances. However, the periastron time, $T_{p}$, and argument of periastron, $\omega$, are highly degenerate, because the orbit is nearly circular (the eccentricity, $e$, is consistent with 0); the orbit has no well-defined periastron, and hence $T_{p}$ and $\omega$ are undefined. All systems with low eccentricities therefore have large uncertainties in $\omega$ and in $T_p$, even when the meaningful parameters of the orbit are well-constrained. 

\begin{table*}
\begin{center}
	\begin{tabular}{cccccccccc}
		\hline\\[-2ex]
		2MASS ID & $N_{\rm epochs}$ & $P$ &$T_{p}$ & $e$ & $\omega$ & $K_1$ & $K_2$ & $\gamma$ & $U_{N}V_{N}$  \\
		 & & [days] & [BMJD] &  & [radians] & [$\rm km\,s^{-1}$] & [$\rm km\,s^{-1}$] & [$\rm km\,s^{-1}$] & \\ [1ex]
		\hline
		\hline\\[-2ex]
06212323+1701485 & 8 & 42.24 & 56657.15 & 0.3166 & 0.684 & 35.27 & 41.40 & 1.37 & 0.79 \\
& & $\pm_{0.17}^{0.19}$ & $\pm_{0.23}^{0.21}$ & $\pm_{0.0092}^{0.0091}$ & $\pm_{0.034}^{0.029}$ & $\pm_{0.29}^{0.31}$ & $\pm_{0.31}^{0.32}$ & $\pm_{0.18}^{0.19}$ & \\
\hline
08544465+1130053 & 21 & 39.23855 & 55933.620 & 0.6932 & 1.0931 & 59.79 & 62.93 & -6.818 & 0.92 \\
& & $\pm_{0.00093}^{0.00097}$ & $\pm_{0.016}^{0.016}$ & $\pm_{0.0014}^{0.0014}$ & $\pm_{0.0035}^{0.0035}$ & $\pm_{0.19}^{0.20}$ & $\pm_{0.21}^{0.20}$ & $\pm_{0.069}^{0.069}$ & \\
\hline
04030722+5150045 & 9 & 69.973 & 55906.20 & 0.569 & 4.026 & 31.5 & 33.4 & -7.176 & 0.76 \\
& & $\pm_{0.093}^{0.076}$ & $\pm_{0.33}^{0.38}$ & $\pm_{0.036}^{0.045}$ & $\pm_{0.026}^{0.024}$ & $\pm_{1.9}^{2.9}$ & $\pm_{2.1}^{3.1}$ & $\pm_{0.091}^{0.089}$ & \\
\hline
21313924+1307507 & 41 & 1.5567964 & 55731.18 & 0.0016 & 1.03 & 59.51 & 71.84 & -52.199 & 0.92 \\
& & $\pm_{0.0000015}^{0.0000016}$ & $\pm_{0.19}^{1.10}$ & $\pm_{0.0012}^{0.0016}$ & $\pm_{0.76}^{4.30}$ & $\pm_{0.12}^{0.12}$ & $\pm_{0.13}^{0.13}$ & $\pm_{0.061}^{0.069}$ & \\	
\hline
18470667-0226077$^{\rm a}$ & 32 & 7.52676 & 55823.81 & 0.0136 & 5.38 & 45.86 & 55.85 & 15.99 & 0.93 \\
& & $\pm_{0.00014}^{0.00015}$ & $\pm_{0.37}^{0.39}$ & $\pm_{0.0050}^{0.0046}$ & $\pm_{0.30}^{0.32}$ & $\pm_{0.26}^{0.27}$ & $\pm_{0.26}^{0.26}$ & $\pm_{0.15}^{0.15}$ & \\
\hline
07355296+2135482 & 15 & 15.3645 & 55879.8 & 0.0065 & 4.62 & 31.47 & 72.11 & 15.50 & 0.87 \\
& & $\pm_{0.0011}^{0.0010}$ & $\pm_{1.5}^{1.5}$ & $\pm_{0.0034}^{0.0033}$ & $\pm_{0.62}^{0.60}$ & $\pm_{0.18}^{0.17}$ & $\pm_{0.27}^{0.27}$ & $\pm_{0.10}^{0.11}$ & \\
\hline
15010903+3702218 & 7 & 17.5079 & 56090.854 & 0.2996 & 2.5573 & 41.962 & 57.13 & -47.938 & 0.61 \\
& & $\pm_{0.0011}^{0.0011}$ & $\pm_{0.045}^{0.046}$ & $\pm_{0.0013}^{0.0014}$ & $\pm_{0.0088}^{0.0092}$ & $\pm_{0.092}^{0.100}$ & $\pm_{0.14}^{0.14}$ & $\pm_{0.036}^{0.040}$ & \\
\hline
08541894+1239291 & 22 & 1.3046792 & 55904.366 & 0.042 & 2.96 & 130.2 & 130.2 & 7.87 & 0.91 \\
& & $\pm_{0.0000089}^{0.0000087}$ & $\pm_{0.052}^{0.054}$ & $\pm_{0.010}^{0.010}$ & $\pm_{0.24}^{0.25}$ & $\pm_{1.7}^{1.7}$ & $\pm_{1.8}^{1.8}$ & $\pm_{0.80}^{0.86}$ & \\
\hline
19303146+4210508$^{\rm b}$ & 24 & 5.55412 & 56444.01 & 0.0120 & 6.05 & 43.20 & -- & -58.76 & -- \\
& & $\pm_{0.00014}^{0.00013}$ & $\pm_{0.22}^{0.25}$ & $\pm_{0.0030}^{0.0029}$ & $\pm_{0.24}^{0.28}$ & $\pm_{0.13}^{0.13}$ &  & $\pm_{0.100}^{0.092}$ & \\
\hline
08464223+1205302 & 17 & 15.0232 & 56654.146 & 0.2980 & 0.290 & 24.76 & 26.54 & -6.493 & 0.93 \\
& & $\pm_{0.0024}^{0.0023}$ & $\pm_{0.023}^{0.023}$ & $\pm_{0.0035}^{0.0034}$ & $\pm_{0.011}^{0.010}$ & $\pm_{0.12}^{0.13}$ & $\pm_{0.14}^{0.14}$ & $\pm_{0.055}^{0.058}$ & \\
\hline 
$\cdots$ & $\cdots$ &$\cdots$ &$\cdots$ &$\cdots$ &$\cdots$ &$\cdots$ &$\cdots$ &$\cdots$ & $\cdots$ \\
\hline
\end{tabular}
	\caption{Orbital solutions for double-line spectroscopic binaries. We report the median and middle 68\% of the marginalized posterior samples for each parameter. $U_N$ and $V_N$ quantify the phase and velocity coverage of the observations (see Appendix~\ref{sec:orbit_convergence}); systems with $U_N V_N \lesssim 0.5-0.6$ are susceptible to erroneous bad fits. $^{\rm a}$System is an SB2 within a hierarchical triple (10 systems). $^{\rm b}$System is an SB1 within a hierarchical triple (3 systems). This table is available in its entirety (with orbital solutions for 64 systems) in machine-readable form.}
    \label{tab:orbits}
\end{center}
\end{table*}

In Table~\ref{tab:orbits}, we provide best-fit orbital parameters and marginalized uncertainties for 64 systems for which an orbital solution could be obtained. Most of these systems are ordinary double-lined binaries (SB2s), similar to the system shown in Figure~\ref{fig:single_visit}. However, we also include solutions for several close SB2s within hierarchical triples (similar to Figure~\ref{fig:triple}), as well as SB1s within hierarchical triples with hidden third components (similar to Figure~\ref{fig:bin_dv}). These are fit very similarly to pure SB2 systems, with the only difference being that $v_{\rm Helio}$ measurements for individual stars at each visit are obtained from the ``SB3'' and ``SB2 with an unseen 3rd object'' models. We only attempt to fit such systems if they are consistent with the third component having constant velocity over the observed baseline. For ``SB2 with an unseen 3rd object'' systems, the fit is to a single radial velocity curve, so $K_2$ is not measurable. The rms velocity residual for all orbital solutions ranges been 0.04 and $1\,\rm km\,s^{-1}$. The systems with the largest velocities errors have a) lower average S/N and b) higher $T_{\rm eff}$ and $v_{\rm macro}$, both of which make it more difficult to accurately measure radial velocities.

The statistics $U_N$ and $V_N$ in the last column of Table~\ref{tab:orbits} quantify the uniformity of coverage in orbital phase and velocity by the measured radial velocity data. We calculate these statistics following \citet{Troup_2016} (their Equations 22 and 23); both $U_N$ and $V_N$ are bounded between 0 and 1, with values near 1 corresponding to uniform phase and velocity coverage. \citet{Troup_2016} estimated that orbital parameters are unreliable for SB1s if $U_N V_N < 0.5$; of course, the probability of recovering the correct orbit also depends on the number of radial velocity measurements and their uncertainties. In Appendix~\ref{sec:orbit_convergence}, we carry out tests with synthetic radial velocity data to determine the number of epochs and phase + velocity coverage required for reliable orbit recovery of SB2s with radial velocity data similar to that obtained for real binaries. 

\begin{figure}
\includegraphics[width=\columnwidth]{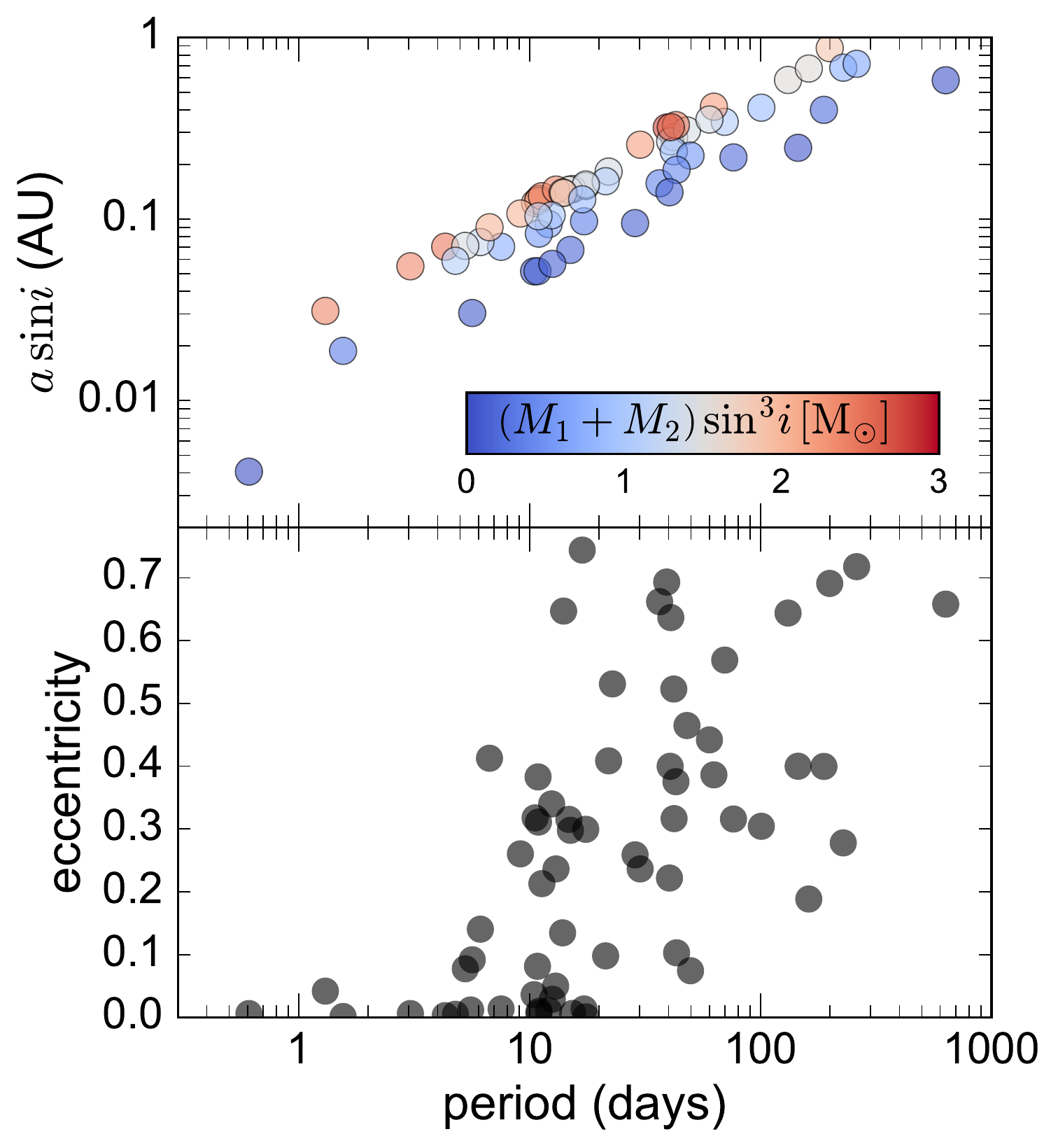}
\caption{\textbf{Top}: Distribution of periods, semi-major axes, and masses for the 64 double-lined binary systems for which we derive an orbital solution. Due to APOGEE's relatively rapid cadence (most targets have maximum baselines of a few months), these systems are heavily biased toward short periods. 
\textbf{Bottom}: Period-eccentricity distribution. Most systems with $P\lesssim 10$ days have low eccentricity due to tidal circularization.}
\label{fig:period_ecc}
\end{figure}

In the top panel of Figure~\ref{fig:period_ecc}, we plot constraints on the semimajor axes and component masses derived from the orbital parameters of all systems for which we present an orbital solution.\footnote{We calculate $a\sin i$, $M_1 \sin^3 i$, and $M_2 \sin^3 i$ using the standard formulas from \citet{Allen_2000}. It is not possible to measure $a$ or $M$ directly from radial velocity data alone; we note that future astrometric constraints can break the degeneracy between these quantities and orbital inclination for nearby systems \citep{Halbwachs_2017}.} Our sample contains systems with periods ranging from 0.6 days (short enough that the two stars are nearly touching, with $a\sin i \sim 0.8\,R_{\odot}$) to $\sim 600$ days. Dynamical constraints on the absolute masses of stars in individual binaries are weak due to the degeneracy with $\sin i$, but the highest lower limit on the mass of an individual component is $\sim 1.5\,M_{\odot}$. This corresponds to $T_{\rm eff} \gtrsim 6600\,\rm K$ for a solar-metallicity star on our adopted \texttt{MIST} isochrones; reassuringly, none of our dynamical mass constraints imply $T_{\rm eff} > 7000\,\rm K$, which is the upper limit adopted for our spectral model. 

We plot the periods and eccentricities of binary systems in the bottom panel of Figure~\ref{fig:period_ecc}. Most of the short period ($P\lesssim 10$ days) systems have nearly circular orbits, likely due tidal dissipation processes \citep[e.g.][]{Koch_1981}. However, a few systems with short periods do have eccentricity constraints that are inconsistent with zero. Some of these systems are short-period binaries within a hierarchical triple; such systems are known to be susceptible to eccentricity boosts via three-body interactions \citep{Kozai_1962}.\footnote{The only short-period system which is distinctly noncircular and is not best fit by a triple model is \texttt{2M21320320+1107560}, with $P=6.70$ days, $e=0.41$, and 41 epochs. It may well also be part of a hierarchical triple in which the third (long-period) component is too faint to appreciably contribute to the spectrum. Such a system would not be identifiable as having an unseen companion, as only systems in which the unseen companion is in the short-period sub-binary have velocities inconsistent with being an isolated binary.}

\subsection{Are binaries gravitationally bound?}
\label{sec:chance_align}
Our method finds binaries and triples by identifying targets in which more than one star falls within a single APOGEE fiber and contributes to the observed spectrum. In all cases where the velocities of the components of a suspected binary or triple system are not observed to vary in a correlated way, there is no guarantee that all the components are gravitationally bound: chance alignments of stars at different distances that fall within the same fiber can produce spectra consistent with binarity. 

To estimate the false-positive rate due to such ``optical binaries'' that are not gravitationally bound, we analyze mock photometric catalogs created with \texttt{Galaxia} \citep{Sharma_2011}. \texttt{Galaxia} implements the Besan\c{c}on model of stellar population synthesis \citep{Robin_2003} to populate the Galactic distribution function and produce realistic mock surveys along arbitrary lines of sight. Using a Galactic dust extinction map computed by \texttt{mwdust} \citep{Bovy_2016}, we produced mock catalogs complete to $J=14$ mag along 3 lines of sight representative of the range of stellar densities spanned by different APOGEE fields: one towards the bulge with $(\ell,b) = (0\,\rm deg, 5\,deg)$, one toward the Galactic anticenter with $(\ell,b) = (180\,\rm deg, 0\,deg)$, and one at high latitude with $(\ell,b) = (0\,\rm deg, 60\,deg)$, where $\ell$ and $b$ represent Galactic longitude and latitude. We then checked, for each star in a mock catalog, whether any other stars fall within a circular aperture of diameter 2 arcseconds centered on that star, representing a single APOGEE fiber. If more than one star was found in a given aperture (including both dwarfs and giants), we classified all stars in that aperture as a single optical binary. 

Toward the Galactic Bulge, we find a 0.2\% probability that a star is an optical binary. The same probability is 0.05\% toward the Galactic anticenter and 0.005\% at high latitude. As these probabilities are all much smaller than our detected binary fraction of $\sim 15\%$, we conclude that optical binaries are unlikely to be a large source of false positives, though a small fraction of the systems we detected in fields toward the Bulge may be chance alignments masquerading as binaries.  

Our model requires all components of a multiple-star system to have identical distances and abundances and fall on a single isochrone. This is likely a reasonable assumption for true, gravitationally bound binaries \citep{Desidera_2004, Andrews_2017}, but it is unlikely to hold for chance alignments. One could thus distinguish between true binaries and chance alignments by allowing the stellar parameters and abundances, and/or relative distance, of the secondary to vary freely and identifying cases where the best-fit model assigns significantly different abundances or distances to the different components. We defer such analysis to future work.

\section{Discussion and Conclusions}
\label{sec:discussion}

\subsection{Comparison to previous work}
\label{sec:prev_work}
\citet{Chojnowski_2015} compiled a catalog of double-lined spectroscopic binaries and triples in APOGEE by identifying targets whose cross-correlation function exhibited multiple peaks.\footnote{Their catalog is available at {\href{http://astronomy.nmsu.edu/drewski/apogee-sb2/apSB2.html}{\url{http://astronomy.nmsu.edu/drewski/apogee-sb2/apSB2.html}}}} Of the 610 targets in their catalog that were also in our initial sample, 574 were also classified as multiple systems by our pipeline, 5 were classified as SB1s, and 31 were classified as consistent with being single stars. Of the 574 stars classified as multiple systems by both pipelines, 514 are in our ``potential close binary'' subsample, which contains variable-velocity targets and binaries with large velocity offsets between the two components (Section~\ref{sec:visit}). This $\sim$95\% agreement rate is encouraging, given the very different approaches of the two pipelines. A primary advantage of the method developed in this work is its increased sensitivity to long-period systems with negligible velocity offsets.

Recently, \citet{Badenes_2017} studied the occurrence rate of short-period, velocity variable binaries in APOGEE. They found the multiplicity fraction for main-sequence stars to be a factor of $\sim$2 higher in the lowest-metallicity tercile of their sample than in the highest-metallicity tercile. We find a similar result: for short-period systems (those with velocity shifts of at least 10 $\rm km\,s^{-1}$ between epochs), the multiplicity fraction is $\sim$60\% higher for the lowest-metallicity tercile of our sample ([Fe/H] < -0.21) than for the highest-metallicity tercile ([Fe/H] > -0.02). \citet{Badenes_2017} studied systems with metallicities as low as [Fe/H] = -2.5; given the smaller range of metallicities in our sample ([Fe/H] > -1), these results are likely consistent. For long-period systems, we find the binary fraction to be consistent with being constant with metallicity. Some theoretical models \citep[e.g.][]{Machida_2008} predict that low-metallicity clouds should preferentially form short-period binaries, consistent with this result. However, we caution against over-interpreting this finding, as we have not attempted to quantify or correct for changes in the completeness of our method at lower metallicity. 

Modeling approaches similar to the method developed in this work have previously been used on a case-by-case basis to fit ``composite spectrum binaries'', a term that refers specifically to binaries containing a cool giant primary and a hot subgiant or main-sequence secondary \citep[e.g.][]{Gonzalez_2006, Griffin_2010}. Similar techniques have also been employed to spectroscopically detect and characterize unresolved binaries composed of very low-mass stars or brown dwarfs with different spectral types \citep{Burgasser_2007, Burgasser_2008}. Although this work focuses on modeling the spectra of binaries in which both components are main-sequence stars, the method we develop is flexible and can be straightforwardly extended to identify other flavors of binaries. The primary requirement is a robust training set spanning the range of single-star spectral types found in the dataset of interest.

For systems known to be double-lined binaries, a wide variety of techniques have been developed to disentangle the spectra of the two component stars in order to measure their individual velocities and stellar labels \citep[e.g.][]{Bagnuolo_1991, Simon_1994, Hadrava_1995, Pavlovski_2010, Czekala_2017}. These techniques can reliably separate the spectra of the individual components of a binary even when lines are blended, but they generally require multi-epoch spectroscopy that captures the combined binary spectrum at several orbital configurations. 

If only single-epoch spectroscopy is available or the binary is sufficiently wide that the orbital velocities of the two components do not change much between visits, the most common approach for measuring radial velocities is cross-correlation with a composite template spectrum \citep{Zucker_1994, Halbwachs_2017b}; this requires first estimating the labels of the individual stars. In such cases, most previous works have attempted to first model the primary star with a synthetic or empirical template, then subtract this template from the composite spectrum, and finally fit a model for the second star to the residual spectrum. However, it is difficult to ensure with this approach that the optimal binary model has been found, as the single-star model spectrum that best fits the combined binary spectrum does not in general correspond to the true best-fit parameters of the primary star \citepalias[see][]{ElBadry_2017}. The method introduced in this work, which fits for the stellar parameters and velocities of both components simultaneously, avoids these complications. 

Recently, a few works have shown that double-lined binaries can also be detected non-parametrically by identifying systems with peculiar spectra that are clustered outliers in a high-dimensional space of arbitrary summary statistics computed for all spectra collected by a survey \citep{Traven_2017, Reis_2017}. While such methods thus far primarily identify binaries with large velocity offsets, we note that non-parametric methods can likely be further optimized for binary identification by searching for targets which are precisely the kind of outliers expected to result from binarity; for example, one could identify systems that cannot be well-described by a single combination of spectral PCA components but can be well-described by two sums of components with different velocities. 

\subsection{Future prospects}
\label{sec:future_pros}

\begin{figure*}
\includegraphics[width=\textwidth]{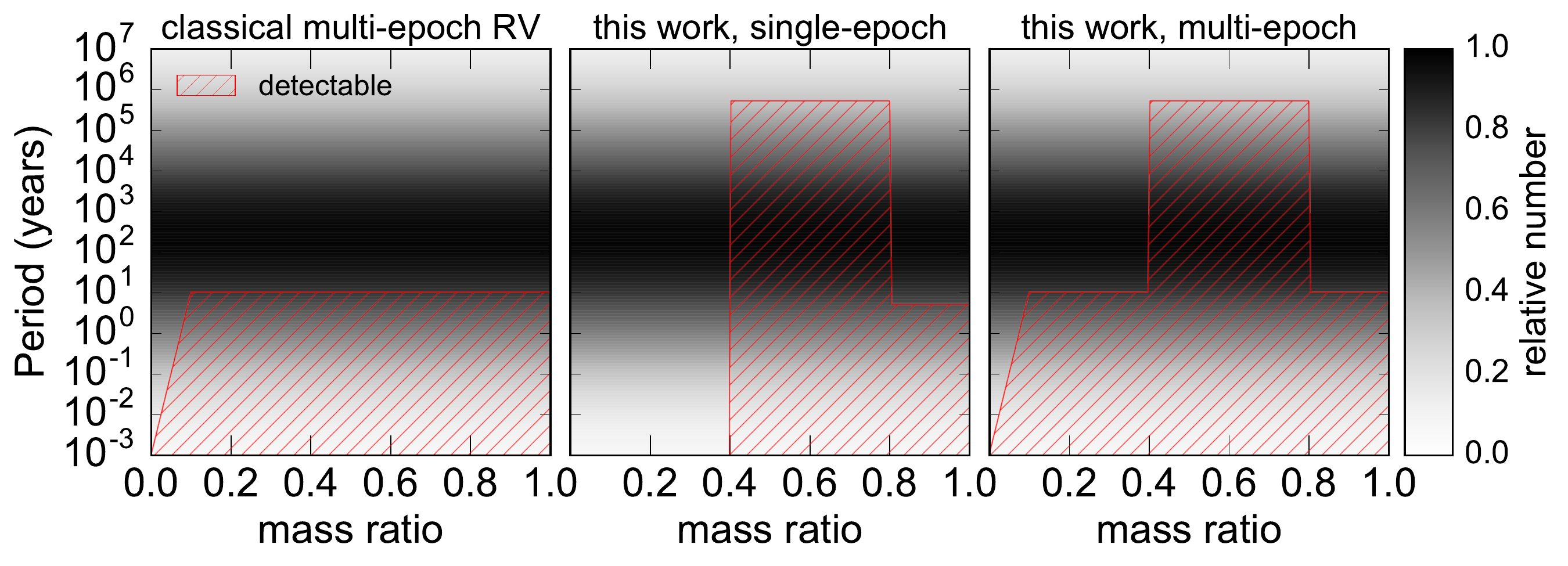}
\caption{Schematic illustration of the range of binary periods and mass ratios that can be detected with different methods. Gray shading (identical in all panels) shows the distribution of periods and mass ratios for solar-type stars; hatches show the regions of parameter space that can be probed by radial velocity variability (left) and fitting a binary model to single- and multi-epoch spectra (middle, right). Conventional multi-epoch radial velocity surveys are sensitive to essentially all mass ratios, but only for short-period binaries, which represent roughly a third of the observed lognormal period distribution for solar-type stars. The binary spectral model introduced in this work is sensitive to all but the longest periods (as long as both stars fall within one spectroscopic fiber) with single-epoch observations, but only for intermediate mass ratio systems. When multi-epoch spectra are available, fitting a binary model can also detect all systems with variable radial velocities as SB1s.}
\label{fig:period_schematic}
\end{figure*}

\subsubsection{Improving the model}
\label{sec:improvements}
A straightforward way to make our model sensitive to a larger fraction of the binary population is to  extend the single-star model to cooler temperatures. As discussed in Appendix~\ref{sec:semi_empirical}, the lower limit of $T_{\rm eff} = 4200\,\rm K$, which is due to shortcomings of  ab-initio spectral models for main-sequence stars at lower temperatures, limits the model to only detecting binaries with mass ratios near $q = 1$ at low temperatures and  prevents us from fitting spectra of the coolest stars altogether. Fitting cooler stars does not required any modification of our general approach, only a robust training set at lower temperatures, which currently does not readily exist. Due to the increased importance of molecular opacity from many species at lower $T_{\rm eff}$, it may be helpful to include more abundances in the spectral model for cooler stars. 

Our model could also be improved by fitting for the projected rotation velocity $v \sin i$ explicitly instead of subsuming it under the Gaussian broadening of $v_{\rm macro}$, since the single-star model currently performs worst for rapidly rotating stars. This is in principle simple to accomplish: rotation velocities for stars in the training set can be obtained straightforwardly in post-processing \citep[e.g.][]{Diaz_2011}, and the inferred $v \sin i$ can then be added as an additional label to the model. Rotation is currently not an important problem for most of the targets in our sample because stars with $T_{\rm eff} \lesssim 6500\,\rm K$ typically lose most of their angular momentum to magnetic braking and do not rotate rapidly on the main sequence \citep{Glebocki_2000, Schatzman_1962}, and stars with $T_{\rm eff} \geq 6500\,\rm K$ represent less than 4\% of our dataset. However, an improved treatment of rotation would make it possible to better model hot stars and would likely decrease the false-positive rate (see Appendix~\ref{sec:false_positive}). This is particularly true for young stars, which can rotate significantly even at cooler temperatures \citep[e.g.][]{Terrien_2014}.

\subsubsection{Hierarchical modeling}
\label{sec:hierarchical}
Beyond the Solar neighborhood, previous spectroscopic studies of the Galactic binary population have been limited to studying the short-period tail of the binary population. Because the model presented in this work does not depend on radial velocity variability or a line-of-sight velocity offset to detect binaries, it has the potential to substantially improve on existing constraints on the binary population of the Milky Way and/or its satellites when combined with a model for detection completeness and the survey selection function. 

Existing radial velocity surveys of the Milky Way and nearby dwarf galaxies are sensitive to binaries with periods less than $\sim (1-10)$ years \citep[e.g.][]{Matijevic_2011, Minor_2013, Hettinger_2015, Gao_2017, Badenes_2017}. For the log-normal period distribution for solar-type stars found in the Solar neighborhood \citep{Duchene_2013}, $\sim$73\% percent of binaries have $P>10$ years; most of these systems will be missed by such surveys. The most probable period for solar-type binaries is $\sim$300 years; assuming random orbit orientations, the typical line-of-sight velocity separation for such systems is $\Delta v_{\rm los}\sim 2\,{\rm km\,s^{-1}}$, and the average radial velocity change over a one-year baseline is $\sim 0.02\,{\rm km\,s^{-1}}$. This is an order of magnitude below the detectability thresholds of existing large spectroscopic surveys, though such weak radial velocity trends in SB1s may be marginally detectable with high-dispersion spectrographs typically used to study exoplanets \citep{Konacki_2005, Katoh_2013}.

Irrespective of radial velocity variability, long-period binaries with favorable mass ratios can be detected with our model as long as both components fall within a single spectroscopic fiber. At a distance of 1 kpc, more than 80\% of solar-type binaries will have projected separations of less than 1 arcsecond, so that both stars would fall with a single 2-arcsecond fiber; this fraction increases at larger distances. On the other hand, for long-period systems, the binary spectral model is sensitive only to intermediate mass ratio systems ($0.4 \lesssim q \lesssim 0.8$), in which the primary and secondary have qualitatively different spectral types, but the secondary still contributes a non-negligible fraction of the total light (see Appendix~\ref{sec:semi_empirical} and \citetalias{ElBadry_2017}). The distribution of mass ratios for solar-type binaries is approximately flat down to $q = 0.1$ \citep{Duchene_2013}, so the binary model will miss many high and low mass ratio systems with long periods. 

We summarize the sensitivity of our method, as well as standard binary-detection methods based on velocity variability, to systems with different periods and mass ratios in Figure~\ref{fig:period_schematic}. Radial velocity variability can probe essentially all mass ratios, but only for the short-period tail of the binary population. On the other hand, fitting a binary spectral model to single-epoch observations can probe most of the period distribution, but only for a restricted subset of mass ratios. We thus expect that the method developed here can be fruitfully combined with existing multi-epoch radial velocity measurements from SB1s, such as the APOGEE constraints on the short-period binary fraction presented in \citet{Badenes_2017} and measurements of the binary fractions of nearby dwarf galaxies presented by \citet{Minor_2013}. This would enable a full hierarchical model for binary populations that is sensitive to an unprecedented range of periods and mass ratios. 

An immediate advantage of our method is that it is sensitive to a large fraction of the binary population even when only single-epoch observations are available. With multi-epoch observations, our model can detect short-period systems as SB1s, with similar sensitivity to traditional methods. Our modeling approach can also be straightforwardly applied to spectra from other surveys. The precise range of mass-ratios to which it is sensitive will vary with wavelength coverage: surveys at optical wavelengths will be more sensitive to binaries with higher mass ratios ($0.8\lesssim q \lesssim 0.9$; see \citetalias{ElBadry_2017}) due to the increased spectral information content at shorter wavelengths, but they will be less sensitive at low $q$ because a cooler secondary star contributes a greater fraction of a binary system's total light in the near-infrared than at optical wavelengths. 

In this work, we fit normalized spectra and only used the CMD to assess the reliability of our spectral model. A promising avenue for future work is to fit spectra and photometry simultaneously, or to place a photometric prior on $q$. This would make it possible to detect systems with $q\sim 1$ and negligible velocity offsets, which are twice as luminous as they would be if they were a single star. Particularly with improved parallaxes from future \textit{Gaia} data releases, photometric constraints could substantially extend the fraction of the binary population to which our method is sensitive. 

\subsection{Summary}
\label{sec:summary}
We have developed a flexible data-driven method for identifying and fitting the spectra of multiple-star systems and have applied it to $\sim$ 20,000 main-sequence targets from the APOGEE survey. Unlike most previous work, our model performs well even for long-period systems in which the line-of-sight velocity offset between components is negligible, substantially expanding the fraction of the binary population that can be probed by observations. Our method is mostly automated and can be straightforwardly applied to other spectroscopic surveys with modest adjustments. 
Our main results are as follows:
\begin{enumerate}
\item \textit{Spectral identification of long- and short-period binaries:} Unresolved binaries can be identified as systems whose spectrum can be better-fit by a sum of two single-star model spectra falling on a single isochrone than any single-star model (Figure~\ref{fig:hotstar}). For systems with mass ratios $0.4 \lesssim q \lesssim 0.8$, in which the two stars have different spectral types, binaries can be identified spectroscopically even in the limit of no velocity offset and with only single-epoch observations. Spectral signatures of binarity are strengthened in the presence of a velocity offset of order one resolution element or greater (Figure~\ref{fig:three_dv_specs}); thus, close binaries can be detected even in the limit of $q\sim 1$. 

\item \textit{Photometric test of the model:} Nearly all spectroscopically identified binaries with accurate distance measurements fall above the main sequence on the CMD, as is predicted for true binaries, and triple systems fall above most binaries (Figure~\ref{fig:cmd}). Photometry does not enter our binary identification procedure, so this agreement with theoretical predictions provides independent validation of our spectral model.  

\item \textit{Dynamical mass ratios:} For short-period binaries in which the velocities of the two components change substantially between visits, it is possible to obtain a dynamical measurement of the mass ratio from the relative changes in the stars' radial velocities between visits (Figure~\ref{fig:single_visit}). This provides a constraint on the mass ratio that is independent of the spectral mass ratio, which determines the contribution of the secondary star to the spectrum. We find good agreement between spectral and dynamical mass ratios, with a median difference of 0.048 and even better agreement for systems with high S/N spectra (Figure~\ref{fig:q_spec_q_dyn}). 

\item \textit{Triple systems:} We identify 114 systems in which the contributions of three stars can be identified in the spectrum (Figure~\ref{fig:triple}) and an additional 108 in which only two stars contribute significantly to the spectrum, but the presence of a third component can be inferred from its gravitational effects (Figure~\ref{fig:bin_dv}). Most identified triples are hierarchical, consisting of a close binary orbited by a third component with a much longer period; we have verified that these systems are all likely gravitationally bound (Figure~\ref{fig:triple_vsys}). 

\item \textit{Orbital solutions:} For double-lined systems with a sufficient number of epochs and well-sampled radial velocity curves, we derive full Keplerian orbital solutions (Figure~\ref{fig:mcmc_orbit}); some of these systems are close binaries within hierarchical triples. We derive orbital solutions for 64 binaries with periods ranging from $\sim$0.6 days to $\sim$2 years and semimajor axes ranging from $\sim R_{\odot}$ to $\sim 1$ AU. Consistent with previous studies, we find that most binaries with $P\lesssim 10\,\rm days$ have eccentricity consistent with 0 due to tidal circularization processes (Figure~\ref{fig:period_ecc}).
\end{enumerate}
We make catalogs of best-fit labels for all identified multiple-star systems publicly available; these are described in Appendix~\ref{sec:data}.
\section*{Acknowledgements}

We are grateful to the anonymous referee for a constructive report. We thank Gaspard Duch\^{e}ne, Keith Hawkins, Jessica Lu, Hans-G{\"u}nter Ludwig, Adrian Price-Whelan, and Silvia Toonen for helpful conversations. We are grateful to Jan Rybizki for assistance in creating mock catalogs with \texttt{Galaxia}, and to Anna Ho for help with \textit{the Cannon}.
This project was developed in part at the 2017 Heidelberg Gaia Sprint, hosted by the Max-Planck-Institut f{\"u}r Astronomie, Heidelberg.
K.E. acknowledges support from the SFB 881 program (A3), a Berkeley Fellowship, a Hellman award for graduate study, and an NSF graduate research fellowship. 
H.W.R. received support from the European Research Council under the European Union's Seventh Framework Programme (FP 7) ERC Grant Agreement n. [321035]. 
Y.S.T is supported by the Australian Research Council Discovery Program DP160103747, the Carnegie-Princeton Fellowship and the Martin A. and Helen Chooljian Membership from the Institute for Advanced Study at Princeton.
E.Q. is supported in part by a Simons Investigator Award from the Simons Foundation.
D.R.W. is supported by a fellowship from the Alfred P. Sloan Foundation. 
C.C. acknowledges support from NASA grant NNX15AK14G, NSF grant AST-1313280, and the Packard Foundation.
The analysis in this paper relied on the python packages \texttt{NumPy} \citep{vanderwalt_2011}, \texttt{Matplotlib} \citep{Hunter_2007}, and \texttt{AstroPy} \citep{Astropy_2013}.



\bibliographystyle{mnras}

\begin{thebibliography}{}
\makeatletter
\relax
\def\mn@urlcharsother{\let\do\@makeother \do\$\do\&\do\#\do\^\do\_\do\%\do\~}
\def\mn@doi{\begingroup\mn@urlcharsother \@ifnextchar [ {\mn@doi@}
  {\mn@doi@[]}}
\def\mn@doi@[#1]#2{\def\@tempa{#1}\ifx\@tempa\@empty \href
  {http://dx.doi.org/#2} {doi:#2}\else \href {http://dx.doi.org/#2} {#1}\fi
  \endgroup}
\def\mn@eprint#1#2{\mn@eprint@#1:#2::\@nil}
\def\mn@eprint@arXiv#1{\href {http://arxiv.org/abs/#1} {{\tt arXiv:#1}}}
\def\mn@eprint@dblp#1{\href {http://dblp.uni-trier.de/rec/bibtex/#1.xml}
  {dblp:#1}}
\def\mn@eprint@#1:#2:#3:#4\@nil{\def\@tempa {#1}\def\@tempb {#2}\def\@tempc
  {#3}\ifx \@tempc \@empty \let \@tempc \@tempb \let \@tempb \@tempa \fi \ifx
  \@tempb \@empty \def\@tempb {arXiv}\fi \@ifundefined
  {mn@eprint@\@tempb}{\@tempb:\@tempc}{\expandafter \expandafter \csname
  mn@eprint@\@tempb\endcsname \expandafter{\@tempc}}}

\bibitem[\protect\citeauthoryear{{Andrews}, {Chanam{\'e}}  \&
  {Ag{\"u}eros}}{{Andrews} et~al.}{2017}]{Andrews_2017}
{Andrews} J.~J.,  {Chanam{\'e}} J.,   {Ag{\"u}eros} M.~A.,  2017, preprint,
  \href {http://adsabs.harvard.edu/abs/2017arXiv171004678A} {} (\mn@eprint
  {arXiv} {1710.04678})

\bibitem[\protect\citeauthoryear{{Astropy Collaboration} et~al.,}{{Astropy
  Collaboration} et~al.}{2013}]{Astropy_2013}
{Astropy Collaboration} et~al., 2013, \mn@doi [\aap]
  {10.1051/0004-6361/201322068}, \href
  {http://adsabs.harvard.edu/abs/2013A%26A...558A..33A} {558, A33}

\bibitem[\protect\citeauthoryear{{Badenes} et~al.,}{{Badenes}
  et~al.}{2017}]{Badenes_2017}
{Badenes} C.,  et~al., 2017, preprint, \href
  {http://adsabs.harvard.edu/abs/2017arXiv171100660B} {} (\mn@eprint {arXiv}
  {1711.00660})

\bibitem[\protect\citeauthoryear{{Bagnuolo} \& {Gies}}{{Bagnuolo} \&
  {Gies}}{1991}]{Bagnuolo_1991}
{Bagnuolo} Jr. W.~G.,  {Gies} D.~R.,  1991, \mn@doi [\apj] {10.1086/170276},
  \href {http://adsabs.harvard.edu/abs/1991ApJ...376..266B} {376, 266}

\bibitem[\protect\citeauthoryear{{Baluev}}{{Baluev}}{2009}]{Baluev_2009}
{Baluev} R.~V.,  2009, \mn@doi [\mnras] {10.1111/j.1365-2966.2008.14217.x},
  \href {http://adsabs.harvard.edu/abs/2009MNRAS.393..969B} {393, 969}

\bibitem[\protect\citeauthoryear{{Bovy}}{{Bovy}}{2016}]{Bovy_2016b}
{Bovy} J.,  2016, \mn@doi [\apj] {10.3847/0004-637X/817/1/49}, \href
  {http://adsabs.harvard.edu/abs/2016ApJ...817...49B} {817, 49}

\bibitem[\protect\citeauthoryear{{Bovy}, {Rix}, {Green}, {Schlafly}  \&
  {Finkbeiner}}{{Bovy} et~al.}{2016}]{Bovy_2016}
{Bovy} J.,  {Rix} H.-W.,  {Green} G.~M.,  {Schlafly} E.~F.,   {Finkbeiner}
  D.~P.,  2016, \mn@doi [\apj] {10.3847/0004-637X/818/2/130}, \href
  {http://adsabs.harvard.edu/abs/2016ApJ...818..130B} {818, 130}

\bibitem[\protect\citeauthoryear{Branch, Coleman  \& Li}{Branch
  et~al.}{1999}]{Branch_1999}
Branch M.~A.,  Coleman T.~F.,   Li Y.,  1999, \mn@doi [SIAM Journal on
  Scientific Computing] {10.1137/S1064827595289108}, 21, 1

\bibitem[\protect\citeauthoryear{{Burgasser}}{{Burgasser}}{2007}]{Burgasser_2007}
{Burgasser} A.~J.,  2007, \mn@doi [\aj] {10.1086/520878}, \href
  {http://adsabs.harvard.edu/abs/2007AJ....134.1330B} {134, 1330}

\bibitem[\protect\citeauthoryear{{Burgasser}, {Liu}, {Ireland}, {Cruz}  \&
  {Dupuy}}{{Burgasser} et~al.}{2008}]{Burgasser_2008}
{Burgasser} A.~J.,  {Liu} M.~C.,  {Ireland} M.~J.,  {Cruz} K.~L.,   {Dupuy}
  T.~J.,  2008, \mn@doi [\apj] {10.1086/588379}, \href
  {http://adsabs.harvard.edu/abs/2008ApJ...681..579B} {681, 579}

\bibitem[\protect\citeauthoryear{{Choi}, {Dotter}, {Conroy}, {Cantiello},
  {Paxton}  \& {Johnson}}{{Choi} et~al.}{2016}]{Choi_2016}
{Choi} J.,  {Dotter} A.,  {Conroy} C.,  {Cantiello} M.,  {Paxton} B.,
  {Johnson} B.~D.,  2016, \mn@doi [\apj] {10.3847/0004-637X/823/2/102}, \href
  {http://adsabs.harvard.edu/abs/2016ApJ...823..102C} {823, 102}

\bibitem[\protect\citeauthoryear{{Chojnowski} et~al.,}{{Chojnowski}
  et~al.}{2015}]{Chojnowski_2015}
{Chojnowski} S.~D.,  et~al., 2015, in American Astronomical Society Meeting
  Abstracts. p. 340.05

\bibitem[\protect\citeauthoryear{{Cox}}{{Cox}}{2000}]{Allen_2000}
{Cox} A.~N.,  2000, {Allen's astrophysical quantities}

\bibitem[\protect\citeauthoryear{{Czekala}, {Mandel}, {Andrews}, {Dittmann},
  {Ghosh}, {Montet}  \& {Newton}}{{Czekala} et~al.}{2017}]{Czekala_2017}
{Czekala} I.,  {Mandel} K.~S.,  {Andrews} S.~M.,  {Dittmann} J.~A.,  {Ghosh}
  S.~K.,  {Montet} B.~T.,   {Newton} E.~R.,  2017, \mn@doi [\apj]
  {10.3847/1538-4357/aa6aab}, \href
  {http://adsabs.harvard.edu/abs/2017ApJ...840...49C} {840, 49}

\bibitem[\protect\citeauthoryear{{Desidera} et~al.,}{{Desidera}
  et~al.}{2004}]{Desidera_2004}
{Desidera} S.,  et~al., 2004, \mn@doi [\aap] {10.1051/0004-6361:20041242},
  \href {http://adsabs.harvard.edu/abs/2004A%26A...420..683D} {420, 683}

\bibitem[\protect\citeauthoryear{{D{\'{\i}}az}, {Gonz{\'a}lez}, {Levato}  \&
  {Grosso}}{{D{\'{\i}}az} et~al.}{2011}]{Diaz_2011}
{D{\'{\i}}az} C.~G.,  {Gonz{\'a}lez} J.~F.,  {Levato} H.,   {Grosso} M.,  2011,
  \mn@doi [\aap] {10.1051/0004-6361/201016386}, \href
  {http://adsabs.harvard.edu/abs/2011A%26A...531A.143D} {531, A143}

\bibitem[\protect\citeauthoryear{{Duch{\^e}ne} \& {Kraus}}{{Duch{\^e}ne} \&
  {Kraus}}{2013}]{Duchene_2013}
{Duch{\^e}ne} G.,  {Kraus} A.,  2013, \mn@doi [\araa]
  {10.1146/annurev-astro-081710-102602}, \href
  {http://adsabs.harvard.edu/abs/2013ARA%26A..51..269D} {51, 269}

\bibitem[\protect\citeauthoryear{{Duquennoy} \& {Mayor}}{{Duquennoy} \&
  {Mayor}}{1991}]{Duquennoy_1991}
{Duquennoy} A.,  {Mayor} M.,  1991, \aap, \href
  {http://adsabs.harvard.edu/abs/1991A%26A...248..485D} {248, 485}

\bibitem[\protect\citeauthoryear{{El-Badry}, {Rix}, {Ting}, {Weisz},
  {Bergemann}, {Cargile}, {Conroy}  \& {Eilers}}{{El-Badry}
  et~al.}{2018}]{ElBadry_2017}
{El-Badry} K.,  {Rix} H.-W.,  {Ting} Y.-S.,  {Weisz} D.~R.,  {Bergemann} M.,
  {Cargile} P.,  {Conroy} C.,   {Eilers} A.-C.,  2018, \mn@doi [\mnras]
  {10.1093/mnras/stx2758}, \href
  {http://adsabs.harvard.edu/abs/2018MNRAS.473.5043E} {473, 5043}

\bibitem[\protect\citeauthoryear{{Fernandez} et~al.,}{{Fernandez}
  et~al.}{2017}]{Fernandez_2017}
{Fernandez} M.~A.,  et~al., 2017, \mn@doi [\pasp] {10.1088/1538-3873/aa77e0},
  \href {http://adsabs.harvard.edu/abs/2017PASP..129h4201F} {129, 084201}

\bibitem[\protect\citeauthoryear{{Ford}, {Kozinsky}  \& {Rasio}}{{Ford}
  et~al.}{2000}]{Ford_2000}
{Ford} E.~B.,  {Kozinsky} B.,   {Rasio} F.~A.,  2000, \mn@doi [\apj]
  {10.1086/308815}, \href {http://adsabs.harvard.edu/abs/2000ApJ...535..385F}
  {535, 385}

\bibitem[\protect\citeauthoryear{{Foreman-Mackey}, {Hogg}, {Lang}  \&
  {Goodman}}{{Foreman-Mackey} et~al.}{2013}]{ForemanMackey_2013}
{Foreman-Mackey} D.,  {Hogg} D.~W.,  {Lang} D.,   {Goodman} J.,  2013, \mn@doi
  [\pasp] {10.1086/670067}, \href
  {http://adsabs.harvard.edu/abs/2013PASP..125..306F} {125, 306}

\bibitem[\protect\citeauthoryear{{Gao}, {Zhao}, {Yang}  \& {Gao}}{{Gao}
  et~al.}{2017}]{Gao_2017}
{Gao} S.,  {Zhao} H.,  {Yang} H.,   {Gao} R.,  2017, \mn@doi [\mnras]
  {10.1093/mnrasl/slx048}, \href
  {http://adsabs.harvard.edu/abs/2017MNRAS.469L..68G} {469, L68}

\bibitem[\protect\citeauthoryear{{Garc{\'{\i}}a P{\'e}rez}
  et~al.,}{{Garc{\'{\i}}a P{\'e}rez} et~al.}{2016}]{GarciaPerez_2016}
{Garc{\'{\i}}a P{\'e}rez} A.~E.,  et~al., 2016, \mn@doi [\aj]
  {10.3847/0004-6256/151/6/144}, \href
  {http://adsabs.harvard.edu/abs/2016AJ....151..144G} {151, 144}

\bibitem[\protect\citeauthoryear{{Glebocki}, {Gnacinski}  \&
  {Stawikowski}}{{Glebocki} et~al.}{2000}]{Glebocki_2000}
{Glebocki} R.,  {Gnacinski} P.,   {Stawikowski} A.,  2000, \actaa, \href
  {http://adsabs.harvard.edu/abs/2000AcA....50..509G} {50, 509}

\bibitem[\protect\citeauthoryear{{Gonz{\'a}lez} \& {Levato}}{{Gonz{\'a}lez} \&
  {Levato}}{2006}]{Gonzalez_2006}
{Gonz{\'a}lez} J.~F.,  {Levato} H.,  2006, \mn@doi [\aap]
  {10.1051/0004-6361:20053177}, \href
  {http://adsabs.harvard.edu/abs/2006A%26A...448..283G} {448, 283}

\bibitem[\protect\citeauthoryear{{Griffin} \& {Griffin}}{{Griffin} \&
  {Griffin}}{2010}]{Griffin_2010}
{Griffin} R.~E.~M.,  {Griffin} R.~F.,  2010, \mn@doi [\mnras]
  {10.1111/j.1365-2966.2009.15800.x}, \href
  {http://adsabs.harvard.edu/abs/2010MNRAS.402.1675G} {402, 1675}

\bibitem[\protect\citeauthoryear{{Hadrava}}{{Hadrava}}{1995}]{Hadrava_1995}
{Hadrava} P.,  1995, \aaps, \href
  {http://adsabs.harvard.edu/abs/1995A%26AS..114..393H} {114, 393}

\bibitem[\protect\citeauthoryear{{Halbwachs}, {Kiefer}, {Arenou}, {Famaey},
  {Guillout}, {Ibata}, {Mazeh}  \& {Pourbaix}}{{Halbwachs}
  et~al.}{2017b}]{Halbwachs_2017b}
{Halbwachs} J.-L.,  {Kiefer} F.,  {Arenou} F.,  {Famaey} B.,  {Guillout} P.,
  {Ibata} R.,  {Mazeh} T.,   {Pourbaix} D.,  2017b, preprint, \href
  {http://adsabs.harvard.edu/abs/2017arXiv171002301H} {} (\mn@eprint {arXiv}
  {1710.02301})

\bibitem[\protect\citeauthoryear{{Halbwachs} et~al.,}{{Halbwachs}
  et~al.}{2017a}]{Halbwachs_2017}
{Halbwachs} J.-L.,  et~al., 2017a, preprint, \href
  {http://adsabs.harvard.edu/abs/2017arXiv171002017H} {} (\mn@eprint {arXiv}
  {1710.02017})

\bibitem[\protect\citeauthoryear{{Hettinger}, {Badenes}, {Strader}, {Bickerton}
   \& {Beers}}{{Hettinger} et~al.}{2015}]{Hettinger_2015}
{Hettinger} T.,  {Badenes} C.,  {Strader} J.,  {Bickerton} S.~J.,   {Beers}
  T.~C.,  2015, \mn@doi [\apjl] {10.1088/2041-8205/806/1/L2}, \href
  {http://adsabs.harvard.edu/abs/2015ApJ...806L...2H} {806, L2}

\bibitem[\protect\citeauthoryear{{Ho} et~al.,}{{Ho} et~al.}{2017}]{Ho_2017}
{Ho} A.~Y.~Q.,  et~al., 2017, \mn@doi [\apj] {10.3847/1538-4357/836/1/5}, \href
  {http://adsabs.harvard.edu/abs/2017ApJ...836....5H} {836, 5}

\bibitem[\protect\citeauthoryear{{Holtzman} et~al.,}{{Holtzman}
  et~al.}{2015}]{Holtzman_2015}
{Holtzman} J.~A.,  et~al., 2015, \mn@doi [\aj] {10.1088/0004-6256/150/5/148},
  \href {http://adsabs.harvard.edu/abs/2015AJ....150..148H} {150, 148}

\bibitem[\protect\citeauthoryear{Hunter}{Hunter}{2007}]{Hunter_2007}
Hunter J.~D.,  2007, \mn@doi [Computing In Science \& Engineering]
  {10.1109/MCSE.2007.55}, 9, 90

\bibitem[\protect\citeauthoryear{{Hurley} \& {Tout}}{{Hurley} \&
  {Tout}}{1998}]{Hurley_1998}
{Hurley} J.,  {Tout} C.~A.,  1998, \mn@doi [\mnras]
  {10.1046/j.1365-8711.1998.01981.x}, \href
  {http://adsabs.harvard.edu/abs/1998MNRAS.300..977H} {300, 977}

\bibitem[\protect\citeauthoryear{{Iglesias-Marzoa}, {L{\'o}pez-Morales}  \&
  {Jes{\'u}s Ar{\'e}valo Morales}}{{Iglesias-Marzoa}
  et~al.}{2015}]{IglesiasMarzoa_2015}
{Iglesias-Marzoa} R.,  {L{\'o}pez-Morales} M.,   {Jes{\'u}s Ar{\'e}valo
  Morales} M.,  2015, \mn@doi [\pasp] {10.1086/682056}, \href
  {http://adsabs.harvard.edu/abs/2015PASP..127..567I} {127, 567}

\bibitem[\protect\citeauthoryear{{Katoh}, {Itoh}, {Toyota}  \& {Sato}}{{Katoh}
  et~al.}{2013}]{Katoh_2013}
{Katoh} N.,  {Itoh} Y.,  {Toyota} E.,   {Sato} B.,  2013, \mn@doi [\aj]
  {10.1088/0004-6256/145/2/41}, \href
  {http://adsabs.harvard.edu/abs/2013AJ....145...41K} {145, 41}

\bibitem[\protect\citeauthoryear{{Koch} \& {Hrivnak}}{{Koch} \&
  {Hrivnak}}{1981}]{Koch_1981}
{Koch} R.~H.,  {Hrivnak} B.~J.,  1981, \mn@doi [\aj] {10.1086/112902}, \href
  {http://adsabs.harvard.edu/abs/1981AJ.....86..438K} {86, 438}

\bibitem[\protect\citeauthoryear{{Konacki}}{{Konacki}}{2005}]{Konacki_2005}
{Konacki} M.,  2005, \mn@doi [\apj] {10.1086/429880}, \href
  {http://esoads.eso.org/abs/2005ApJ...626..431K} {626, 431}

\bibitem[\protect\citeauthoryear{{Kozai}}{{Kozai}}{1962}]{Kozai_1962}
{Kozai} Y.,  1962, \mn@doi [\aj] {10.1086/108790}, \href
  {http://adsabs.harvard.edu/abs/1962AJ.....67..591K} {67, 591}

\bibitem[\protect\citeauthoryear{{Kurucz}}{{Kurucz}}{1970}]{Kurucz_1970}
{Kurucz} R.~L.,  1970, SAO Special Report, \href
  {http://adsabs.harvard.edu/abs/1970SAOSR.309.....K} {309}

\bibitem[\protect\citeauthoryear{{Kurucz}}{{Kurucz}}{1979}]{Kurucz_1979}
{Kurucz} R.~L.,  1979, \mn@doi [\apjs] {10.1086/190589}, \href
  {http://adsabs.harvard.edu/abs/1979ApJS...40....1K} {40, 1}

\bibitem[\protect\citeauthoryear{{Kurucz}}{{Kurucz}}{1993}]{Kurucz_1993}
{Kurucz} R.~L.,  1993, {SYNTHE spectrum synthesis programs and line data}

\bibitem[\protect\citeauthoryear{{Li}, {de Grijs}  \& {Deng}}{{Li}
  et~al.}{2013}]{Li_2013}
{Li} C.,  {de Grijs} R.,   {Deng} L.,  2013, \mn@doi [\mnras]
  {10.1093/mnras/stt1669}, \href
  {http://adsabs.harvard.edu/abs/2013MNRAS.436.1497L} {436, 1497}

\bibitem[\protect\citeauthoryear{{Machida}}{{Machida}}{2008}]{Machida_2008}
{Machida} M.~N.,  2008, \mn@doi [\apjl] {10.1086/590109}, \href
  {http://adsabs.harvard.edu/abs/2008ApJ...682L...1M} {682, L1}

\bibitem[\protect\citeauthoryear{{Majewski} et~al.,}{{Majewski}
  et~al.}{2017}]{Majewski_2017}
{Majewski} S.~R.,  et~al., 2017, \mn@doi [\aj] {10.3847/1538-3881/aa784d},
  \href {http://adsabs.harvard.edu/abs/2017AJ....154...94M} {154, 94}

\bibitem[\protect\citeauthoryear{{Matijevi{\v c}} et~al.,}{{Matijevi{\v c}}
  et~al.}{2010}]{Matijevic_2010}
{Matijevi{\v c}} G.,  et~al., 2010, \mn@doi [\aj]
  {10.1088/0004-6256/140/1/184}, \href
  {http://adsabs.harvard.edu/abs/2010AJ....140..184M} {140, 184}

\bibitem[\protect\citeauthoryear{{Matijevi{\v c}} et~al.,}{{Matijevi{\v c}}
  et~al.}{2011}]{Matijevic_2011}
{Matijevi{\v c}} G.,  et~al., 2011, \mn@doi [\aj]
  {10.1088/0004-6256/141/6/200}, \href
  {http://adsabs.harvard.edu/abs/2011AJ....141..200M} {141, 200}

\bibitem[\protect\citeauthoryear{{Merle} et~al.,}{{Merle}
  et~al.}{2017}]{Merle_2017}
{Merle} T.,  et~al., 2017, preprint, \href
  {http://adsabs.harvard.edu/abs/2017arXiv170701720M} {} (\mn@eprint {arXiv}
  {1707.01720})

\bibitem[\protect\citeauthoryear{{Michalik}, {Lindegren}  \&
  {Hobbs}}{{Michalik} et~al.}{2015}]{Michalik_2015}
{Michalik} D.,  {Lindegren} L.,   {Hobbs} D.,  2015, \mn@doi [\aap]
  {10.1051/0004-6361/201425310}, \href
  {http://adsabs.harvard.edu/abs/2015A%26A...574A.115M} {574, A115}

\bibitem[\protect\citeauthoryear{{Minor}}{{Minor}}{2013}]{Minor_2013}
{Minor} Q.~E.,  2013, \mn@doi [\apj] {10.1088/0004-637X/779/2/116}, \href
  {http://adsabs.harvard.edu/abs/2013ApJ...779..116M} {779, 116}

\bibitem[\protect\citeauthoryear{{Moe} \& {Di Stefano}}{{Moe} \& {Di
  Stefano}}{2017}]{Moe_2017}
{Moe} M.,  {Di Stefano} R.,  2017, \mn@doi [\apjs] {10.3847/1538-4365/aa6fb6},
  \href {http://adsabs.harvard.edu/abs/2017ApJS..230...15M} {230, 15}

\bibitem[\protect\citeauthoryear{{Murray} \& {Correia}}{{Murray} \&
  {Correia}}{2010}]{Murray_2010}
{Murray} C.~D.,  {Correia} A.~C.~M.,  2010, {Keplerian Orbits and Dynamics of
  Exoplanets}.
pp 15--23

\bibitem[\protect\citeauthoryear{{Naoz}, {Farr}, {Lithwick}, {Rasio}  \&
  {Teyssandier}}{{Naoz} et~al.}{2013}]{Naoz_2013}
{Naoz} S.,  {Farr} W.~M.,  {Lithwick} Y.,  {Rasio} F.~A.,   {Teyssandier} J.,
  2013, \mn@doi [\mnras] {10.1093/mnras/stt302}, \href
  {http://adsabs.harvard.edu/abs/2013MNRAS.431.2155N} {431, 2155}

\bibitem[\protect\citeauthoryear{{Ness}, {Hogg}, {Rix}, {Ho}  \&
  {Zasowski}}{{Ness} et~al.}{2015}]{Ness_2015}
{Ness} M.,  {Hogg} D.~W.,  {Rix} H.-W.,  {Ho} A.~Y.~Q.,   {Zasowski} G.,  2015,
  \mn@doi [\apj] {10.1088/0004-637X/808/1/16}, \href
  {http://adsabs.harvard.edu/abs/2015ApJ...808...16N} {808, 16}

\bibitem[\protect\citeauthoryear{{Nidever} et~al.,}{{Nidever}
  et~al.}{2015}]{Nidever_2015}
{Nidever} D.~L.,  et~al., 2015, \mn@doi [\aj] {10.1088/0004-6256/150/6/173},
  \href {http://adsabs.harvard.edu/abs/2015AJ....150..173N} {150, 173}

\bibitem[\protect\citeauthoryear{{Pavlovski} \& {Hensberge}}{{Pavlovski} \&
  {Hensberge}}{2010}]{Pavlovski_2010}
{Pavlovski} K.,  {Hensberge} H.,  2010, in {Pr{\v s}a} A.,  {Zejda} M.,  eds,
  Astronomical Society of the Pacific Conference Series Vol. 435, Binaries -
  Key to Comprehension of the Universe. p.~207 (\mn@eprint {arXiv} {0909.3246})

\bibitem[\protect\citeauthoryear{{Pourbaix} et~al.,}{{Pourbaix}
  et~al.}{2004}]{Pourbaix_2004}
{Pourbaix} D.,  et~al., 2004, \mn@doi [\aap] {10.1051/0004-6361:20041213},
  \href {http://adsabs.harvard.edu/abs/2004A%26A...424..727P} {424, 727}

\bibitem[\protect\citeauthoryear{{Price-Whelan}, {Hogg}, {Foreman-Mackey}  \&
  {Rix}}{{Price-Whelan} et~al.}{2017}]{PriceWhelan_2017}
{Price-Whelan} A.~M.,  {Hogg} D.~W.,  {Foreman-Mackey} D.,   {Rix} H.-W.,
  2017, \mn@doi [\apj] {10.3847/1538-4357/aa5e50}, \href
  {http://adsabs.harvard.edu/abs/2017ApJ...837...20P} {837, 20}

\bibitem[\protect\citeauthoryear{{Raghavan} et~al.,}{{Raghavan}
  et~al.}{2010}]{Raghavan_2010}
{Raghavan} D.,  et~al., 2010, \mn@doi [\apjs] {10.1088/0067-0049/190/1/1},
  \href {http://adsabs.harvard.edu/abs/2010ApJS..190....1R} {190, 1}

\bibitem[\protect\citeauthoryear{{Reid} \& {Gizis}}{{Reid} \&
  {Gizis}}{1997}]{Reid_1997}
{Reid} I.~N.,  {Gizis} J.~E.,  1997, \mn@doi [\aj] {10.1086/118436}, \href
  {http://adsabs.harvard.edu/abs/1997AJ....113.2246R} {113, 2246}

\bibitem[\protect\citeauthoryear{{Reis}, {Poznanski}, {Baron}, {Zasowski}  \&
  {Shahaf}}{{Reis} et~al.}{2017}]{Reis_2017}
{Reis} I.,  {Poznanski} D.,  {Baron} D.,  {Zasowski} G.,   {Shahaf} S.,  2017,
  preprint, \href {http://adsabs.harvard.edu/abs/2017arXiv171100022R} {}
  (\mn@eprint {arXiv} {1711.00022})

\bibitem[\protect\citeauthoryear{{Robin}, {Reyl{\'e}}, {Derri{\`e}re}  \&
  {Picaud}}{{Robin} et~al.}{2003}]{Robin_2003}
{Robin} A.~C.,  {Reyl{\'e}} C.,  {Derri{\`e}re} S.,   {Picaud} S.,  2003,
  \mn@doi [\aap] {10.1051/0004-6361:20031117}, \href
  {http://adsabs.harvard.edu/abs/2003A%26A...409..523R} {409, 523}

\bibitem[\protect\citeauthoryear{{Schatzman}}{{Schatzman}}{1962}]{Schatzman_1962}
{Schatzman} E.,  1962, Annales d'Astrophysique, \href
  {http://esoads.eso.org/abs/1962AnAp...25...18S} {25, 18}

\bibitem[\protect\citeauthoryear{{Sharma}, {Bland-Hawthorn}, {Johnston}  \&
  {Binney}}{{Sharma} et~al.}{2011}]{Sharma_2011}
{Sharma} S.,  {Bland-Hawthorn} J.,  {Johnston} K.~V.,   {Binney} J.,  2011,
  \mn@doi [\apj] {10.1088/0004-637X/730/1/3}, \href
  {http://adsabs.harvard.edu/abs/2011ApJ...730....3S} {730, 3}

\bibitem[\protect\citeauthoryear{{Simon} \& {Sturm}}{{Simon} \&
  {Sturm}}{1994}]{Simon_1994}
{Simon} K.~P.,  {Sturm} E.,  1994, \aap, \href
  {http://adsabs.harvard.edu/abs/1994A%26A...281..286S} {281, 286}

\bibitem[\protect\citeauthoryear{{Simons}, {Henry}  \& {Kirkpatrick}}{{Simons}
  et~al.}{1996}]{Simons_1996}
{Simons} D.~A.,  {Henry} T.~J.,   {Kirkpatrick} J.~D.,  1996, \mn@doi [\aj]
  {10.1086/118176}, \href {http://adsabs.harvard.edu/abs/1996AJ....112.2238S}
  {112, 2238}

\bibitem[\protect\citeauthoryear{{Terrien} et~al.,}{{Terrien}
  et~al.}{2014}]{Terrien_2014}
{Terrien} R.~C.,  et~al., 2014, \mn@doi [\apj] {10.1088/0004-637X/782/2/61},
  \href {http://adsabs.harvard.edu/abs/2014ApJ...782...61T} {782, 61}

\bibitem[\protect\citeauthoryear{{Ting}, {Rix}, {Conroy}, {Ho}  \&
  {Lin}}{{Ting} et~al.}{2017}]{Ting_2017b}
{Ting} Y.-S.,  {Rix} H.-W.,  {Conroy} C.,  {Ho} A.~Y.~Q.,   {Lin} J.,  2017,
  preprint, \href {http://adsabs.harvard.edu/abs/2017arXiv170801758T} {}
  (\mn@eprint {arXiv} {1708.01758})

\bibitem[\protect\citeauthoryear{{Toonen}, {Hamers}  \& {Portegies
  Zwart}}{{Toonen} et~al.}{2016}]{Toonen_2016}
{Toonen} S.,  {Hamers} A.,   {Portegies Zwart} S.,  2016, \mn@doi
  [Computational Astrophysics and Cosmology] {10.1186/s40668-016-0019-0}, \href
  {http://adsabs.harvard.edu/abs/2016ComAC...3....6T} {3, 6}

\bibitem[\protect\citeauthoryear{{Traven} et~al.,}{{Traven}
  et~al.}{2017}]{Traven_2017}
{Traven} G.,  et~al., 2017, \mn@doi [\apjs] {10.3847/1538-4365/228/2/24}, \href
  {http://adsabs.harvard.edu/abs/2017ApJS..228...24T} {228, 24}

\bibitem[\protect\citeauthoryear{{Troup} et~al.,}{{Troup}
  et~al.}{2016}]{Troup_2016}
{Troup} N.~W.,  et~al., 2016, \mn@doi [\aj] {10.3847/0004-6256/151/3/85}, \href
  {http://adsabs.harvard.edu/abs/2016AJ....151...85T} {151, 85}

\bibitem[\protect\citeauthoryear{{Van Der Walt}, {Colbert}  \&
  {Varoquaux}}{{Van Der Walt} et~al.}{2011}]{vanderwalt_2011}
{Van Der Walt} S.,  {Colbert} S.~C.,   {Varoquaux} G.,  2011, preprint, \href
  {http://adsabs.harvard.edu/abs/2011arXiv1102.1523V} {} (\mn@eprint {arXiv}
  {1102.1523})

\bibitem[\protect\citeauthoryear{{Zasowski} et~al.,}{{Zasowski}
  et~al.}{2013}]{Zasowski_2013}
{Zasowski} G.,  et~al., 2013, \mn@doi [\aj] {10.1088/0004-6256/146/4/81}, \href
  {http://adsabs.harvard.edu/abs/2013AJ....146...81Z} {146, 81}

\bibitem[\protect\citeauthoryear{{Zucker} \& {Mazeh}}{{Zucker} \&
  {Mazeh}}{1994}]{Zucker_1994}
{Zucker} S.,  {Mazeh} T.,  1994, \mn@doi [\apj] {10.1086/173605}, \href
  {http://adsabs.harvard.edu/abs/1994ApJ...420..806Z} {420, 806}

\makeatother
\end{thebibliography}



\appendix

\section{Neural network spectral model}
\label{sec:neural_network}
As mentioned in Section~\ref{sec:single_star}, we use a neural network to predict the normalized flux density at a given wavelength pixel as a function of stellar labels. As applied in this work, a neural network is essentially a flexible function produced through the composition of simple functions. It takes as its argument a vector of labels ($\vec{\ell}$; Equation~\ref{eqn:single}) and returns the normalized flux density predicted at a particular wavelength pixel. 

For the neural network used in this work, which contains a single hidden layer with 5 neurons, the normalized flux density at wavelength pixel $\lambda$ is given by
\begin{align}
\label{eqn:NN}
\hat{f}_{\lambda}=\tilde{w}_{\lambda}^{i}\sigma\left(w_{\lambda i}^{k}\hat{\ell}_{k}+b_{\lambda i}\right)+\tilde{b}_{\lambda}
\end{align}
with implied summation over $k=1\dots N_{{\rm labels}}$ and $i=1\dots N_{{\rm neurons}}$. Here $\hat{\ell}=(\vec{\ell}-\vec{\ell}_{{\rm min}})/(\vec{\ell}_{{\rm max}}-\vec{\ell}_{{\rm min}})-0.5$ is a scaled label vector, $\vec{\ell}_{{\rm max}}$ and $\vec{\ell}_{{\rm min}}$ are vectors of the maximum and minimum values of each label in the training set, and $\sigma(z)=1/\left(1+e^{-z}\right)$ is the ``sigmoid'' activation function. The \textit{weights}, $w$ and $\tilde{w}$, and \textit{biases}, $b$ and $\tilde{b}$, parameterize the neural network; these are the free parameters that are adjusted during training. 

In order to treat spectra with different line-of-sight velocities, all spectra are shifted to rest-frame and linearly interpolated onto a common wavelength grid. Training the model consists of minimizing a \textit{loss function}, comparable to the $\chi^2$ statistic, that quantifies how well the model can fit the training set. We use an L1 loss function, which minimizes the total absolute difference between fluxes predicted by the neural network and those in the training set. We expect this to perform better than e.g. the $\chi^2$ statistic during the iterative cleaning of the training set and re-training, since it is less sensitive to outliers. During training, we mask all pixels with S/N $< 50$, bad or missing pixels, and pixels with poor sky subtraction. 

We implement and train the neural network using the python package \texttt{PyTorch}. We tested a wide range of network architectures, varying the network depth, width, and activation function, with both data-driven and synthetic spectral models. We find that using a small neural network and a large training set is the most straightforward way to prevent overfitting; using a substantially larger network with more neurons or hidden layers causes the model to reach lower losses (i.e., fit the training set better) but perform worse in cross-validation. We verified that our spectral model performs equally well on the training and cross-validation sets at fixed S/N, so it does not overfit the training set. 

An advantage of using a neural network spectral model is that the neural network's flexibility makes it possible to model a wide range of stellar parameters in a single model, rather than stitching together multiple models covering different regions of label space. However, our basic approach of constructing a binary spectral model does not depend critically on use of a neural network; a comparable binary model could likely be built from other forms of single-star model (e.g., \textit{The Cannon}). 

\section{Model selection}
\label{sec:model_selection}
Because the single-star model is a special case of the binary model, it is \textit{always} possible to obtain a binary model that fits a data spectrum at least as well as does the best-fit single-star model. As the binary model is more complex than the single-star model, with three additional free parameters, one might expect to find a better fit, in a $\chi^2$ sense, with the binary model even for targets which are true single stars. It is therefore necessary to formulate a heuristic to determine ``how much'' better a fit with a binary model is required to constitute reliable evidence in favor of the binary model. 

The primary statistic used for model selection is the $\chi^2$ difference, $\Delta \chi^2=\chi^2_{\rm single} - \chi^2_{\rm binary}$, which simply quantifies how much better a fit is obtained by the binary model. We also calculate a second statistic, the ``improvement fraction'' $f_{\rm imp}$, to quantify how much better a fit the binary model achieves \textit{relative to how different it is from the single-star model.} The basic idea here is that if a binary model spectrum is very different from the single-star model, but only achieves a slightly better fit to the data, this constitutes weaker evidence in favor of the binary model than a case with comparable $\Delta \chi^2$ in which most of the difference between the best-fit binary and single-star model goes toward improving the fit. The improvement fraction is defined as
\begin{align}
\label{eqn:f_imp}
f_{{\rm imp}}=\frac{\sum\left\{ \left(\left|\hat{f}_{\lambda,{\rm single}}-\hat{f}_{\lambda}\right|-\left|\hat{f}_{\lambda,{\rm binary}}-\hat{f}_{\lambda}\right|\right)/\hat{\sigma}_{\lambda}\right\} }{\sum\left\{ \left|\hat{f}_{\lambda,{\rm single}}-\hat{f}_{\lambda,{\rm binary}}\right|/\hat{\sigma}_{\lambda}\right\} },
\end{align}
where $\hat{f}_{\lambda}$ and $\hat{\sigma}_{\lambda}$ are the normalized flux density and corresponding uncertainty, $\hat{f}_{\lambda,{\rm single}}$ and $\hat{f}_{\lambda,{\rm binary}}$ are the best-fit normalized single-star and binary model spectra, and the sum is over all wavelength pixels. 

Our full acceptance criterion for preferring the binary model is given in Table~\ref{tab:criteria}. These thresholds were motivated in part by the $\Delta \chi^2$ and $f_{\rm imp}$ values calculated for semi-empirical binaries as described below, and in part by validation with the CMD (Figure~\ref{fig:cmd}). The adopted thresholds are conservative, and prevent us from identifying some binaries whose spectra can only be marginally better fit by a binary model; however, setting them to substantially lower values causes the model to begin categorizing more targets near the main sequence of the CMD as binaries, indicating a non-negligible false-positive rate. We have not tuned the minimum $f_{\rm imp}$ value for each $\Delta \chi^2$; we set the intuitive requirement that a higher $f_{\rm imp}$ should be required for systems with a lower $\Delta \chi^2$. 

Different $\Delta \chi^2$ thresholds are required for systems identified as potential close binaries, because (a) for these systems, we fit multiple visit spectra simultaneously, and (b) we fit a total of 5 different models (see Section~\ref{sec:summary_labels}). For these systems, we begin with a fiducial threshold of $\Delta \chi^2 = 300\times N_{\rm epochs}$ for each increase in model complexity, where $N_{\rm epochs}$ is the number of visit spectra fit simultaneously; i.e., we require $\chi^2_{\rm single\,star}-\chi^2_{\rm SB1} > 300 N_{\rm epochs}$ for a system to be initially classified as an SB1; $\chi^2_{\rm SB1}-\chi^2_{\rm SB2} > 300 N_{\rm epochs}$ for a system to be initially classified as an SB2, etc. We then inspected the spectra of these targets individually and reclassified suspected false positives (see Appendix~\ref{sec:false_positive}).  

We experimented with generating single-star spectra directly from the spectral model, adding noise, and fitting them with both a binary and single-star model. This produces typical $\Delta \chi^2$ values of order unity, which is smaller than we find for the majority of APOGEE targets. This occurs because spectra generated in this way can necessarily be perfectly fit by the single-star spectral model, which is not necessarily the case for real spectra. We quantify the $\Delta \chi^2$ values expected for real binary spectra in the next section. 

\begin{table}
\begin{center}
\begin{tabular}{ |c|c|} 
 \hline
$\Delta \chi^2 = \chi^2_{\rm single} - \chi^2_{\rm binary}$ & minimum $f_{\rm imp}$ \\ 
\hline 
$\Delta \chi^2 \geq 3000$ & 0 \\ 
$2500 \leq \Delta \chi^2 < 3000$ & 0.05 \\ 
$2000 \leq \Delta \chi^2 < 2500$ & 0.075 \\ 
$1500 \leq \Delta \chi^2 < 2000$ & 0.1 \\ 
$1000 \leq \Delta \chi^2 < 1500$ & 0.125 \\ 
$750  \leq \Delta \chi^2 < 1000$ & 0.15 \\ 
$600  \leq \Delta \chi^2 < 750 $ & 0.175 \\ 
$450  \leq \Delta \chi^2 < 600 $ & 0.2 \\ 
$300 \leq \Delta \chi^2 <  450 $ & 0.225 \\ 
 \hline
\end{tabular}
\end{center}
	\caption{Minimum $\Delta \chi^2$ and improvement fraction $f_{\rm imp}$ (Equation~\ref{eqn:f_imp}) for a target to be classified as a binary. All systems with $\Delta \chi^2<300$, and all systems falling below the minimum $f_{\rm imp}$ for a given $\Delta \chi^2$, are classified as inconclusive; i.e., showing no strong evidence of binarity.}
    \label{tab:criteria}
\end{table}

\subsection{Tests with semi-empirical synthetic binary spectra}
\label{sec:semi_empirical}

\begin{figure*}
\includegraphics[width=\textwidth]{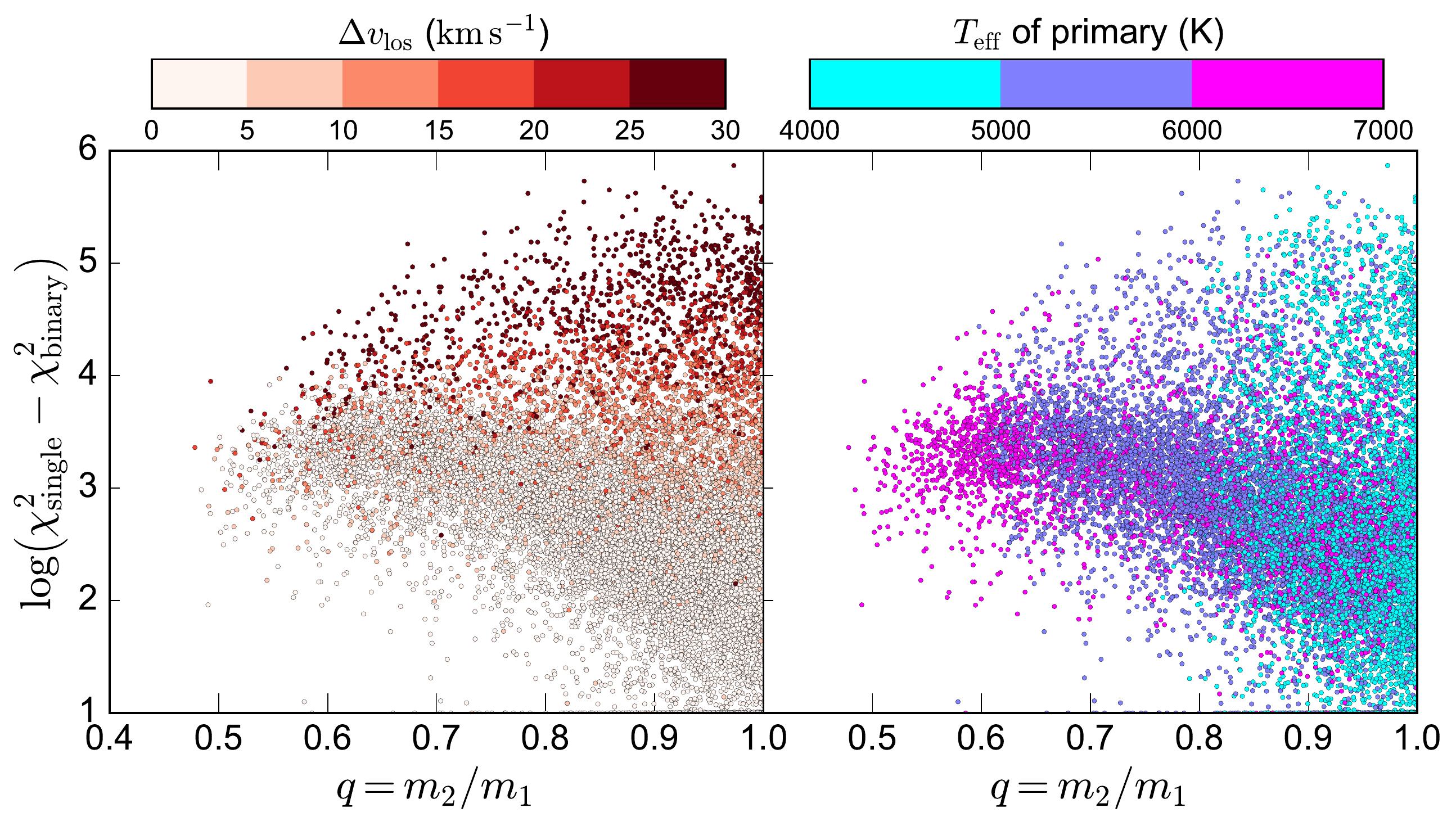}
\caption{Results of fitting semi-empirical binary spectra with single-star and binary models. Semi-empirical binary spectra are created by adding together flux-calibrated APOGEE spectra of two stars with similar abundances. At $q\lesssim 0.8$, most semi-empirical binary spectra can be significantly better fit with a binary model, with $\Delta \chi^2 \gtrsim 1000$. The $\chi^2$ difference is nearly always larger for systems with large velocity offsets; for systems with $q\sim 1$, only binaries with $\Delta v_{\rm los}\gtrsim 10\,{\rm km\,s^{-1}}$ have $\Delta \chi^2 \gtrsim 1000$. At fixed $q$ and $\Delta v_{\rm los}$, the typical $\Delta \chi^2$ is larger for systems with cooler primaries. Due to our single-star spectral model's minimum $T_{\rm eff}$ of 4200 K, low mass ratio systems can only be modeled for hot primaries.}
\label{fig:semi-empirical}
\end{figure*}

To assess the accuracy and potential systematics of our method, and to measure the expected $\Delta \chi^2$ values for true binaries with a particular mass ratio, we construct a library of $\sim$15,000 ``semi-empirical'' synthetic binary spectra. These are created by combining randomly chosen pairs of APOGEE spectra in unnormalized space, following the method outlined in Section~\ref{sec:binary_spectra}. We then normalize and fit these spectra following the same procedure used to fit real spectra. An advantage of constructing synthetic binary spectra by combining real spectra (as opposed to simply generating binary spectra from the data-driven model) is that this accounts for the possibility that the model does not capture all the variance in the real spectra; this is likely the case for our model, which only contains 5 labels. 

\begin{figure*}
\includegraphics[width=\textwidth]{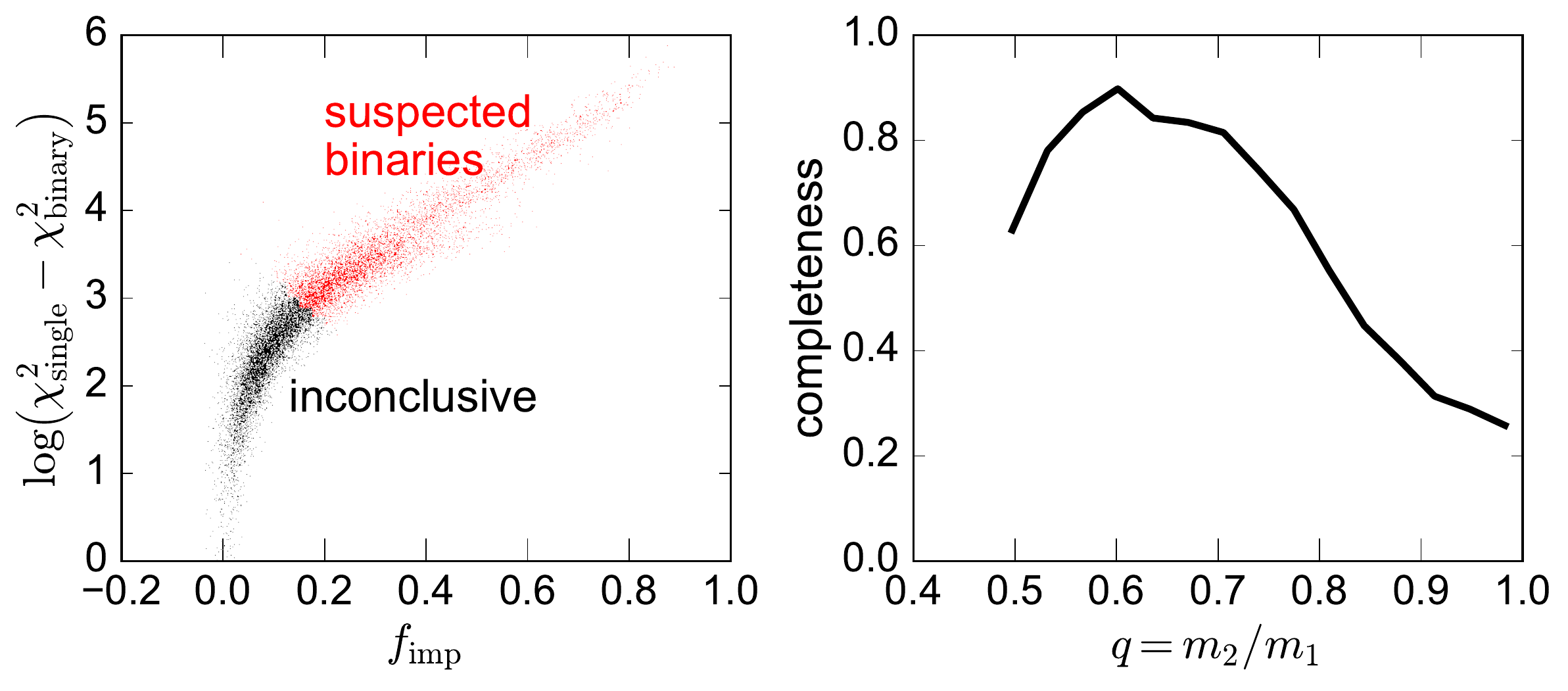}
\caption{Results for fitting semi-empirical synthetic binary spectra. Left panel shows the $\chi^2$ difference and improvement fraction (Equation~\ref{eqn:f_imp}); systems passing the adopted acceptance criterion for a binary candidate to be considered legitimate are plotted in red. Right panel shows the resulting completeness function; i.e., the fraction of semi-empirical binary systems at a given $q$ that are successfully identified as binaries.}
\label{fig:completeness}
\end{figure*}

\begin{figure}
\includegraphics[width=\columnwidth]{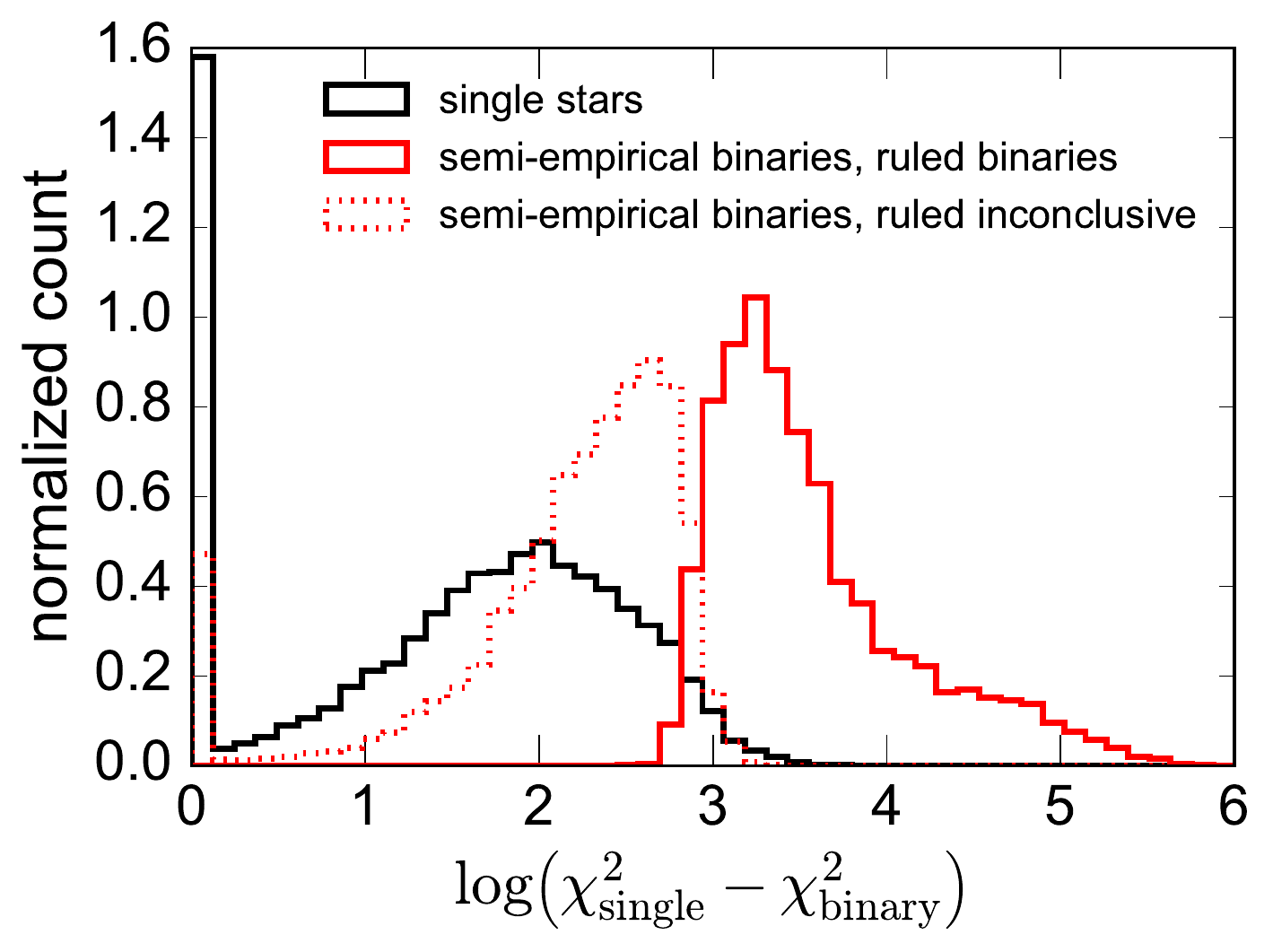}
\caption{$\chi^2$ differences between best-fit single and binary models. Black histogram shows suspected single stars from which semi-empirical binary spectra are constructed. Solid (dotted) red histogram shows semi-empirical binary spectra which pass (fail) the $\Delta \chi^2$ and $f_{\rm imp}$ binary acceptance criterion.}
\label{fig:semi_empirical_hist}
\end{figure}

We require that the two stars used to construct a semi-empirical binary spectrum have similar abundances (within 0.05 dex in [Fe/H] and [$\alpha$/Fe]) and fall within 0.03 dex in $\log g$ of a single isochrone. We only combine spectra consistent with being single stars; i.e., those which cannot be significantly better-fit by a binary model than a single-star model according to the thresholds in Table~\ref{tab:criteria}. We assign realistic orbital parameters to each system following \citetalias{ElBadry_2017}, drawing orbital periods from the log-normal period distribution for solar-type stars from \citet{Duchene_2013} and assuming random orbit orientations and phases. Results from fitting these semi-empirical binary spectra are shown in Figures~\ref{fig:semi-empirical},~\ref{fig:completeness},~\ref{fig:semi_empirical_hist}, and~\ref{fig:semi_empirical_acc}. 

In Figure~\ref{fig:semi-empirical}, we show the $\chi^2$ difference in favor of the binary model, $\Delta \chi^2 = \chi^2_{\rm single} - \chi^2_{\rm binary}$, as a function of the mass ratio $q$, line-of-sight velocity offset $\Delta v_{\rm los}$, and $T_{\rm eff}$ of the primary star. As expected, $\Delta \chi^2$ is a strong function of $q$: most binaries with $q \lesssim 0.75$ have $\Delta \chi^2 > 1000$, while most systems with $q\sim 1$ have much lower $\Delta \chi^2$. This is expected, because the two stars in binaries with $q\sim 1$ will have similar spectra, making the combined binary spectrum indistinguishable from that of either single star unless there is a sizable velocity offset between the two stars. The left panel of Figure~\ref{fig:semi-empirical} shows that $\Delta \chi^2$ is also a strong function of the velocity offset $\Delta v_{\rm los}$: at fixed $q$, systems in which the velocity offset is larger nearly always have larger $\Delta \chi^2$. In particular, most binaries with $\Delta v_{\rm los}\gtrsim 10\,\rm km\,s^{-1}$ have $\Delta \chi^2 > 1000$, even at $q\sim 1$. Indeed, among binaries with large velocity offsets, the typical $\Delta \chi^2$ is largest for systems with $q\sim 1$; in such systems, both stars contribute significantly to the spectrum, and absorption lines are obviously split. 

The right panel of Figure~\ref{fig:semi-empirical} shows the dependence of $\Delta \chi^2$, and the range of mass ratios to which our method is sensitive, on $T_{\rm eff}$ of the primary. At fixed mass ratio and $\Delta v$, the median $\Delta \chi^2$ is slightly lower for hot stars ($T_{\rm eff} > 6000\,\rm K$), particularly for systems with $q\sim 1$ and large $\Delta v$. This occurs because lines are on average weaker and more rotationally broadened in hot stars, reducing the information content of the spectrum. This panel also shows that, due to the minimum $T_{\rm eff}$ of $4200\,\rm K$ of the single-star spectral model, the minimum mass ratio that can be modeled varies with $T_{\rm eff}$ of the primary. This means that our completeness is higher for hot stars than for cool stars. 

To determine whether our binary model fit has converged on the true globally optimal model rather than a local $\chi^2$ minimum, we check whether the $\chi^2$ value of the best-fit binary model is at least as low as that corresponding to the binary model with the true labels of the system. We find that our fit converges on the globally optimal solution for $\sim$99\% of all semi-empirical binaries. About half of the systems in which the fit converges on a local minimum are binaries with $q\sim 1$ in which the velocity assignments of the primary and secondary star are switched; in these cases, the derived stellar labels are still reasonably accurate.

In Figure~\ref{fig:completeness}, we show how our adopted model selection threshold translates to the range of mass ratios to which the model is sensitive. In the left panel, we plot the distribution of semi-empirical binaries in $\Delta \chi^2 - f_{\rm imp}$ space. $\Delta \chi^2$ and $f_{\rm imp}$ are correlated: most systems whose spectrum can be significantly better fit by a binary model (high $\Delta \chi^2$) also have high $f_{\rm imp}$. In the right panel, we plot the fraction of semi-empirical binaries at a given mass ratio that pass our adopted model selection criteria to be considered reliable binary candidates. As expected, this ``completeness'' function is a strong function of mass ratio. Most binaries with $0.55 \lesssim q \lesssim 0.75$ have spectra that are sufficiently different from any single-star model that they can be unambiguously identified as binaries; at higher mass ratios, the spectra of the two component stars become similar, so only the $\sim 20\%$ of binaries with $\Delta v_{\rm los} \gtrsim 10\,\rm km\,s^{-1}$ can be detected. At sufficiently low mass ratios, the secondary contributes a negligible fraction of the total light. 

We emphasize that this completeness function does \textit{not} represent our global completeness function for all binaries, for two reasons. First, the population of binaries in our semi-empirical library is not statistically representative of the Galactic binary population: because our single-star spectral model cannot model stars with $T_{\rm eff} < 4200\,\rm K$, we cannot currently model low mass ratio binaries in which the primary is cool, as the secondary will have $T_{\rm eff} < 4200\,\rm K$ (see Section~\ref{sec:binary_spectra}). Second, we have made no attempt to model the APOGEE selection function, which would complicate the distribution of $T_{\rm eff}$ at a given $q$. For the particular set of semi-empirical binaries analyzed here, $\sim 50\%$ of all binary systems pass the model selection threshold to be characterized as binaries. However, this is not representative of the global sensitivity of our model, since we make no attempt to use a realistic distribution of mass ratios in the semi-empirical library: the distribution of mass ratios in our semi-empirical binary library is skewed toward $q=1$, which is precisely the regime in which the model performs poorly.

In Figure~\ref{fig:semi_empirical_hist}, we compare the distribution of $\Delta \chi^2$ values for the suspected single stars used in constructing the semi-empirical binary library to those for semi-empirical synthetic binary spectra. Systems for which $0 < \Delta \chi^2 < 1$ are assigned $\log \Delta \chi^2 = 0$ on this plot; each histogram is normalized separately. The median $\Delta \chi^2$ is $\sim$50 for single stars, $\sim$200 for semi-empirical binaries that fail the detection criteria to be considered reliable binaries, and $\sim$2600 for semi-empirical binaries that pass the threshold to be considered reliable. There is some overlap in the distribution of $\chi^2$ values for single stars and semi-empirical binaries, since for systems with small $\Delta v_{\rm los}$, the binary model in principle transitions smoothly to the single-star model both as $q\to 1$ and $q\to 0$. However, for binaries with favorable mass ratios, the typical $\Delta \chi^2$ is more than an order of magnitude greater than for single stars. 

\subsubsection{Cross validation}
\label{sec:cross_validation}
\begin{figure*}
\includegraphics[width=\textwidth]{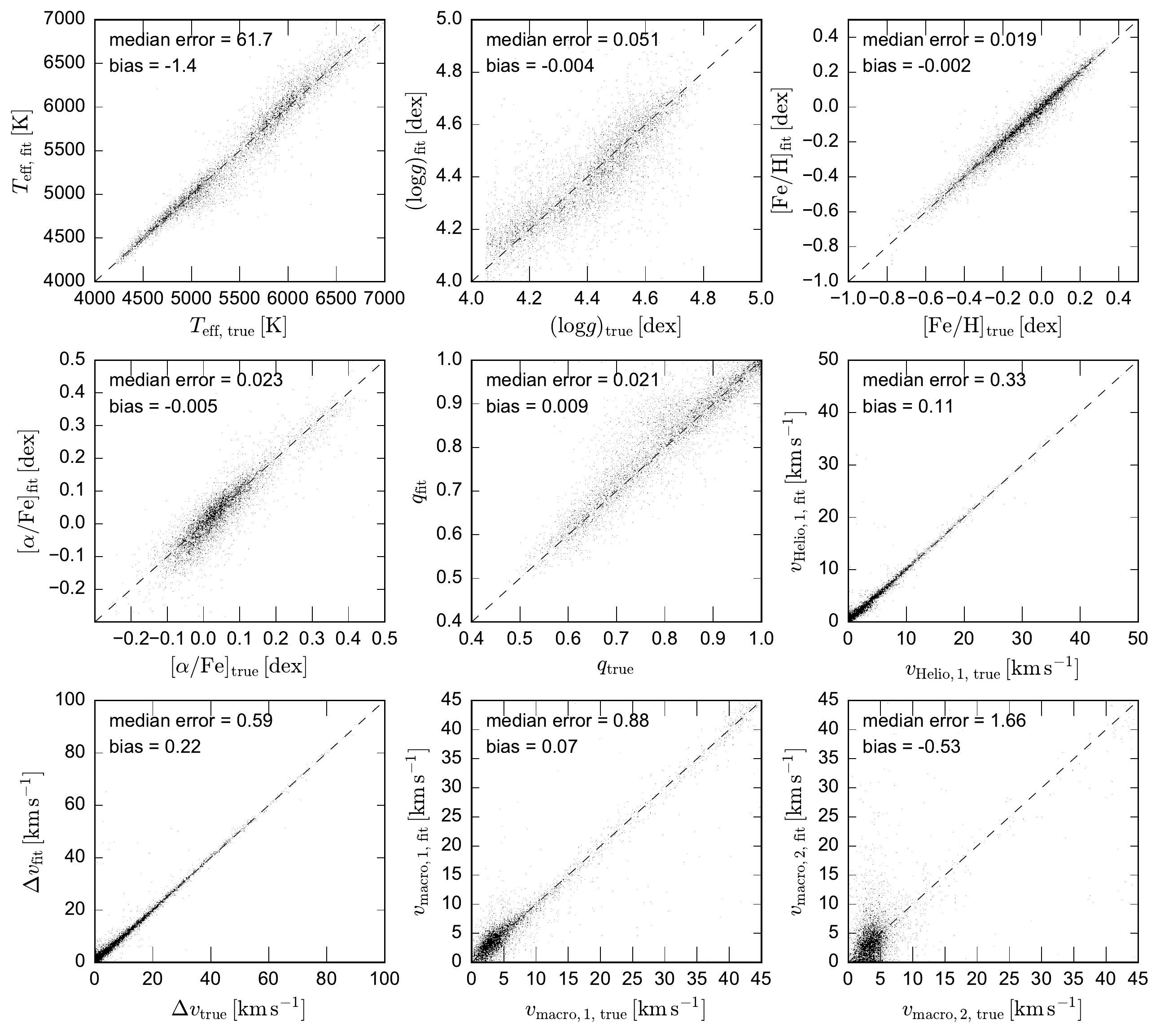}
\caption{Label recovery diagnostic for semi-empirical binary spectra that pass our $\Delta \chi^2$ and $f_{\rm imp}$ threshold to be considered real detections. For each label $\ell_i$, inset text indicates the median error, ${\rm med}(|\ell_{i,\,{\rm fit}} - \ell_{i,\,{\rm true}}|)$, and bias, ${\rm med}(\ell_{i,\,{\rm fit}} - \ell_{i,\,{\rm true}})$.}
\label{fig:semi_empirical_acc}
\end{figure*}

In Figure~\ref{fig:semi_empirical_acc}, we compare the best-fit labels for all semi-empirical binaries that pass the binary detection threshold to the true labels used in constructing the semi-empirical binary spectra. We set the ``true'' abundance for each binary as the average of the abundances of the two single-stars, which we required to be within 0.05 dex for [Fe/H] and [$\alpha$/Fe]. $T_{\rm eff}$ and $\log g$ refer to the primary. These semi-empirical binaries were not used to train the spectral model, so this experiment constitutes cross validation of the binary spectral model. In each panel, we indicate the median signed error (bias) and absolute error (scatter) in the best-fit label.  
Overall, this experiment reveals that labels inferred from fitting our binary model are reasonably precise: the true and best-fit labels fall near the one-to-one line, with small scatter. The only label for which this is not obviously true is $v_{\rm macro,2}$; this occurs primarily because only hot stars have $v_{\rm macro} \gtrsim 10\,\rm km\,s^{-1}$, so only the few binaries in which both stars are hot have non-negligible $v_{\rm macro,2}$. 

The median error in the best-fit $q$ for the semi-empirical binaries is 0.021, which is smaller than the median difference of 0.048 between $q_{\rm dyn}$ and $q_{\rm spec}$ found for real binaries in Figure~\ref{fig:q_spec_q_dyn}. This is not unexpected, because (a) $q_{\rm dyn}$ also has nonzero uncertainty, and (b) ``$q_{\rm true}$'' for the semi-empirical binaries is calculated with the same isochrones used in the model from which $q_{\rm fit}$ is obtained. That is, the median difference of 0.021 does not account for uncertainties in the isochrones; the larger difference of 0.048 does, because $q_{\rm dyn}$ is independent of isochrones. 

We emphasize that while Figure~\ref{fig:semi_empirical_acc} shows our derived stellar labels to be \textit{precise}, this does not guarantee that they are \textit{accurate}. The reason for this is that the uncertainties in stellar parameters obtained from spectral fitting are often dominated by systematic uncertainties in the model, which enter primarily from errors in the ``ground truth'' labels of the training and validation sets, and are not accounted for in cross-validation. The cross-validation errors therefore are reasonable estimates of the precision of the model, but represent lower limits on the absolute uncertainties. 

\section{False positives}
\label{sec:false_positive}
As discussed above, there are some targets for which our fitting and model selection formally prefers a binary model but visual inspection of the spectrum reveals that the evidence in favor of the binary model is weak and a single-star model should likely be preferred. After visually inspecting all $\sim$ 3300 targets for which our model selection thresholds preferred a model other than a single star, we flagged $\sim$ 300 targets as probable false-positives. The vast majority of such cases are hot stars ($T_{\rm eff} \gtrsim 6500\,\rm K$) exhibiting significant rotational broadening. 

\begin{figure*}
    \includegraphics[width=\textwidth]{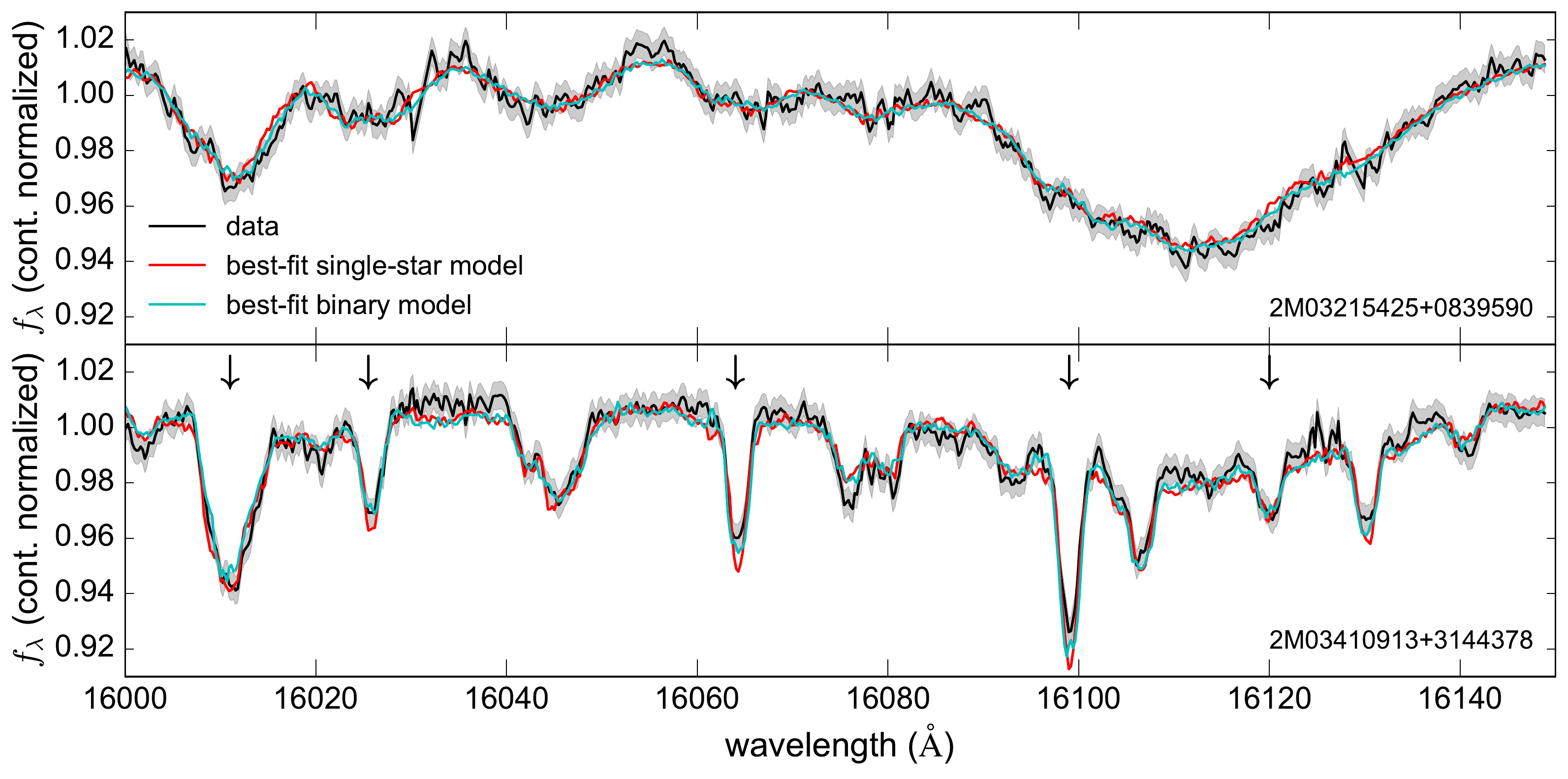}
    \caption{Spectra of two stars flagged as candidate binaries in our initial fitting that are likely false positives. \textbf{Top}: Nearly featureless spectrum of a hot ($T_{\rm eff}\approx \rm 7000\,K$), rotationally broadened star. The main difference between the single-star and binary model (which has $q=1$ and $\Delta v_{\rm los} = 40\,\rm km\,s^{-1}$) is that the binary model produces slightly broader lines. This target is an APOGEE telluric standard. \textbf{Bottom}: Although the binary model (with $q=1$ and $\Delta v_{\rm los} = 25\,\rm km\,s^{-1}$) is formally preferred, it has many marginally split lines, marked with arrows, which are not present in the data spectrum. This target is a young star in an embedded cluster.}
    \label{fig:false_positive}
\end{figure*}

In Figure~\ref{fig:false_positive}, we show example spectra of two stars flagged as false positives. The top panel shows the spectrum of a hot, rotating star. Such spectra are less informative than spectra of cooler stars with less rotational broadening: most metallic lines are intrinsically weaker, since more species are ionized at higher $T_{\rm eff}$, and the profiles of individual lines are smeared out due to rotation. Although the binary model is formally preferred, with $\Delta \chi^2 \sim 1000$, it does not obviously fit the profiles of individual lines better than the single-star model. Indeed, the main difference between the binary and single-star models is that the binary model produces slightly wider blended lines. 

Most of the stars we flag as false positives have spectra similar to this target and large $v_{\rm macro}$. There is in principle no reason to expect the method model to generically fail for such targets; the most likely explanation is that our 5-label single-star spectral model is not sufficiently complex to fully characterize the results of rotation. More than half of these targets are APOGEE telluric standard stars, which are selected to be used for telluric correction because they have nearly featureless spectra. Because hot, rotationally-broadened stars constitute only a small fraction of our initial sample, we simply visually inspected and removed questionable targets by hand. A straightforward method for automating this procedure in future work would be to remove all telluric standards and/or targets for which a Fourier transform of the spectrum reveals little power at high frequencies. 

The bottom panel of Figure~\ref{fig:false_positive} shows the spectrum of a false-positive candidate that is not featureless. The binary spectral model achieves a fit that is formally better than the single-star model ($\Delta \chi^2 \sim 1500$), but unlike the single-star model, the binary model spectrum has noticeably split lines as a result of the velocity offset between the two components. Because the data spectrum does not show such split line profiles, the binary model fit is likely erroneous and formally preferred only because it produces wider and shallower lines than the single-star model can accommodate. False-positives like this one are rare; only a few dozen spectra are identified in which the binary model produces a less realistic line profile than the single-star model despite achieving a formally better fit. A significant fraction of these are targets in young embedded clusters. Young stars often exhibit spectral features that are uncommon in older stars, such as chromospheric emission and increased rotation in cooler stars. Our method is more susceptible to incorrectly preferring a binary model if the spectrum cannot be well accommodated by the single-star model; this is likely the case for these targets. In future studies, such false-positives can potentially be eliminated by using a more complex spectral model for single stars and ensuring that different varieties of ``unusual'' spectra are represented in the training set. 

Besides reclassifying stars flagged as false-positives upon visual inspection to be single stars, we also reclassified some of the potential close binary targets (Section~\ref{sec:visit}) from one multiple-star class to another. For example, we reclassified SB2 systems for which the best-fit velocities of the primary and secondary fall on a one-to-one line with \textit{positive} slope as SB1s, and SB3s with broad lines that were not obviously better bit by the SB3 model than the SB2 model as SB2s. We attempted to be conservative in classifying systems as triples; i.e., some triple systems are likely miscategorized as binaries, but all systems classified as triples are unambiguously better fit by the SB3 model than the SB2 model.

\section{Orbit fitting convergence}
\label{sec:orbit_convergence}

\begin{figure*}
\includegraphics[width=\textwidth]{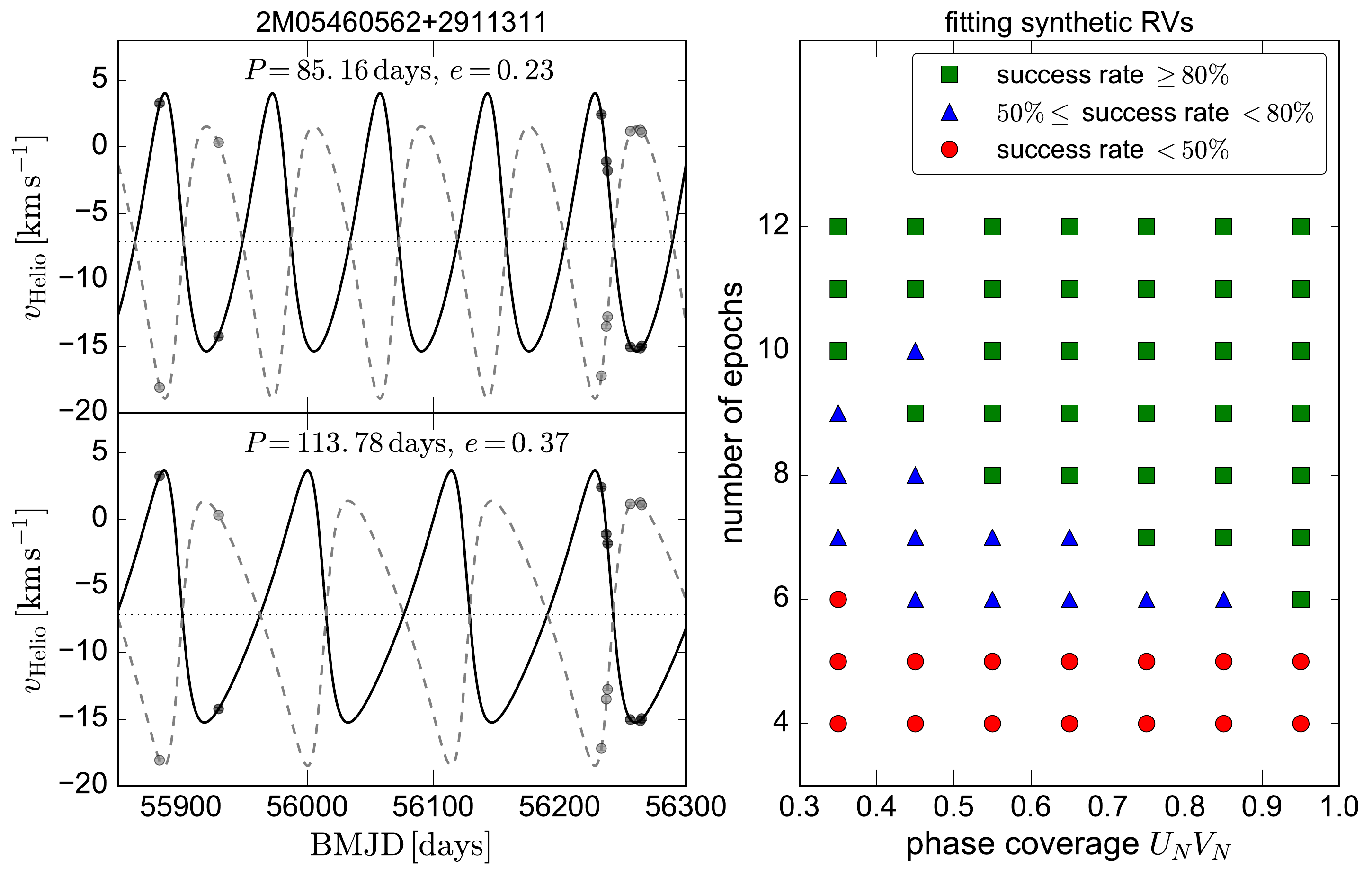}
\caption{\textbf{Left}: Example APOGEE binary system with 8 epochs in which the available velocity data is not sufficient to fully constrain the orbit. Top and bottom panels show two qualitatively different orbital solutions which can both fit the measured velocity data (identical in the two panels) well. Because the measured velocities are sparse, the posterior is multimodal. We do not provide orbital solutions for such systems. \textbf{Right}: Results of fitting synthetic radial velocity data for an SB2 population with realistic orbital parameters and survey cadence. The ``success rate'' indicates the fraction of systems in each cell of $N_{\rm epoch} - U_{N}V_{N}$ space for which our fitting procedure converged on the true orbit from which RV measurements were generated. Orbits can usually be reliably constrained for $N_{\rm epochs} \gtrsim 7$, though systems with poor phase coverage can still converge on incorrect orbital solutions. We only attempt to fit orbits for systems with $N_{\rm epochs} \geq 7$ and $U_{N}V_{N} > 0.5$.}
\label{fig:unconstrained_rv}
\end{figure*}

Sparse radial velocity data can often be well-fit by several families of qualitatively different orbits, particularly when there are few radial velocity measurements and/or phase coverage is poor \citep[e.g.][]{PriceWhelan_2017}. Due to the complex multi-modal structure of the posterior in these cases, standard MCMC techniques fail to fully explore the orbital parameter space in finite time, meaning that there is no guarantee that our orbit-fitting procedure will converge on the true solution or that the orbital parameter uncertainties found by \texttt{emcee} are reliable. In the left panel of Figure~\ref{fig:unconstrained_rv}, we show an example APOGEE binary in which the radial velocity data are not sufficiently constraining to yield an unambiguous orbital solution: (at least) two qualitatively different orbital solutions can fit the measured radial velocities. 

To assess the number of epochs and phase + velocity coverage required for reliable orbit constraints, we generated synthetic radial velocity measurements for a population of synthetic binaries (as described in \citetalias{ElBadry_2017}) with $4 < N_{\rm epochs} < 12$ and phase coverage $0.3 < U_{N}V_{N} < 1$ and fit them using the same procedure described in Section~\ref{sec:orbits}. To ensure realistic survey cadence, we drew observation times from observations of real APOGEE stars. We added Gaussian noise to the synthetic data with $\sigma_{RV} = 0.2\,\rm km\,s^{-1}$, which is typical for our observations. 

The results of this experiment are shown in Figure~\ref{fig:unconstrained_rv}. We label each fit successful if the true orbital parameters fall within the marginalized 90\% credibility regions returned by MCMC fitting. Although it is in principle possible to constrain the orbit of an SB2 with as few as 5 radial velocity epochs,\footnote{Although 7 parameters are required to parameterize a two-body orbit, the system velocity $\gamma$ and mass ratio $q=K_{1}/K_{2}$ can be obtained ``for free:'' if the radial velocities of the primary and secondary fall on a line $v_{{\rm Helio,2}}=\alpha v_{{\rm Helio,1}}+\beta$, the system velocity is $\gamma=\beta/\left(1-\alpha\right)$ and the mass ratio is $q = -1/\alpha$.} reliable constraints for can only be obtained from realistic data with $\gtrsim 7$ epochs. At fixed number of epochs, constraints are more reliable for systems with larger $U_{N}V_{N},$ as expected. Based on the results of this experiment, we only attempt to fit orbits to systems with $N_{\rm epochs} \geq 7$ and $U_{N}V_{N}\geq 0.5$. All but two of the targets for which we provide an orbital solution have $U_{N}V_{N} > 0.6$; we caution that orbital parameters for systems with lower $U_{N}V_{N}$ may be less reliable. 

We stress that the orbit of an SB2 can in general be constrained with fewer radial velocity measurements than that of an SB1. The basic reason for this is that even with only a few epochs, having velocity measurements for both stars pins down the system velocity $\gamma$ exactly. Therefore, many of the families of orbits with different combinations of $P, e$, and $\gamma$ that would be permitted if radial velocity measurements were only available for the primary can be excluded when the secondary is detected. 

Other works have used more conservative limits to determine whether the available velocity data were sufficient to constrain a binary orbit. For example, \citet{Halbwachs_2017} require $N_{\rm epochs} \geq 11$. The true probability of convergence on a local minimum depends on the radial velocity uncertainties, number of epochs, and the uniformity of observational coverage. This experiment indicates that $N_{\rm epochs} \geq 7$ is usually sufficient for radial velocity data similar to what we obtain in this work, but the probability of convergence on an erroneous orbital solution is, of course, lower for systems with a larger number of epochs. 

\section{Data products}
\label{sec:data}
Here we make available the best-fit labels for all targets identified as multiple-star systems. We also provide a list of targets consistent with being single stars in order to make it possible to reconstruct our initial sample of 20,142 targets. 

In Table~\ref{tab:single_stars}, we list all targets consistent with being single stars; labels for these systems will be released by Ting et al. (in prep). 
Labels for targets identified as SB1s are listed in Table~\ref{tab:sb1s}; these are obtained by simultaneously fitting visit spectra. 
Labels for targets classified as binaries are listed in Table~\ref{tab:sb2s}. As described in Section~\ref{sec:fitting}, these are obtained by simultaneously fitting visit spectra for potential close binaries, which are primarily RV-variable systems, and by fitting the combined spectrum for all other systems. For close binaries in which the orbital configuration changes substantially between visits, we list dynamical mass ratios and center-of-mass velocities. 
Labels for targets classified as SB2s in which the gravitational effects of an with an unseen third component can be detected are listed in Table~\ref{tab:sb2_hidden_companion}; these are obtained by simultaneously fitting visit spectra. 
Labels for targets identified as triples in which all three components contribute to the spectrum are listed in Table~\ref{tab:sb3s}; these are obtained by simultaneously fitting visit spectra. 

\begin{table}
\begin{center}
\begin{tabular}{ |c} 
 \hline
APOGEE ID \\ 
\hline 
2M00000233+1452324 \\ 
2M00001701+7052395 \\ 
2M00003475+5723259 \\ 
2M00004578+5654428 \\ 
$\cdots$  \\ 
\hline
\end{tabular}
\end{center}
	\caption{List of targets identified as single stars. This table is available in its entirety (with 16834 rows) in machine-readable form.}
    \label{tab:single_stars}
\end{table}

\begin{table*}
\begin{center}
\begin{tabular}{ |c|c|c|c|c|c|c } 
 \hline
APOGEE ID & $T_{\rm eff}$ & $\log g$ & [Fe/H] & [Mg/Fe] & $v_{\rm macro}$ & $\Delta v_{\rm max}$ \\ 
& [K] & [dex] & [dex] & [dex] &  [$\rm km\,s^{-1}$]&  [$\rm km\,s^{-1}$]\\
\hline 
2M00010204+0049037 & 6008  & 4.26 & -0.41 & 0.00  & 6.48  & 2.93 \\ 
2M00031962-0017109 & 6157  & 4.23 & -0.23 & -0.06 & 42.95 & 65.17\\ 
2M00041803+1519505 & 5941  & 4.92 & 0.15  & 0.17  & 2.21  & 0.78 \\ 
2M00041859+7104111 & 4921  & 4.36 & 0.13  & -0.01 & 5.64  & 20.26\\ 
$\cdots$ & $\cdots$  & $\cdots$ & $\cdots$ & $\cdots$ & $\cdots$ & $\cdots$ \\ 
\hline
\end{tabular}
\end{center}
	\caption{Best-fit labels for targets identified as SB1s. $\Delta v_{\rm max}$ is the maximum change in radial velocity between visits. This table is available in its entirety (with parameters for 663 systems) in machine-readable form.}
    \label{tab:sb1s}
\end{table*}

\begin{table*}
\begin{center}
\begin{tabular}{ |c|c|c|c|c|c|c|c|c|c } 
 \hline
APOGEE ID & $T_{\rm eff}$ & $\log g$ & [Fe/H] & [Mg/Fe] & $q_{\rm spec}$ & $v_{\rm macro,1}$ & $v_{\rm macro,2}$ & $q_{\rm dyn}$ & $\gamma $ \\ 
& [K] & [dex] & [dex] & [dex] &  &  [$\rm km\,s^{-1}$] &  [$\rm km\,s^{-1}$] & &  [$\rm km\,s^{-1}$] \\
\hline 
2M00003968+5722329 & 4517 & 4.62 & 0.05  & -0.04 & 0.88 & 3.09  & 8.19  & ---  & ---\\ 
2M00012717+0128193 & 5048 & 4.51 & 0.07  &  0.06 & 0.75 & 0.00  & 14.57 & ---  & ---\\ 
2M00023179+1521164 & 4589 & 4.37 & -0.20 &  0.10 & 0.99 & 19.29 & 16.15 & 0.84 & -0.92 \\ 
2M00024073+6354560 & 5664 & 4.26 & 0.18  & -0.01 & 0.67 & 1.42  & 3.34  & ---  & ---\\ 
$\cdots$ & $\cdots$  & $\cdots$ & $\cdots$ & $\cdots$ & $\cdots$ & $\cdots$ & $\cdots$ & $\cdots$ & $\cdots$ \\ 
\hline
\end{tabular}
\end{center}
	\caption{Best-fit labels for targets identified as binaries in which both components contribute to the spectrum. $T_{\rm eff}$ and $\log g$ refer to the primary. For the 623 targets in which the orbital configuration changes substantially between visits, we provide the dynamical mass ratios, $q_{\rm dyn}$, and center-of-mass velocity, $\gamma$. This table is available in its entirety (with parameters for 2423 systems) in machine-readable form.}
    \label{tab:sb2s}
\end{table*}

\begin{table*}
\begin{center}
\begin{tabular}{ |c|c|c|c|c|c|c|c } 
 \hline
APOGEE ID & $T_{\rm eff}$ & $\log g$ & [Fe/H] & [Mg/Fe] & $q_{\rm spec}$ & $v_{\rm macro,1}$ & $v_{\rm macro,2}$  \\ 
& [K] & [dex] & [dex] & [dex] &  &  [$\rm km\,s^{-1}$] &  [$\rm km\,s^{-1}$]  \\
\hline 
2M00103470+0043200 & 4200 & 4.50 & -0.21 &  0.21  & 1.00 & 19.98 & 5.40 \\ 
2M00265252+6359169 & 6416 & 4.56 & -0.04 &  -0.29 &  0.85& 10.34 & 14.38\\ 
2M00310678+8508494 & 5742 & 4.44 & 0.01  & -0.07  & 0.62 & 4.91  & 3.11 \\ 
2M01194897+8532293 & 5272 & 4.54 & -0.08 &  0.03  & 0.90 & 2.68  & 4.80 \\ 
$\cdots$ & $\cdots$  & $\cdots$ & $\cdots$ & $\cdots$ & $\cdots$ & $\cdots$ & $\cdots$  \\ 
\hline
\end{tabular}
\end{center}
	\caption{Best-fit labels for targets in which two components contribute to the spectrum but the gravitational effects a third components can be detected (e.g. Figure~\ref{fig:bin_dv}). $T_{\rm eff}$ and $\log g$ refer to the primary. This table is available in its entirety (with parameters for 108 systems) in machine-readable form.}
    \label{tab:sb2_hidden_companion}
\end{table*}

\begin{table*}
\begin{center}
\begin{tabular}{ |c|c|c|c|c|c|c|c|c|c  } 
 \hline
APOGEE ID & $T_{\rm eff}$ & $\log g$ & [Fe/H] & [Mg/Fe] & $q_2$ & $q_3$ & $v_{\rm macro,1}$ & $v_{\rm macro,2}$ & $v_{\rm macro,3}$  \\ 
& [K] & [dex] & [dex] & [dex] &  &  & [$\rm km\,s^{-1}$] &  [$\rm km\,s^{-1}$] &  [$\rm km\,s^{-1}$]  \\
\hline 
2M00182859+6207248 & 5398 & 4.33 & 0.22  & -0.20 & 1.00 &  1.00 & 8.69  & 4.33  & 0.19\\ 
2M00285967+5931138 & 4655 & 4.20 & -0.22 &  0.08 & 0.94 &  0.94 & 1.39  & 30.07 & 36.15\\ 
2M00470197+1751448 & 6386 & 4.07 & -0.47 &  0.03 & 0.89 &  0.88 & 19.03 &  3.50 & 4.46\\ 
2M01103850+6655525 & 4756 & 4.60 & 0.07  & -0.00 & 0.92 &  0.81 & 3.71  & 1.34  & 16.35\\ 
$\cdots$ & $\cdots$  & $\cdots$ & $\cdots$ & $\cdots$ & $\cdots$ & $\cdots$ & $\cdots$ & $\cdots$ & $\cdots$  \\ 
\hline
\end{tabular}
\end{center}
	\caption{Best-fit labels for SB3s, targets in which three components contribute to the spectrum (e.g. Figure~\ref{fig:triple}). $T_{\rm eff}$ and $\log g$ refer to the primary. This table is available in its entirety (with parameters for 114 systems) in machine-readable form.}
    \label{tab:sb3s}
\end{table*}


\bsp	
\label{lastpage}
\end{document}